\documentclass[prd,showpacs, %twocolumn,
nofootinbib, preprintnumbers,floatfix]{revtex4}%

\usepackage{amssymb,hyperref,amsmath}
\usepackage[dvips]{color}
\usepackage[dvips]{graphicx}
\usepackage{epsfig}

\unitlength=1.2mm \preprint{JLAB-THY-08-928}
\begin{document}
\newcommand{\tr}{\mbox{tr}\,}
\newcommand{\Dslash}{{\mathchoice
    {\Dslsh \displaystyle}%
    {\Dslsh \textstyle}%
    {\Dslsh \scriptstyle}%
    {\Dslsh \scriptscriptstyle}}}
\newcommand{\Dslsh}[1]{\ooalign{\(\hfill#1/\hfill\)\crcr\(#1D\)}}
\newcommand{\leftvec}[1]{\vect \leftarrow #1 \,}
\newcommand{\rightvec}[1]{\vect \rightarrow #1 \:}
\renewcommand{\vec}[1]{\vect \rightarrow #1 \:}
\newcommand{\vect}[3]{{\mathchoice
    {\vecto \displaystyle \scriptstyle #1 #2 #3}%
    {\vecto \textstyle \scriptstyle #1 #2 #3}%
    {\vecto \scriptstyle \scriptscriptstyle #1 #2 #3}%
    {\vecto \scriptscriptstyle \scriptscriptstyle #1 #2 #3}}}
\newcommand{\vecto}[5]{\!\stackrel{{}_{{}_{#5#2#3}}}{#1#4}\!}
\newcommand{\vdot}{\!\cdot\!}
\newcommand{\ignore}[1]{}

\bibliographystyle{apsrev}

\title{Strange Baryon Electromagnetic Form Factors and SU(3) Flavor Symmetry Breaking
%in dynamical sea
}

\author{Huey-Wen Lin}
\email{hwlin@jlab.org} \affiliation{Thomas Jefferson National Accelerator Facility, Newport News, VA 23606}
\author{ Kostas Orginos}
\email{kostas@wm.edu} \affiliation{Department of Physics College of William \\ Williamsburg, VA 23187-8795}\affiliation{Thomas Jefferson National Accelerator Facility, Newport News, VA 23606}
%\date{\today}
\date{Dec. 26, 2008}

\begin{abstract}
We study the nucleon, Sigma and cascade octet baryon electromagnetic form factors and the effects of SU(3) flavor symmetry breaking from 2+1-flavor lattice calculations. We find that electric and magnetic radii are similar; the maximum discrepancy is about 10\%. In the pion-mass region we explore, both the quark-component and full-baryon moments have small SU(3) symmetry breaking. We extrapolate the charge radii and the magnetic moments using three-flavor heavy-baryon chiral perturbation theory (HBXPT).  The systematic errors due to chiral and continuum extrapolations
remain significant, giving rise to charge radii for $p$ and $\Sigma^-$ that are 3--4 standard deviations away from the known experimental ones. Within these systematics  the predicted  $\Sigma^+$ and $\Xi^-$ radii are 0.67(5) and 0.306(15)~fm$^2$ respectively. When the next-to-next-to-leading order of HBXPT is included,
the extrapolated magnetic moments are less than 3 standard deviations away from PDG values, and the discrepancy is possibly due to remaining chiral and continuum extrapolation errors.
\end{abstract}

%%%%%%%%%%%%%%%%%%%%%%%%%%%%%%%%%%%%%%%%%%%%%%%%%%%%%%%%%%%%%%%%%%%%%%%%%%%%%%%%%%%%%%
\maketitle

\section{Introduction}

The study of hadron electromagnetic form factors reveals information
important to our understanding of hadronic structure. However,
experimental measurements on baryons with
strange quarks, such as hyperons,  are difficult due to their unstable nature. From the
theoretical perspective, studies in the nonperturbative regime of
quantum chromodynamics (QCD) have been difficult without resorting to
model-dependent calculations or making approximations. In lattice QCD,
we are able to compute the path integral directly via numerical
integration, providing a first-principles calculation of the
consequences of QCD.  Such study of hadronic form
factors together with experiment will serve as valuable theoretical
input for understanding hadronic structure.

The nucleon form factors have been calculated on the lattice by many
groups, and calculations are still
ongoing\cite{Liu:1994dr,Gockeler:2003ay,Alexandrou:2006ru,Orginos:2006zz,Hagler:2007xi,Alexandrou:2007xj,Gockeler:2007hj,Hagler:2007hu,Sasaki:2007gw,Lin:2008uz,Zanotti:2008zm}. However,
few have devoted effort to calculating the form factors for other
members of the octet. The JLab/Adelaide
group\cite{Wang:2008vb,Boinepalli:2006xd,Leinweber:1990dv} performed form factor
calculations for the entire octet, and QCDSF collaboration studied the
spin structure of the Lambda\cite{Gockeler:2002uh}. Both of these
calculations used the quenched approximation, where fermion vacuum-polarization loop contributions are ignored; this introduces an uncontrollable
systematic error into their calculations. We recently performed a
first lattice calculation of the hyperon axial coupling
constants\cite{Lin:2007ap} with dynamical fermions, where the SU(3)
flavor symmetry breaking was investigated and found to be
non-negligible.

The assumption of SU(3) flavor symmetry has been common in studies
involving baryonic observables. For example, in the determination of
$V_{us}$ from hyperon decays reported in the PDG\cite{PDBook} (and
thus in much experimental work following from it), exact SU(3) is
assumed in the $g_1/f_1$ entry. Symmetry-breaking effects in the axial
couplings could therefore impact the world-average $V_{us}$ value. The SU(3) breaking of baryon masses is
relatively small, as in the Gell-Mann--Okubo
relation\cite{GellMann:1962xb,Okubo:1961jc} (which has also been
studied in a full-QCD lattice calculation\cite{Beane:2006pt}) or the
decuplet equal-spacing relation. However, this small breaking is in contrast to
other quantities with substantial breaking, such as the magnetic
moments (as suggested by Coleman and Glashow\cite{Coleman:1961jn}) or
the axial coupling constants\cite{Lin:2007ap,Lin:2008rb}. In lattice QCD
calculations, we can vary the up/down and strange quark masses
%in the calculation
to move away from the SU(3)-symmetric point and explicitly
observe flavor-breaking effects.

In this work, we concentrate on the electromagnetic properties of
octet baryons. This allows us to study SU(3) flavor symmetry within
the octet family. The structure of this paper is as follows: In
Sec.~\ref{sec:LatticeSetup}, we define the operators used for this
calculation and detail how we extract the electromagnetic form factors
from lattice calculations. In Sec.~\ref{sec:Numerical}, we discuss the
momentum dependence of the form factors and examine the validity of
the dipole extrapolations that are commonly used on lattice data. We
also extract the electric charge radii, magnetic radii and magnetic
moments for the nucleon, Sigma and cascade baryons and discuss the
SU(3) flavor-breaking of these quantities. Our conclusions are
presented in Sec.~\ref{sec:conclusion}, and some detailed numbers
obtained from this calculation are listed in the appendix.

\begin{table}
\begin{center}
\begin{tabular}{c|ccccc}
\hline\hline
 	               & m010      &  m020    &  m030     & m040      & m050     \\
\hline
$N_{\rm prop}$   &   612     &   345    &   561     &   320     &   342    \\
$m_\pi$ (MeV)    & 354.2(8)  & 493.6(6) & 594.2(8)  & 685.4(19) & 754.3(16)\\
%$m_\pi/f_\pi$    & 2.316(7)  & 3.035(7) & 3.478(8)  & 3.822(23) & 4.136(20)\\
%$m_K/f_\pi$ 	    & 3.951(14) & 3.969(10)& 4.018(11) & 4.060(26) & 4.107(21)\\
$M_N$ (GeV)      & 1.15(3)   & 1.29(2)  & 1.366(18) & 1.490(14) & 1.558(8) \\
$M_\Sigma$ (GeV) & 1.349(19) & 1.41(2)  & 1.448(12) & 1.524(14) & 1.558(8) \\
$M_\Xi$ (GeV)    & 1.438(11) & 1.475(16)& 1.491(10) & 1.546(12) & 1.558(8) \\
\hline\hline
\end{tabular}
\end{center}
\caption{Quantities associated with the gauge ensembles used in our calculation}\label{tab:numbers}
\end{table}

\ignore{We use 2+1 flavors of improved staggered fermions
(asqtad)\cite{Kogut:1975ag,Orginos:1998ue,Orginos:1999cr} for the
expensive sea quarks (in configuration ensembles generated by the MILC
collaboration\cite{Bernard:2001av}), and domain-wall fermions
(DWF)\cite{Kaplan:1992bt,Kaplan:1992sg,Shamir:1993zy,Furman:1994ky}
for the valence sector. The pion mass ranges from around 350 to
750~MeV in a lattice box of size 2.6~fm. The gauge fields entering the
DWF action are HYP~\cite{HASENFRATZ,LHPC,DURR} smeared to reduce the explicit chiral
symmetry breaking on the lattice, and the baryon interpolating fields use
 gauge invariant Gaussian smeared quark sources  to improve the signal. The source-sink separation is fixed at
10 time units. The number of configurations used from each ensemble
ranges from 350 to 700; for more details, see Table~\ref{tab:numbers}
(or Table~1 in Ref.~\cite{Lin:2007ap}).}

\section{Lattice Setup}\label{sec:LatticeSetup}

For this calculation we use 2+1 flavors of improved staggered fermions
(asqtad)\cite{Kogut:1975ag,Orginos:1998ue,Orginos:1999cr} for the
expensive sea quarks (in configuration ensembles generated by the MILC
collaboration\cite{Bernard:2001av}), and domain-wall fermions
(DWF)\cite{Kaplan:1992bt,Kaplan:1992sg,Shamir:1993zy,Furman:1994ky}
for the valence sector. The pion mass ranges from around 350 to
750~MeV in a lattice box of size 2.6~fm. The gauge fields entering the
DWF action are HYP\cite{Hasenfratz:2001hp,WalkerLoud:2008bp,Durr:2004ta} smeared to reduce the residual chiral symmetry breaking on the lattice, and the baryon
interpolating fields use gauge-invariant Gaussian smeared quark
sources to improve the signal. The source-sink separation is fixed at
10 time units. The number of configurations used from each ensemble
ranges from 350 to 700; for more details, see Table~\ref{tab:numbers}
(or Table~1 in Ref.~\cite{Lin:2007ap}).

The interpolating fields we use for the nucleon, Sigma and cascade octet baryons have the general form
\begin{eqnarray}\label{eq:lat_B-op}
 \chi^B (x) &=&  \epsilon^{abc} [{\phi}_1^{a\mathrm{T}}(x)C\gamma_5\phi_2^b(x)]\phi_1^c(x),
\end{eqnarray}
where $C$ is the charge conjugation matrix, and $\phi_1$ and $\phi_2$ are any of the quarks $\{u,d,s\}$. For example, to create a proton, we want $\phi_1=u$ and $\phi_2=d$; for the $\Xi^{-}$, $\phi_1=s$ and $\phi_2=d$.

The electromagnetic form factors of an octet baryon $B$ can be written as
\begin{eqnarray}\label{eq:lat_ME}
 \langle B(p^\prime)\left|V_{\mu}(q)\right|B(p)\rangle(q)
&=& {\overline u}_{B}(p^\prime)\left[\gamma_{\mu} F_1(q^2) +
\sigma_{\mu \nu}q_{\nu} \frac{F_2(q^2)}{2M_{B}}
\right]u_{B}(p) e^{-iq\cdot x}
\end{eqnarray}
from Lorentz symmetry and vector-current conservation. $F_1$ and $F_2$ are the Dirac and Pauli form factors. On the lattice, we calculate the quark-component inserted current,  $V_{\mu}=\overline{\phi}\gamma_\mu \phi$, with $\phi=u,d$ light-quark current and $\phi=s$ strange-quark vector current. Due to the small chiral symmetry breaking of DWF at finite lattice spacing (on the order of $\left(m_{\rm res} a\right)^2 \ll %{1\mbox{--}2} <<
1\%$\cite{WalkerLoud:2008bp}), the $O(a)$ terms for the vector current are highly suppressed. By calculating two-point and three-point correlators on the lattice, we will be able to extract the form factors from Eq.~\ref{eq:lat_ME}.

The octet two-point correlators measured on the lattice are
\begin{eqnarray}\label{eq:general_2pt}
\Gamma^{(2),T}_{AC}(t_i,t;\vec{p})&=& \left\langle \tr\left(T  (\chi^B_C) (\chi^B_A)^\dagger   \right) \right\rangle
= \sum_n \frac{E_n(\vec{p})+M_{n}}{2E_n(\vec{p})} Z_{n,A}
Z_{n,C} e^{-E_n(\vec{p})(t-t_i)},
\end{eqnarray}
where $A$ and $C$ indicate the smearing parameters, $\langle \cdots \rangle$ indicates the ensemble average, and $E_n(\vec{p})$ is $\sqrt{M_n^2+p^2}$. $\chi^B$ (with $B\in\{N,\Sigma,\Xi\}$) is a baryon interpolating field.
The states in Eq.~\ref{eq:general_2pt} are defined to be normalized as
\begin{eqnarray}
\langle 0 |(\chi^B_C)^\dagger|p,s\rangle_n &=& {Z_{n,C}} u_s(\vec p),
\end{eqnarray}
where the spinors satisfy
\begin{eqnarray}
\sum_{s} u_s(\vec p)\bar u_s(\vec p)&=& \frac{E(\vec p)\gamma^t
-i\vec\gamma\cdot \vec p+ M_B}{2 E(\vec p)}. \label{eq:spinor}
\end{eqnarray}
The projection used is $T=\frac{1}{4}(1+\gamma_4)(1+i\gamma_5\gamma_3)$. Since we are only interested in the ground state of each baryon, we tune the smearing parameters and the source-sink separation so that only the ground-state signal remains relevant at larger $t$. Therefore, $n=0$ in Eq.~\ref{eq:general_2pt}.

\begin{figure}
\includegraphics[width=0.32\textwidth]{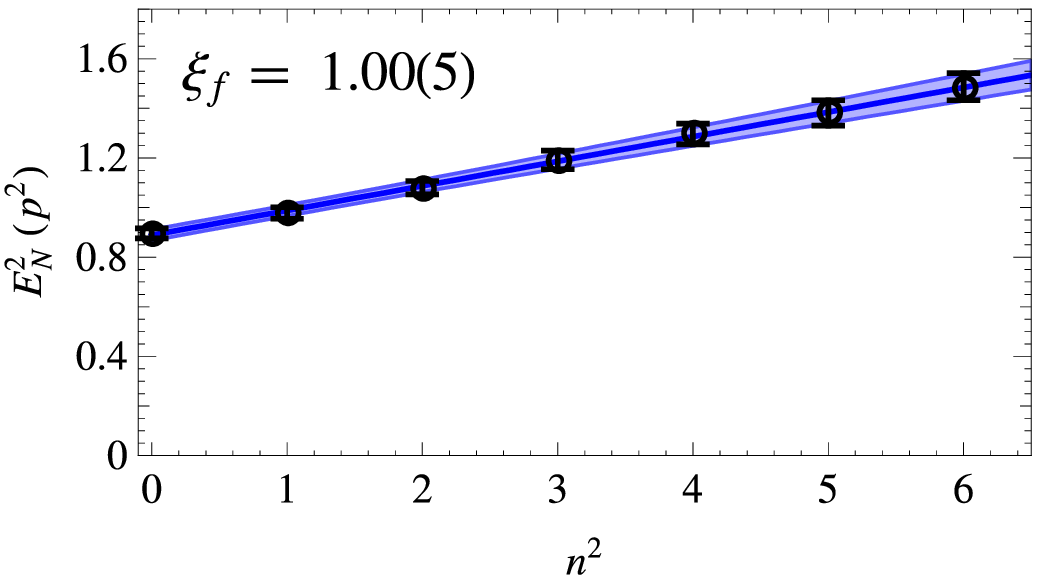}
\includegraphics[width=0.32\textwidth]{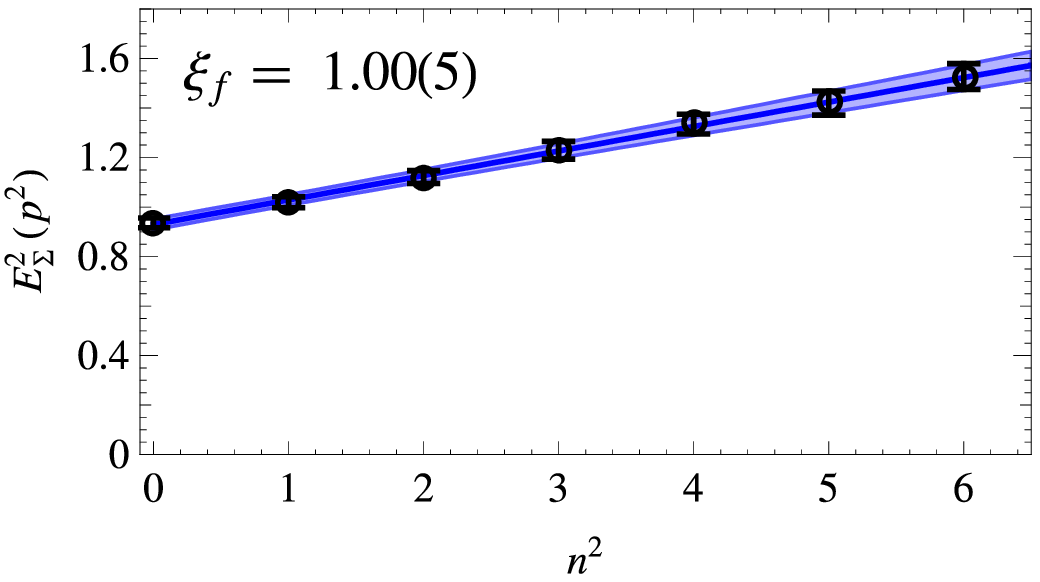}
\includegraphics[width=0.32\textwidth]{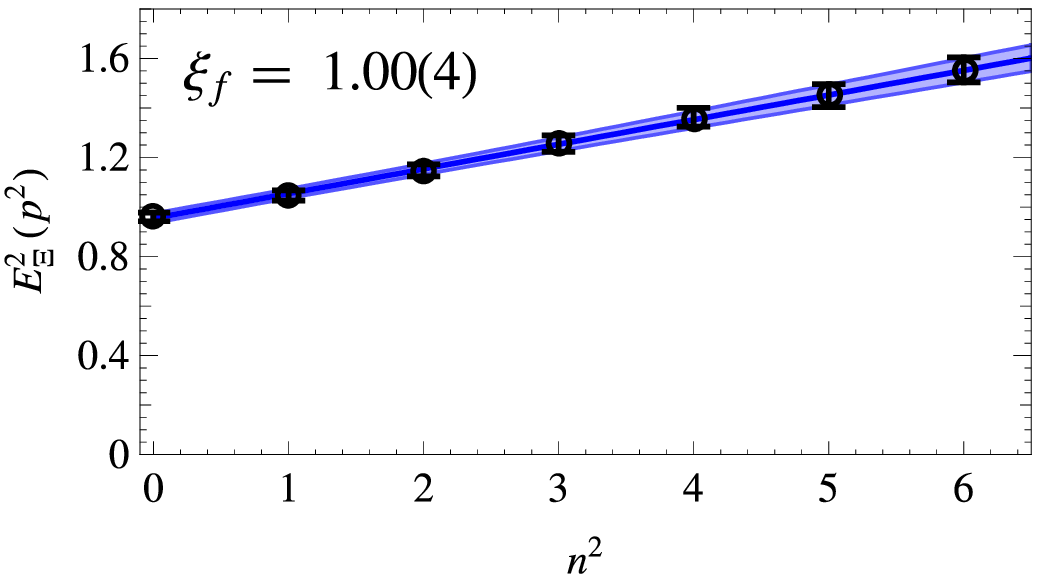}
\caption{The dispersion relation of the $N$ (left), $\Sigma$ (middle) and $\Xi$ (right)
on ensemble ``m040''. The x-axis is in units of $\left(\frac{2\pi}{L}a^{-1}\right)^2$.
}\label{fig:m040_dispersion}
\end{figure}

The energy of each baryon is measured using a single-exponential fit to the larger-time data. The masses are listed in Table~\ref{tab:numbers}. We measure baryons with these momenta:
\begin{eqnarray}\label{eq:n-list}
\vec p_i &=& \frac{2\pi}{L}a^{-1} \vec n \\
\vec n &\in& \left\{
\left( \begin{array}{c}
0\\0\\0\\
\end{array}
\right),
\left( \begin{array}{c}
1\\0\\0\\
\end{array}
\right),
 \left(
\begin{array}{c}
1\\1\\0\\
\end{array}
\right), \left(
\begin{array}{c}
1\\1\\1\\
\end{array}
\right), \left(
\begin{array}{c}
2\\0\\0\\
\end{array}
\right), \left(
\begin{array}{c}
2\\1\\0\\
\end{array}
\right), \left(
\begin{array}{c}
2\\1\\1\\
\end{array}
\right) \right\},
\end{eqnarray}
and their rotational equivalents. We check the dispersion relation $E(\vec{p}) = \sqrt{M_n^2+p^2/\xi_f^2}$ on the ensemble ``m040'' for each of the octet baryons with our lattice data and find that $\xi_f$ is consistent with 1; see Figure~\ref{fig:m040_dispersion}.

Similarly to the two-point function, the lattice three-point function is
\begin{eqnarray}\label{eq:general_3pt}
\Gamma^{(3),T}_{\mu,AC}(t_i,t,t_f;\vec{p}_i,\vec{p}_f) &=&
 \langle\tr \left(T  (\chi^B_C) V_\mu (\chi^B_A)^\dagger  \right) \rangle =
\sum_n \sum_{n^\prime} \frac{Z_{n^\prime,A}(\vec{p}_f) Z_{n,C}(\vec{p}_i)}{Z_V}
 \nonumber \\
\,\,\,&& \times
     e^{-(t_f-t)E_n^\prime(\vec{p}_f)}e^{-(t-t_i)E_n(\vec{p}_i)}
 \times \langle  B |V_\mu |B^\prime \rangle,
\end{eqnarray}
where $n$ and $n^\prime$ index energy states and $Z_V$ is the vector current renormalization constant, which we will set to its nonperturbative value\cite{WalkerLoud:2008bp}. Again, we are only interested in the ground state of each baryon, so $n=n^\prime=0$ in Eqs.~\ref{eq:general_2pt} and
\ref{eq:general_3pt}.

\begin{figure}
\includegraphics[width=0.46\textwidth]{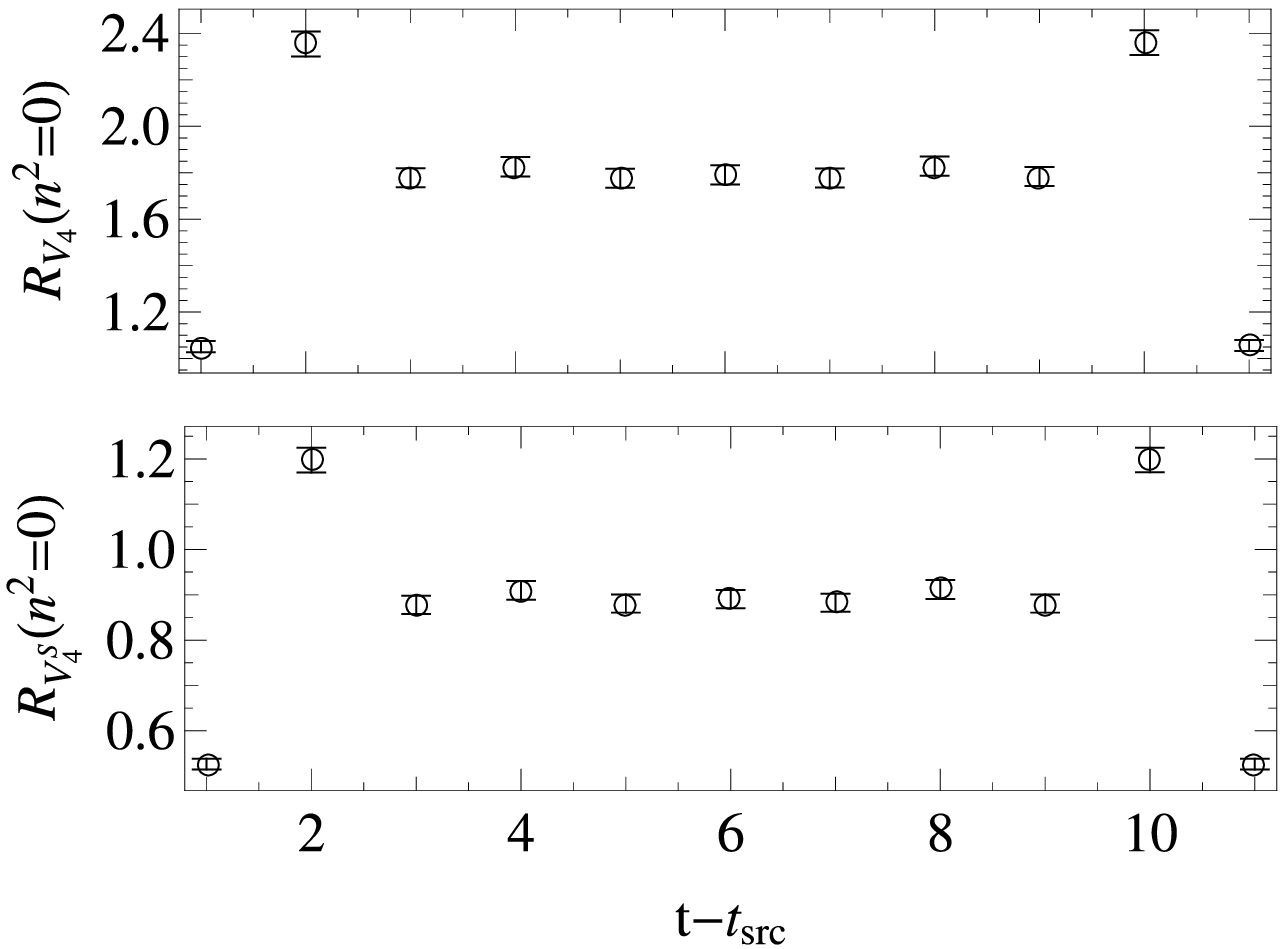}
\includegraphics[width=0.46\textwidth]{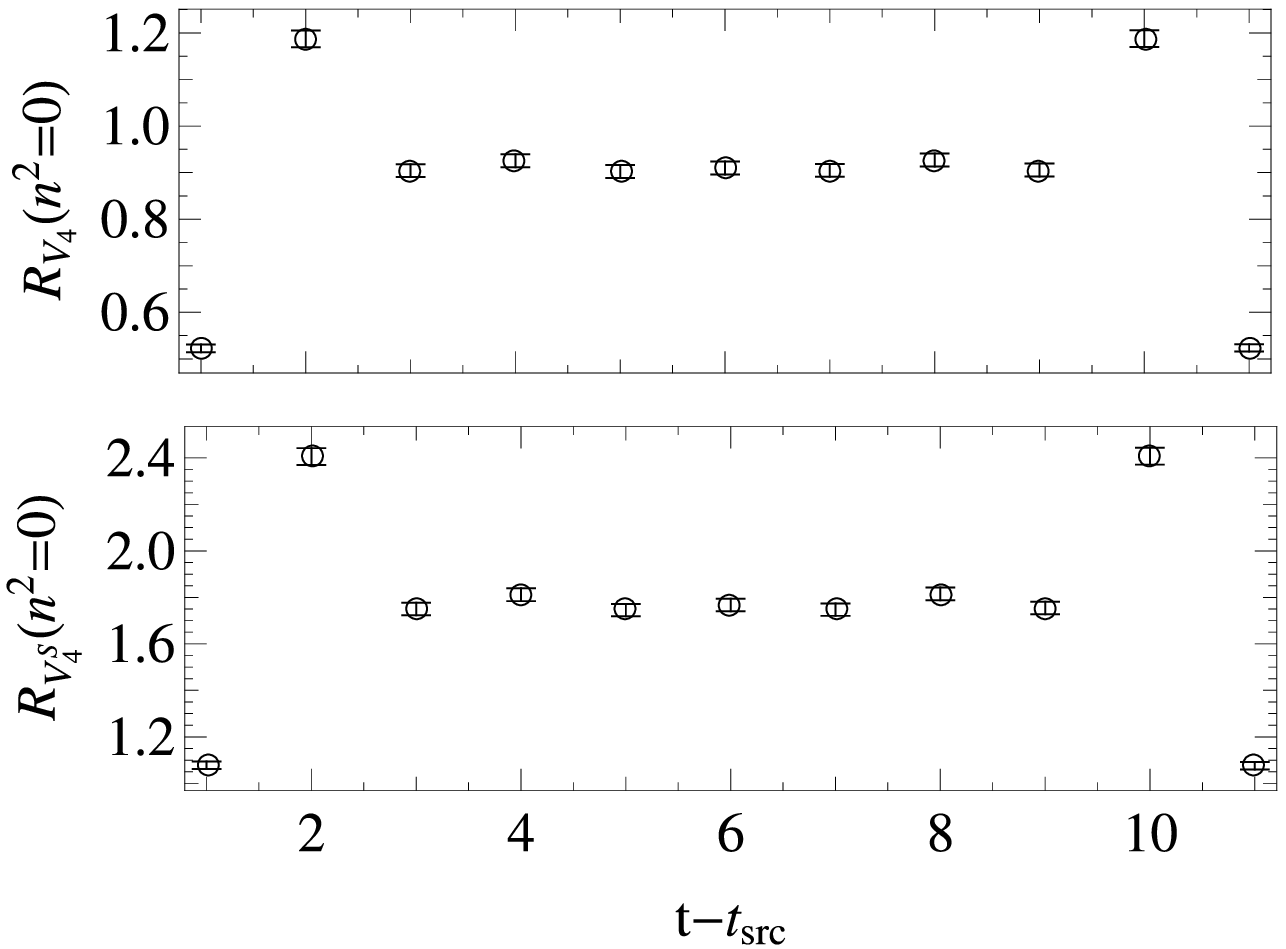}
\caption{$R_{V_\mu}$ from Eq.~\ref{eq:Ratio_GP} for $\langle \Sigma|\bar{\phi}\gamma_4 \phi |\Sigma\rangle$ (left) and $\langle \Xi|\bar{\phi}\gamma_4 \phi |\Xi\rangle$ (right) with $\phi=u/d$ (upper panels) and $\phi=s$ (lower panels) for our lightest-pion ensemble.
}\label{fig:m010-RV4}
\end{figure}

If we denote smearing parameters as $A$, $C$, ..., $I$, we have the freedom to construct a ratio of three- and two-point functions:
\begin{eqnarray}\label{eq:Ratio_Gen}
R_{V_\mu} &=& \frac{Z_V
      \Gamma^{(3),T}_{\mu,AI}(t_i,t,t_f;\vec p_i,\vec p_f)}{\Gamma^{(2),T}_{IC}(t_i,t_f;\vec p_f)}
%      \nonumber \\&\times&
      \sqrt{\frac{\Gamma^{(2),T}_{DE}(t,t_f;\vec p_i)}{\Gamma^{(2),T}_{FH}(t,t_f;\vec
      p_f)}}
      \nonumber \\&\times&
      \sqrt{\frac{\Gamma^{(2),T}_{CC}(t_i,t;\vec p_f)}{\Gamma^{(2),T}_{AA}(t_i,t;\vec
      p_i)}}
%        \nonumber \\ &\times&
      \sqrt{\frac{\Gamma^{(2),T}_{FH}(t_i,t_f;\vec p_f)}{\Gamma^{(2),T}_{DE}(t_i,t_f;\vec
      p_i)}},
\end{eqnarray}
where all the $Z(\vec{p})$ factors are exactly canceled, as well as the remaining time dependence. In this work, we choose $D=F=P$, where $P$ denotes a point source and choose the rest of the smearing parameters to be $G$, denoting Gaussian smearing. Thus,
\begin{eqnarray}\label{eq:Ratio_GP}
R_{V_\mu} &=& \frac{Z_V
      \Gamma^{(3),T}_{\mu,GG}(t_i,t,t_f;\vec p_i,\vec p_f)}{\Gamma^{(2),T}_{GG}(t_i,t_f;\vec p_f)}
%      \nonumber \\     &\times&
 \sqrt{\frac{\Gamma^{(2),T}_{PG}(t,t_f;\vec p_i)}{\Gamma^{(2),T}_{PG}(t,t_f;\vec
      p_f)}}\nonumber \\
 &\times&     \sqrt{\frac{\Gamma^{(2),T}_{GG}(t_i,t;\vec p_f)}{\Gamma^{(2),T}_{GG}(t_i,t;\vec p_i)}}
%        \nonumber \\ &\times&
   \sqrt{\frac{\Gamma^{(2),T}_{PG}(t_i,t_f; \vec p_f)}{\Gamma^{(2),T}_{PG}(t_i,t_f;\vec p_i)}},
\end{eqnarray}
Figure~\ref{fig:m010-RV4} shows data from our lightest-pion ensemble for the $\Sigma$ and $\Xi$ (with at-rest initial and final baryon states) for the strange- and light-quark vector currents. After a few early time slices, the excited states start to die out and leave the data relatively time-independent.

\begin{figure}
\includegraphics[width=0.43\textwidth]{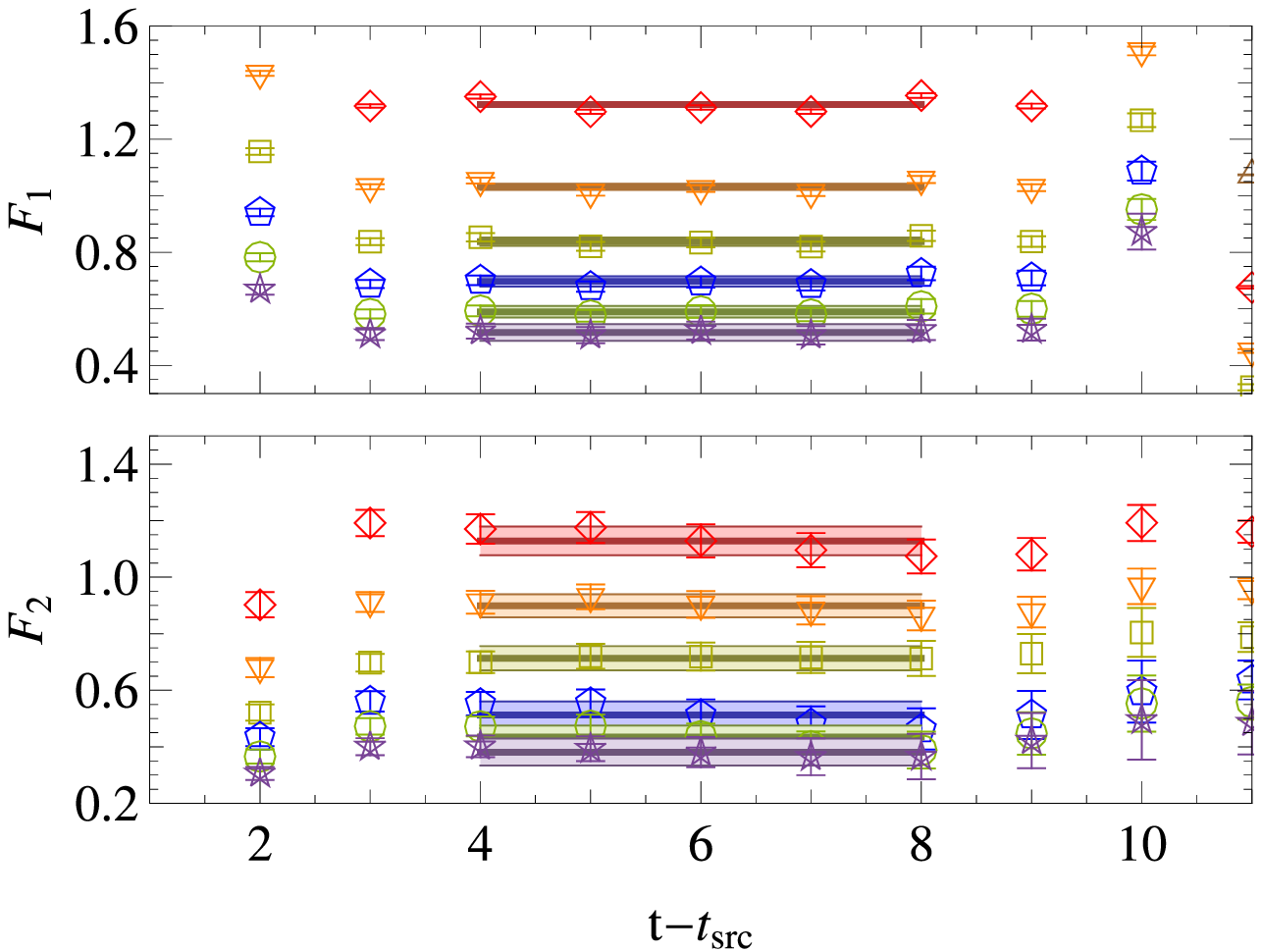}
\includegraphics[width=0.43\textwidth]{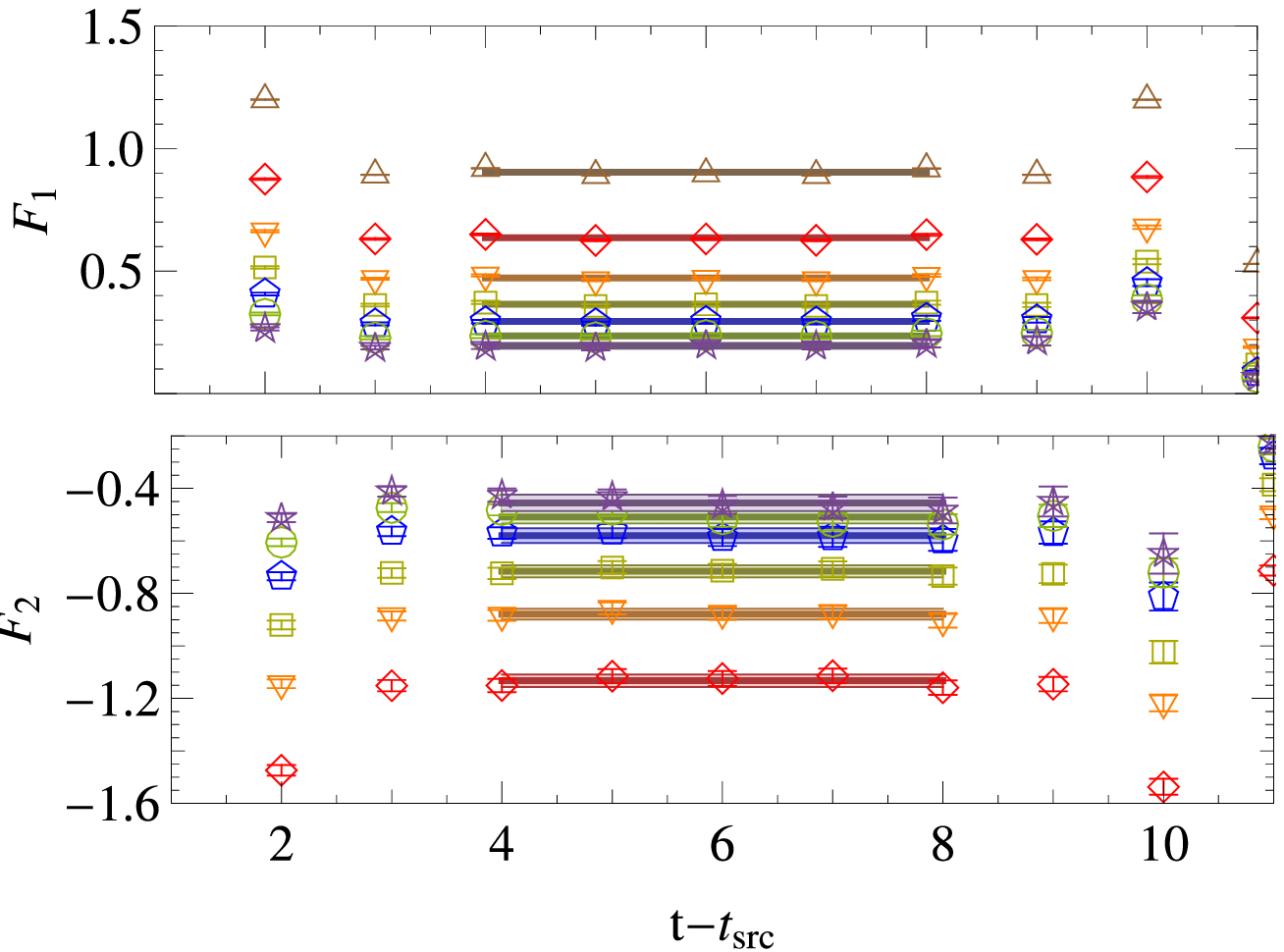}
%\vspace{-0.3in}
\caption{The light-quark vector current inserted $\Sigma$ (left) and $\Xi$ (right) form factor
$F_{1,2}$ for all momenta $\vec{q}$ at $m_\pi\approx600$~MeV. Different symbols represent different transfer momenta: triangles ($n^2=0$), diamonds ($n^2=1$), reverse triangles ($n^2=2$), squares ($n^2=3$), pentagons ($n^2=4$), circles ($n^2=5$), stars ($n^2=6$).
}\label{fig:m030-F12Plau}
\end{figure}

Throughout this work, we fix the sink momentum to $\vec p_f=(0,0,0)$ and vary the initial momentum amongst the list in Eq.~\ref{eq:n-list}. The form factors $F_{1,2}$ in Eq.~\ref{eq:lat_ME} are connected to $R_{V_\mu}$ (Eq.~\ref{eq:Ratio_Gen}) through
\begin{eqnarray}\label{eq:FF-solving}
\left( \begin{array}{c}
 R_{V_1} \\ R_{V_2} \\ R_{V_3} \\R_{V_4}
\end{array} \right)
&=& \left(
\begin{array}{ll}
 \frac{2 M_B p_y-2 i M_B p_x}{2 \sqrt{2} M_B \sqrt{E(M_B,\vec{p}_i )} \sqrt{M_B+E(M_B,\vec{p}_i )}}
 & \frac{-i M_B p_x+i E(M_B,\vec{p}_i ) p_x+2 M_B p_y}{2 \sqrt{2} M_B \sqrt{E(M_B,\vec{p}_i )} \sqrt{M_B+E(M_B,\vec{p}_i )}} \\
 \frac{-2 M_B p_x-2 i M_B p_y}{2 \sqrt{2} M_B \sqrt{E(M_B,\vec{p}_i )} \sqrt{M_B+E(M_B,\vec{p}_i )}}
 & \frac{-2 M_B p_x-i M_B p_y+i p_y E(M_B,\vec{p}_i )}{2 \sqrt{2} M_B \sqrt{E(M_B,\vec{p}_i )} \sqrt{M_B+E(M_B,\vec{p}_i )}} \\
 -\frac{i p_z}{\sqrt{2} \sqrt{E(M_B,\vec{p}_i )} \sqrt{M_B+E(M_B,\vec{p}_i )}}
 & -\frac{i p_z (M_B-E(M_B,\vec{p}_i ))}{2 \sqrt{2} M_B \sqrt{E(M_B,\vec{p}_i )} \sqrt{M_B+E(M_B,\vec{p}_i )}} \\
 \frac{\sqrt{M_B+E(M_B,\vec{p}_i )}}{\sqrt{2} \sqrt{E(M_B,\vec{p}_i )}}
 & \frac{(M_B-E(M_B,\vec{p}_i )) \sqrt{M_B+E(M_B,\vec{p}_i )}}{2 \sqrt{2} M_B \sqrt{E(M_B,\vec{p}_i )}}
\end{array}
\right)
\cdot
\left(\begin{array}{l}F_1\\ F_2 \end{array}\right).
\end{eqnarray}
We solve for $F_{1,2}$ using singular value decomposition (SVD) at each time slice from source to sink with data from all momenta with the same $q^2$ and all $\mu$. (Tables~\ref{tab:NuclFsM1}--\ref{tab:XiFsM4} in the appendix summarize all the results.) For example, Figure~\ref{fig:m030-F12Plau} shows the light vector current inserted $\Sigma$ and $\Xi$ form factor data obtained from Eq.~\ref{eq:FF-solving}. The final form factors are obtained from a fit to the plateau.

Another common set of form factor definitions, widely used in experiments, are the Sachs form factors; these can be related to the Dirac and Pauli form factors through
\begin{eqnarray}
G_E (q^2) &=& F_1(q^2) - \frac{q^2}{4M_B^2}F_2(q^2) \\
G_M (q^2) &=& F_1(q^2) + F_2(q^2).
\end{eqnarray}
In this work, we only calculate the ``connected'' diagram, which means the inserted quark current is contracted with the valence quarks in the baryon interpolating fields.

\begin{figure}
\includegraphics[width=0.43\textwidth]{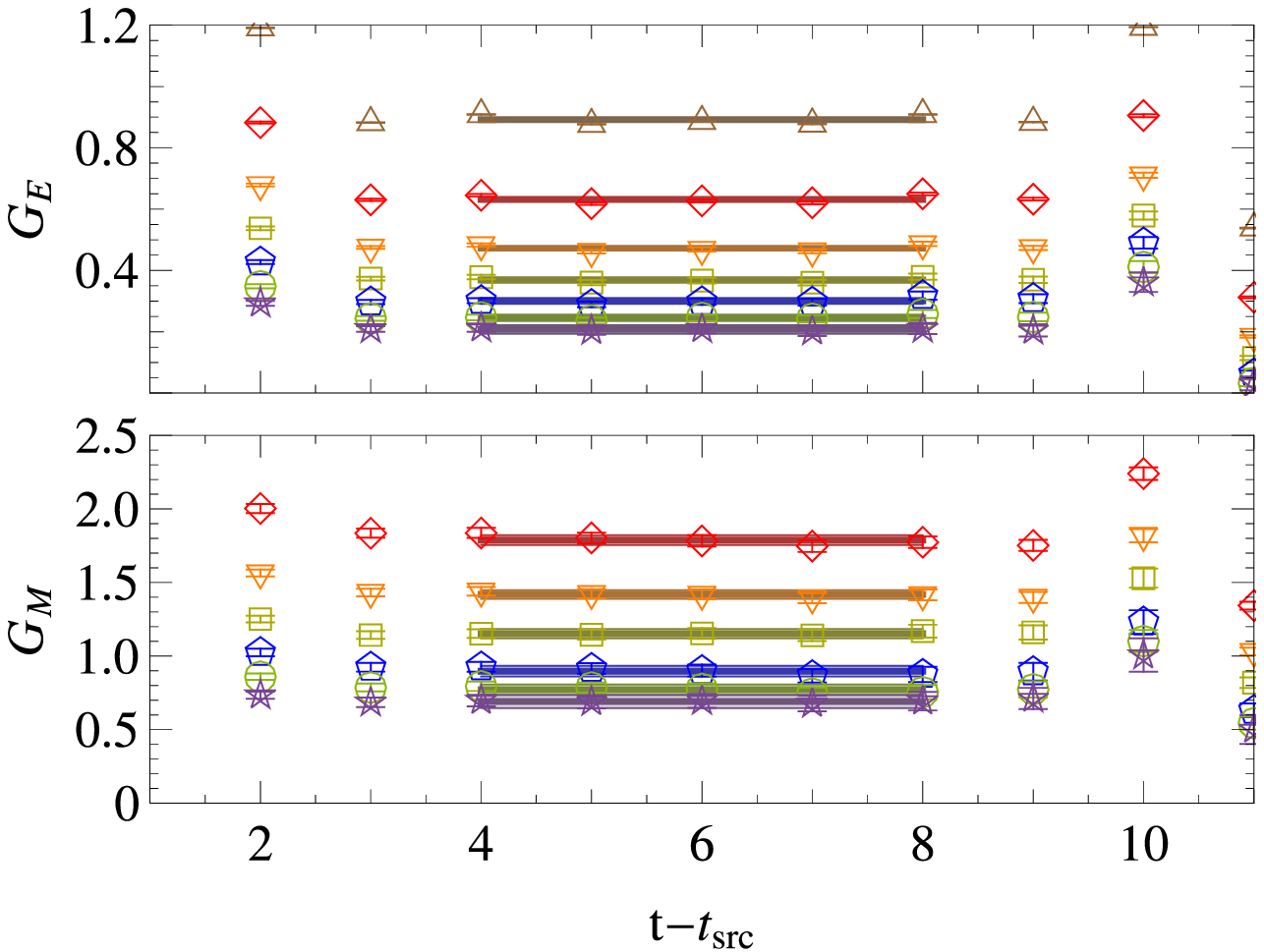}
\includegraphics[width=0.43\textwidth]{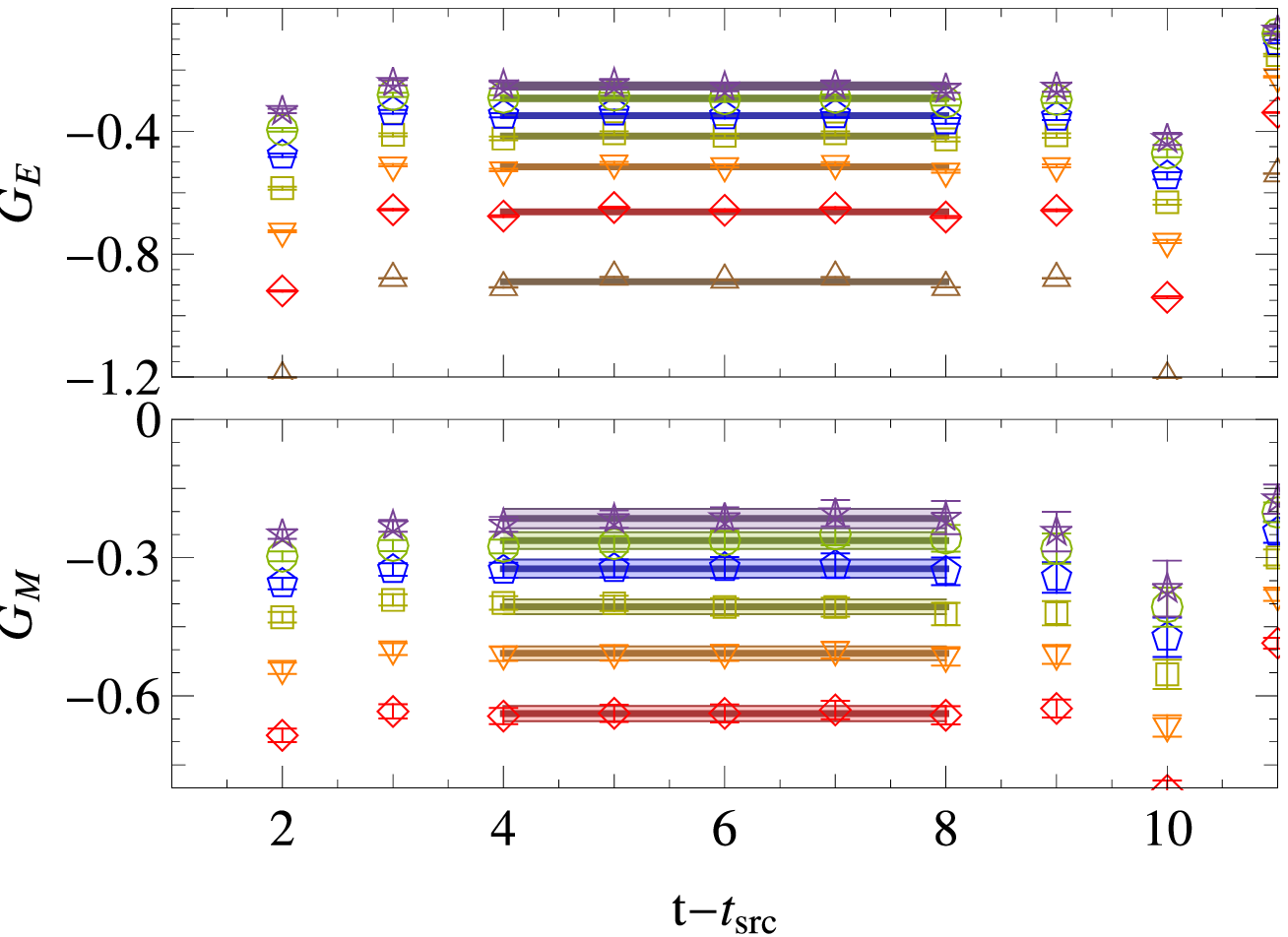}
\caption{The $\Sigma^+$ (left) and $\Xi^-$ (right) form factors $G_{E,M}$ for all momentum-transfer dependence  at $m_\pi\approx 600$~MeV. Symbols are the same as Figure~\ref{fig:m030-F12Plau}. Note that $G_{M}^B$ here are in units of the natural magneton $\frac{e}{2M_B}$, where the $M_B$ is the baryon mass on the 600-MeV pion sea.
}\label{fig:m030-GEMPlau}
\end{figure}

On the lattice, we calculate the matrix element $\langle B| V^\phi |B\rangle$ with $V^\phi=\overline{\phi}\gamma_\mu \phi$. Using SU(2) isospin symmetry, we can connect the proton and neutron matrix elements via
\begin{eqnarray}
u^p &\equiv&  \langle p| V^u|p\rangle =  \langle n| V^d |n\rangle \\
d^p &\equiv& \langle p| V^d |p\rangle = \langle n| V^u |n\rangle.
\end{eqnarray}
Therefore, the proton and neutron form factors are
\begin{eqnarray}
G_{E,M}^p &=& \frac{2}{3} (G_{E,M})_{u^p} - \frac{1}{3} (G_{E,M})_{d^p} \\
G_{E,M}^n &=& -\frac{1}{3} (G_{E,M})_{u^p} + \frac{2}{3} (G_{E,M})_{d^p},
\end{eqnarray}
where $(G_{E,M})_{u^p}$ are the Sachs form factors obtained from the $\langle p| V^u |p\rangle$ matrix element. In the case of the hyperons $\Sigma$ and $\Xi$,
\begin{eqnarray}
l^{\Sigma} &\equiv&    \langle \Sigma^+| V^{u}|\Sigma^+\rangle =  \langle \Sigma^-| V^{d} |\Sigma^-\rangle \\
s^{\Sigma} &\equiv& \langle \Sigma^+| V^{s}|\Sigma^+\rangle = \langle \Sigma^-| V^{s} |\Sigma^-\rangle \\
l^{\Xi} &\equiv&  \langle \Xi^0| V^{u}|\Xi^0\rangle =  \langle \Xi^-| V^{d} |\Xi^-\rangle \\
s^{\Xi} &\equiv& \langle \Xi^0| V^{s}|\Xi^0\rangle = \langle \Xi^-| V^{s} |\Xi^-\rangle.
\end{eqnarray}
Similarly, for $\Sigma$ and $\Xi$ baryons
\begin{eqnarray}
G_{E,M}^{\Sigma^+} &=& \frac{2}{3} (G_{E,M})_{l^\Sigma} - \frac{1}{3} (G_{E,M})_{s^\Sigma} \\
G_{E,M}^{\Sigma^-} &=& -\frac{1}{3} (G_{E,M})_{l^\Sigma} - \frac{1}{3} (G_{E,M})_{s^\Sigma}\\
G_{E,M}^{\Xi^-} &=& -\frac{1}{3} (G_{E,M})_{l^\Xi} - \frac{1}{3} (G_{E,M})_{s^\Xi} \\
G_{E,M}^{\Xi^0} &=& \frac{2}{3} (G_{E,M})_{l^\Xi} - \frac{1}{3} (G_{E,M})_{s^\Xi}.
\end{eqnarray}
Figure~\ref{fig:m030-GEMPlau} shows examples of the plateaus for the each of the transfer momenta obtained with Sachs form factors at $m_\pi\approx600$~MeV for $\Sigma^+$ and $\Xi^-$. (Tables~\ref{tab:NuclGsM1}--\ref{tab:XiGsM4} in the appendix summarize all the results.)

The magnetic form factors are naturally calculated in units of $\frac{e}{2M_B}$, where $M_B$ is the baryon mass calculated in its corresponding pion sea. In experiment, the nuclear magneton $\frac{e}{2M_N}$ is generally used in describing the magnetic moments for all baryons. Therefore, to compare with experimental values, we need a conversion factor of $\frac{M_N}{M_B}$ multiplying
our magnetic form factors and moments.

\section{Numerical Results}\label{sec:Numerical}

In this section, we first discuss the behavior of the form factors as
functions of momentum transfer squared and compare our data with
experimental expectations. Furthermore, we explore the conventional
dipole extrapolation used on the form factors and determine that the
dipole fit could fail for certain quantities. Later in
this section, we discuss the transverse structure of the baryon by
studying the charge and magnetic radii, the magnetic moments and the
SU(3) flavor symmetry breaking in these quantities.

\subsection{Momentum Dependence of Form Factors}\label{subsec:MomDepFF}

Studying the momentum-transfer ($Q^2$) dependence of the elastic
electromagnetic form factors is important in understanding the
structure of hadrons at different scales. There have been many
experimental studies of these form factors on the nucleon. A recent such experiment, the Jefferson
Lab double-polarization experiment (with both a polarized
target and longitudinally polarized beam) revealed a non-trivial
momentum dependence for the ratio $G_E^p/G_M^p$. This contradicts
results from the Rosenbluth separation method, which suggested $\mu_p
G_E^p/G_M^p \approx 1$. The contradiction has been
attributed to systematic errors due to two-photon exchange that
contaminate the Rosenbluth separation method more than the double-polarization. (For details and
further references, see the recent review articles,
Refs.~\cite{Arrington:2007ux,Perdrisat:2006hj,Arrington:2006zm}.) Lattice calculations
can make valuable contributions to the study of nucleon form factors, since they allow access to both the pion-mass and momentum dependence
of such form factors. In addition, lattice calculations can report individual quark contributions to the baryon form factors. Furthermore, by varying the light-quark masses we can study SU(3) flavor symmetry
breaking effects in octet baryons.

\begin{figure}
\includegraphics[width=0.45\textwidth]{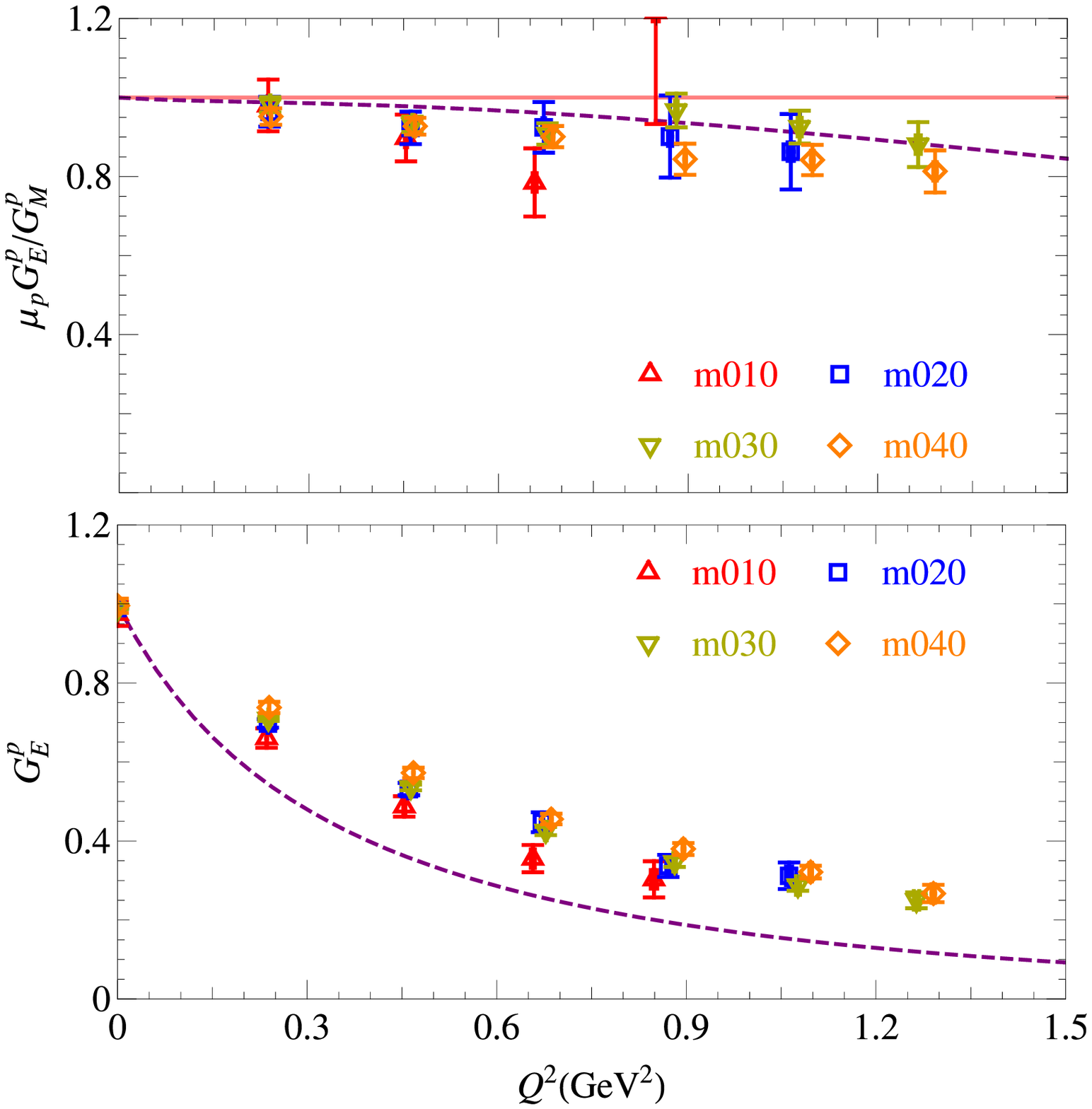}
\includegraphics[width=0.45\textwidth]{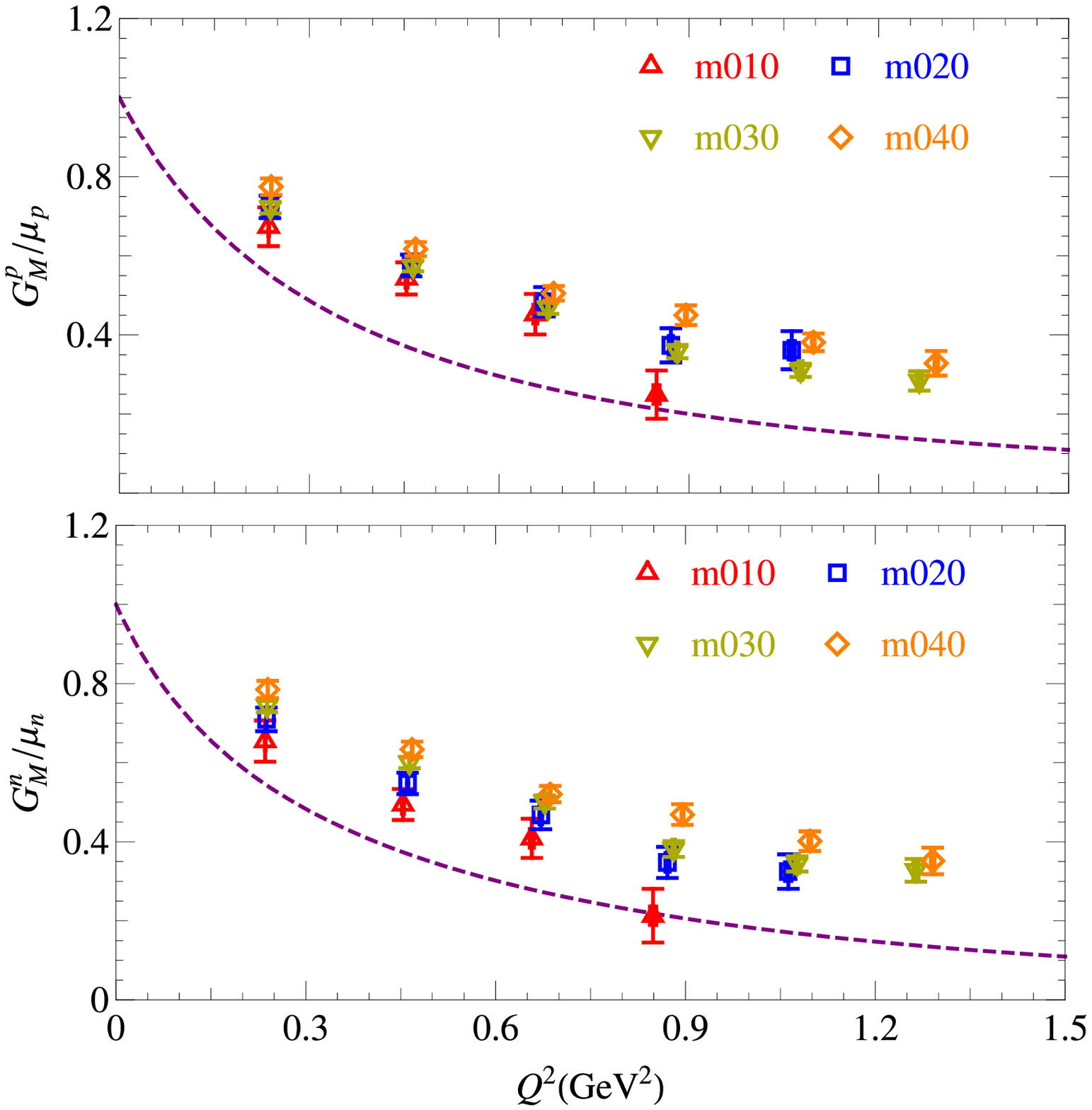}
\caption{Nucleon form factors. Left: The ratios $\mu_pG_{E}^p/G_{M}^p$ (top) and $G_{E}^p$ (bottom). Right: The magnetic form factor for the proton and neutron divided by their magnetic moments. Different symbols represent different pion-mass ensembles: triangles (m010), squares (m020), reverse triangles (m030) and diamonds (m040). The dashed lines are plotted using experimental form-factor fit parameters\cite{Arrington:2007ux,Kelly:2004hm}.
}\label{fig:N_Gs}
\end{figure}

\begin{figure}
\includegraphics[width=0.45\textwidth]{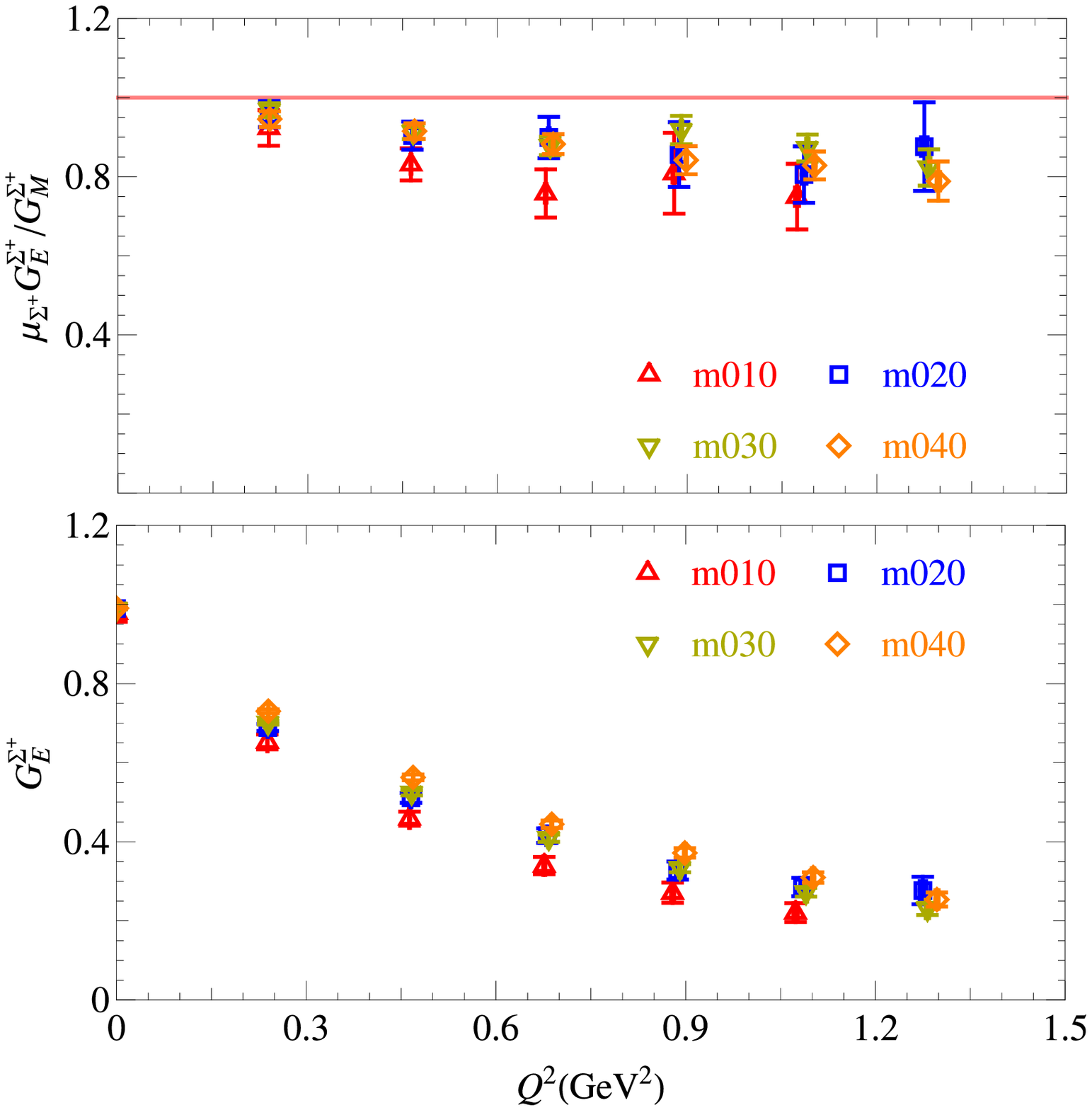}
\includegraphics[width=0.45\textwidth]{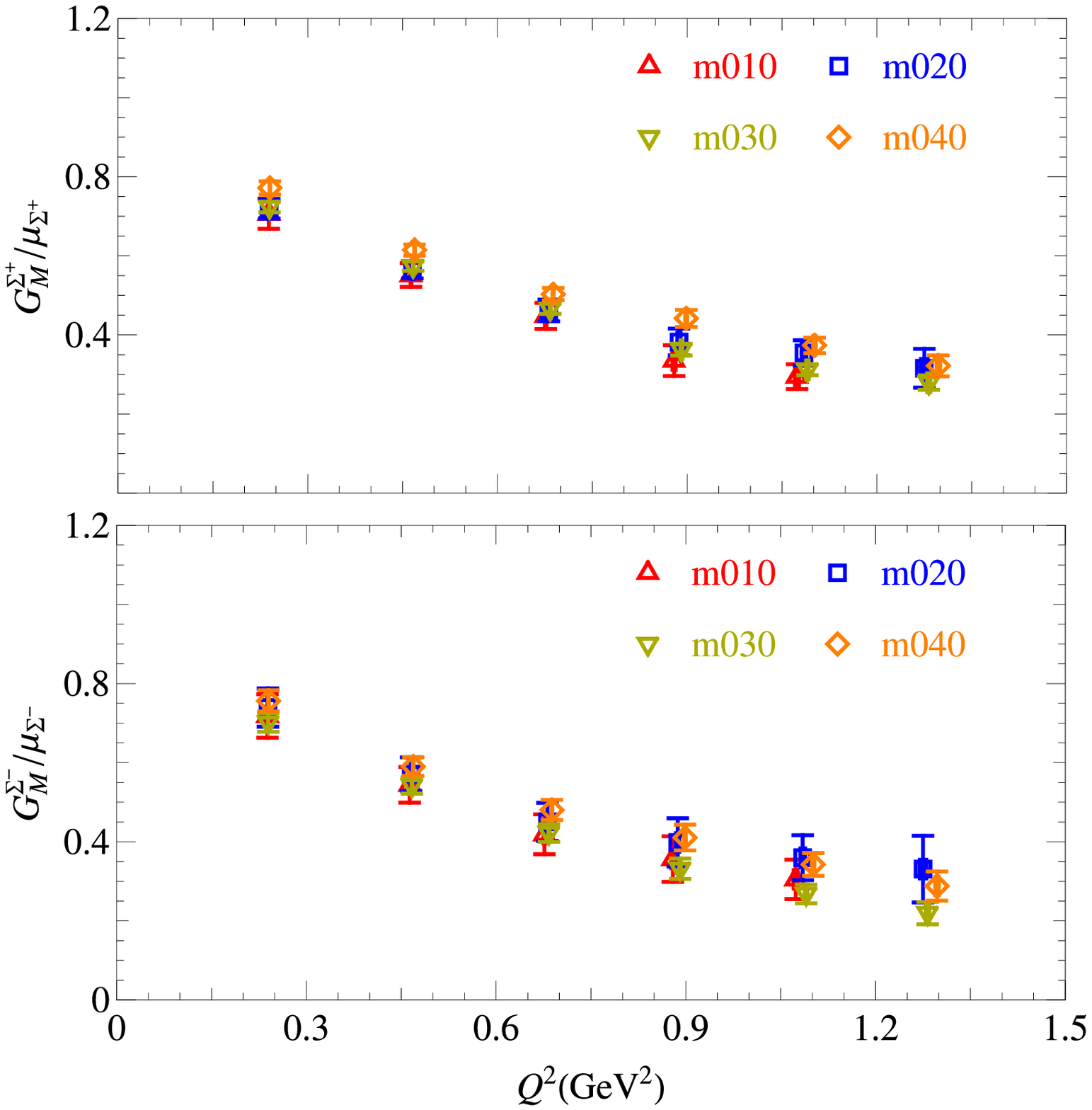}
\caption{Sigma form factors. Left: The ratios $\mu_{\Sigma^+}G_E^{\Sigma^+}/G_M^{\Sigma^+}$ (top) and $G_E^{\Sigma^+}$ (bottom). Right: The magnetic form factors for the $\Sigma^+$ and $\Sigma^-$ divided by their magnetic moments. Symbols are the same as Figure~\ref{fig:N_Gs}.
}\label{fig:Sig_Gs}
\end{figure}

\begin{figure}
\includegraphics[width=0.45\textwidth]{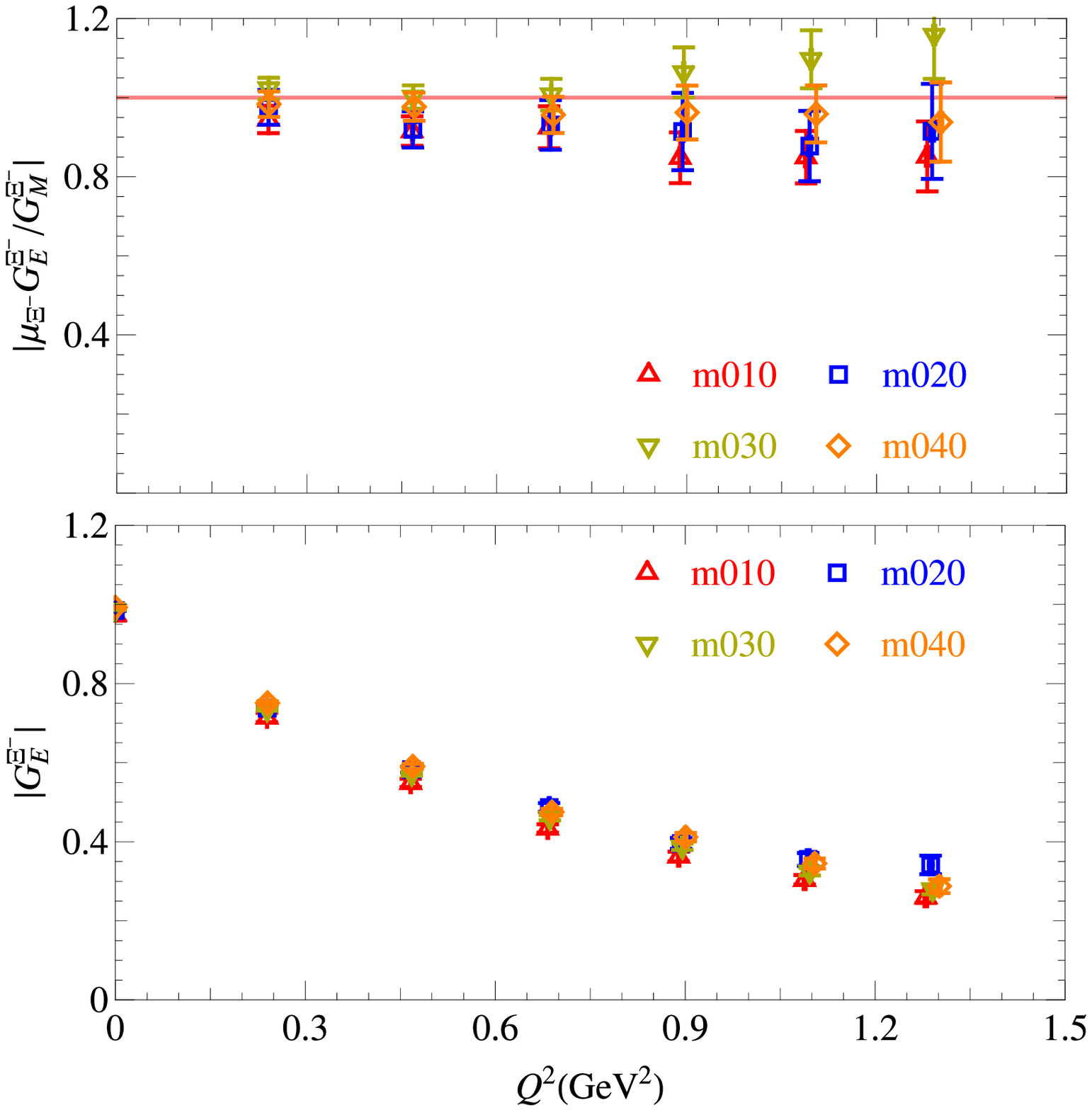}
\includegraphics[width=0.45\textwidth]{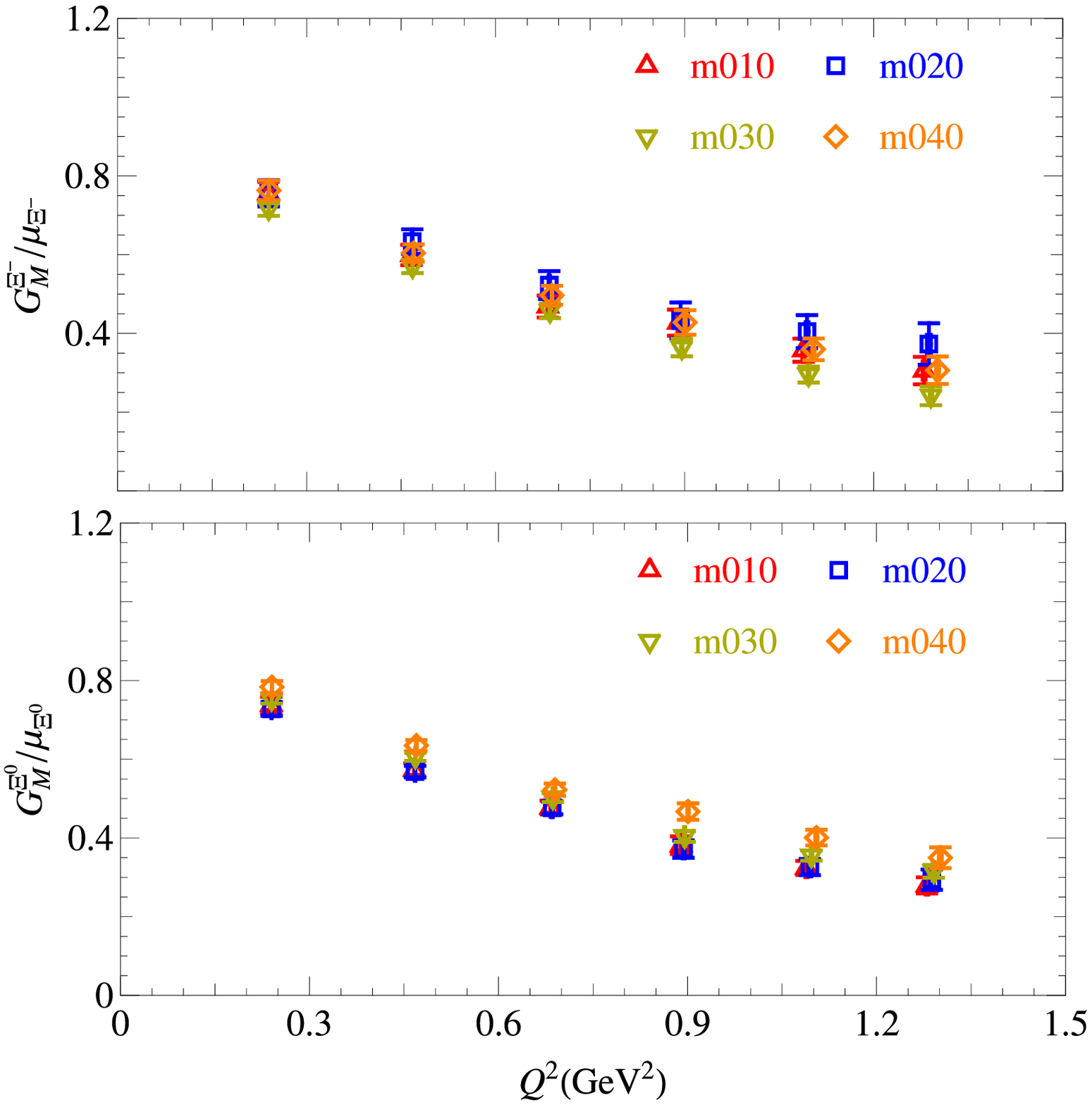}
\caption{Cascade form factors. Left: The ratios $\mu_{\Xi^-}G_E^{\Xi^-}/G_M^{\Xi^-}$ (top) and $G_E^{\Sigma^+}$ (bottom). Right: The magnetic form factors for the $\Xi^-$ and $\Xi^0$ divided by their magnetic moments. Symbols are the same as Figure~\ref{fig:N_Gs}.
}\label{fig:Xi_Gs}
\end{figure}

The upper-left panels in Figures~\ref{fig:N_Gs}, \ref{fig:Sig_Gs} and
\ref{fig:Xi_Gs} show the ratios $\mu_B G_{E}^B/G_{M}^B$, where $\mu_B$ is the magnetic moment on the lattice taken from
Sec.~\ref{subsec:MagneticMoments} ($B$ stands for $p$, $\Sigma^+$ and
$\Xi^-$ respectively). The
straight line on each plot is located at 1, the expected value for the
nucleon. The lower-left panels show $G_E^B$; in each, the $Q^2=0$ points are in
good consistency with 1.
The right-column plots show the magnetic form factors divided by their magnetic moments, $G_M^B/\mu_B$.

First, we focus on the nucleon system, which has been widely studied
by experiments and lattice calculations.\footnote{Here we
  only use a subset of the nucleon data available for these ensembles:
  those which overlap with the hyperon measurements for the Sigma and
  cascade baryons. The Lattice Hadron Physics Collaboration (LHPC)
  published a paper on generalized parton distributions (GPDs), which
  covers slightly more configurations\cite{Hagler:2007xi}, and they
  will publish an analysis using different source-sink separations and
  higher statistics in the near future.} In both $\mu_p
G_{E}^p/G_{M}^p$ and $G_{E}^p$, there is a decreasing trend as
$Q^2$ increases.
The pion-mass dependence is rather mild in the case of $\mu_p G_{E}^p/G_{M}^p$.
The slope of $\mu_p G_{E}^p/G_{M}^p$ is roughly
consistent with those measured in double-polarization experiments, around $-0.14$.
(For example, see the summary plots in Figure~17 of Ref.~\cite{Perdrisat:2006hj}.)
The dashed lines in these plots are the fitted forms of Ref.~\cite{Arrington:2007ux} for the proton and Ref.~\cite{Kelly:2004hm} for the neutron; for this ratio, our points are distributed around the lines, showing there is consistency with the experimental values.
$G_{E}^p$, $/G_{M}^p/\mu_p$ and $/G_{M}^n/\mu_n$ shows distinguishable
but small pion-mass dependence. As the pion mass decreases, our data
appears to trend towards the dashed lines which represent fits to the experimental data.

The hyperon $\Sigma$ and $\Xi$ form factors are more poorly known compared with
the nucleon case. Similar to the ratio of its SU(3) flavor partner, the ratios
$\mu_{\Sigma^+}G_E^{\Sigma^+}/G_M^{\Sigma^+}$ and $\mu_{\Xi^-}G_E^{\Xi^-}/G_M^{\Xi^-}$ are around 1 within our $Q^2$ range,
and there is mild pion-mass dependence in our study. Most of the
$\mu_{\Sigma^+}G_E^{\Sigma^+}/G_M^{\Sigma^+}$ values are slightly below experimental proton
fitted line, while the $\mu_{\Xi^-}G_E^{\Xi^-}/G_M^{\Xi^-}$ points are distributed
around the line, except for those from 685-MeV pion mass. Since the effect of replacing the up/down quark in the proton is likely
suppressed in the ratios of the individual form factors, it is not surprising to
find that these hyperon form factor ratios are not far from experimental proton line.

$G_{E}^{\Sigma^+}$, $G_{M}^{\Sigma^+}/\mu_{\Sigma^+}$ and $G_{M}^{\Sigma^-}/\mu_{\Sigma^-}$ have mild discrepancies from nucleon case.
The single replacement of a light quark in the nucleon to a strange in the $\Sigma$ baryon has a mild
change on the $Q^2$ dependence of the form factors. The cascade form factors ($|G_{E}^{\Xi^-}|$, $G_{M}^{\Xi^-}/\mu_{\Xi^+}$ and $G_{M}^{\Xi^0}/\mu_{\Xi^0}$) are larger
than the nucleon case, more dramatically for the lightest quark. The pion-mass dependence is small since the dominant quark flavor (strange) is less sensitive to changes of the up/down masses in the sea and valence sectors. Overall, the hyperon form factors are slightly larger than the nucleon ones, up to 15\% in certain
cascade channels.

\subsection{Validity of the Dipole Extrapolation}\label{subsec:Dipole}

A widely adopted momentum extrapolation in lattice calculations for electromagnetic form factors is the dipole form
\begin{eqnarray}\label{eq:DipoleFitForm}
{\cal F}(Q^2) &=& \frac{{\cal F}(0)}{\left(1+\frac{Q^2}{M_D^2}\right)^2},
\end{eqnarray}
where $M_D$ is the dipole mass.
To demonstrate how well dipole form works for Dirac and Pauli form factors, we can look at
\begin{eqnarray}\label{eq:DipoleConstantForm}
{r}^\prime &=&\frac{12}{Q^2} \left(\sqrt{\frac{{\cal F}(0)}{{\cal F}(Q^2)}}-1\right).
\end{eqnarray}
If the dipole form describes the momentum dependence of the form factor, there will be no momentum dependence in $r^\prime$.

\begin{figure}
\includegraphics[width=0.32\textwidth]{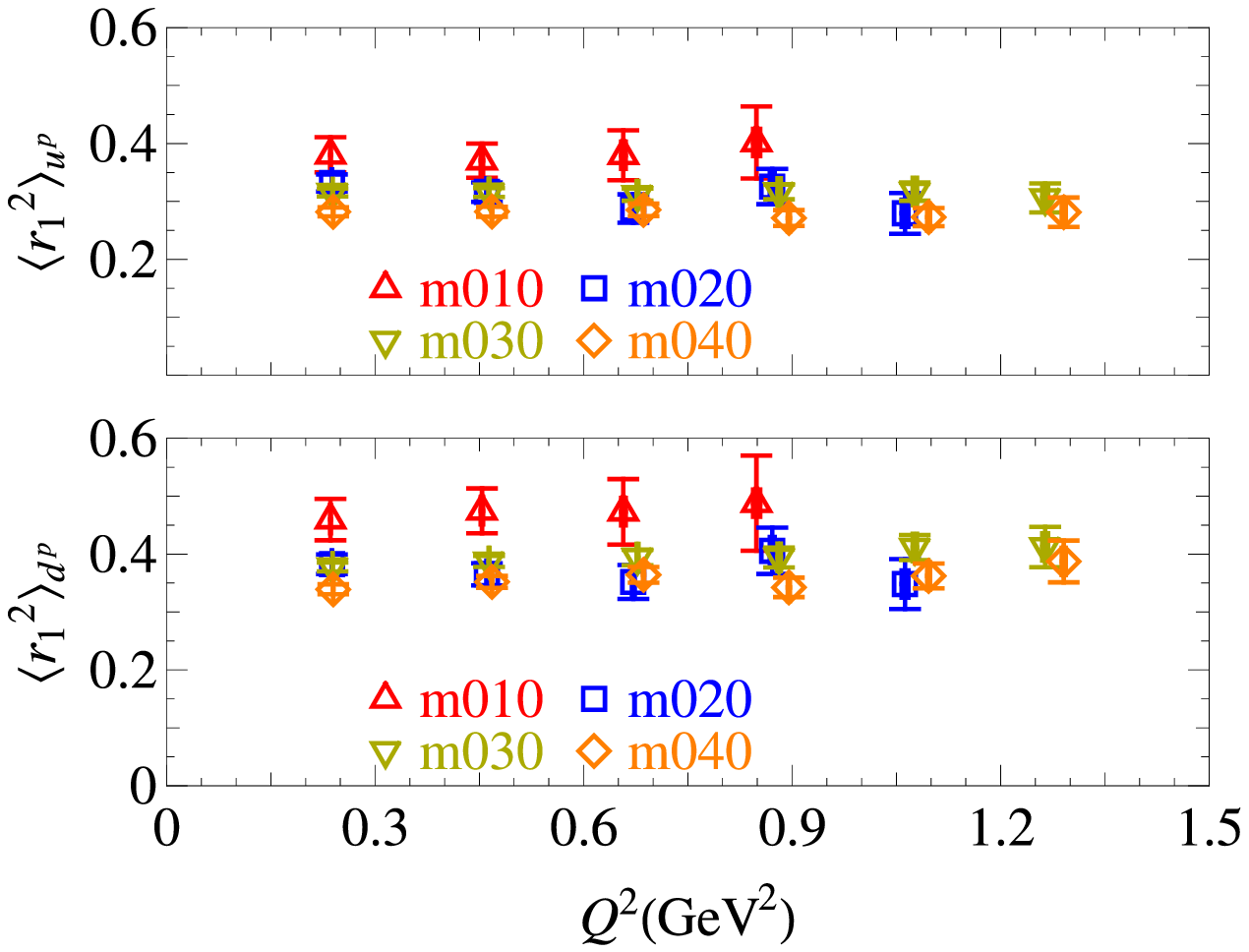}
\includegraphics[width=0.32\textwidth]{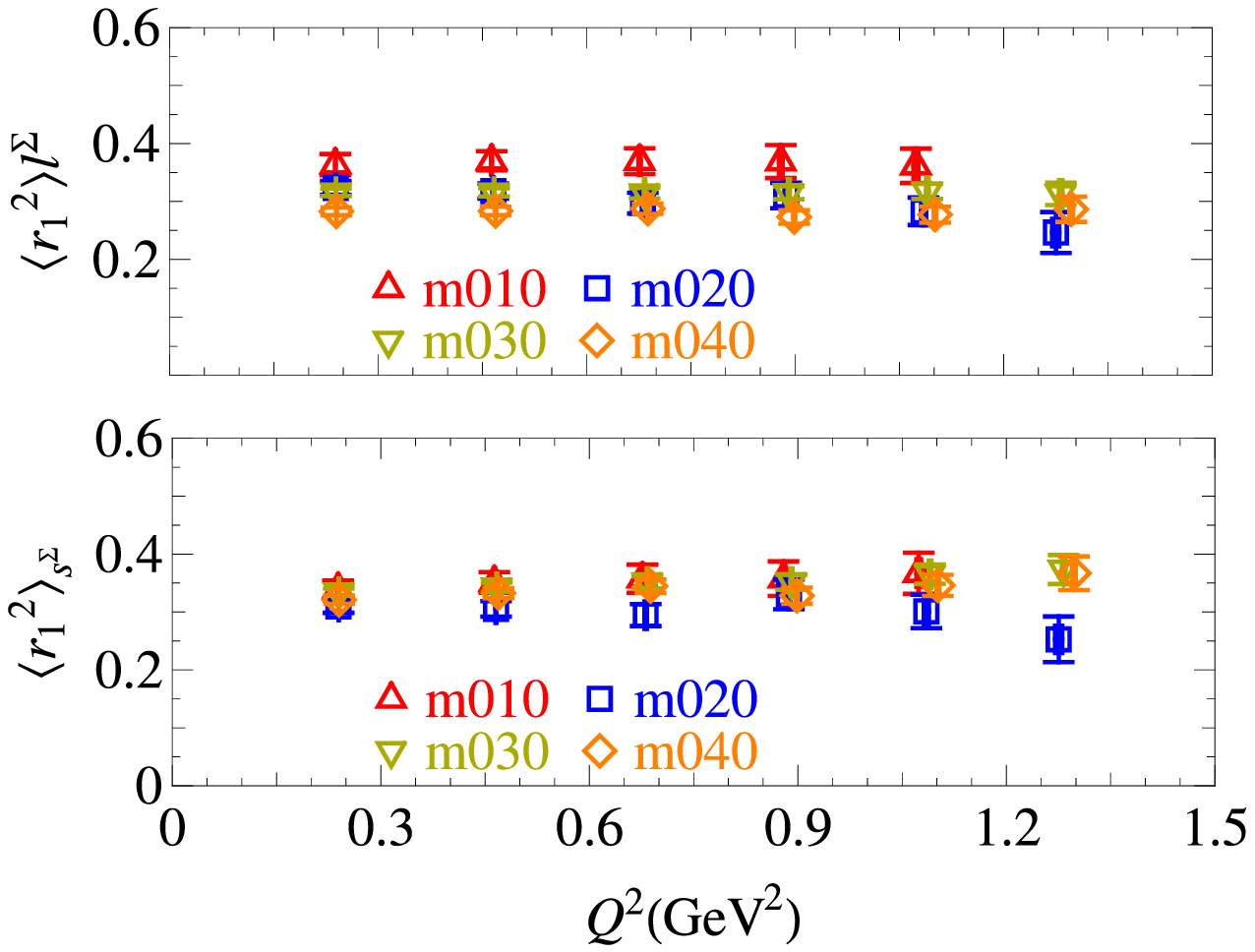}
\includegraphics[width=0.32\textwidth]{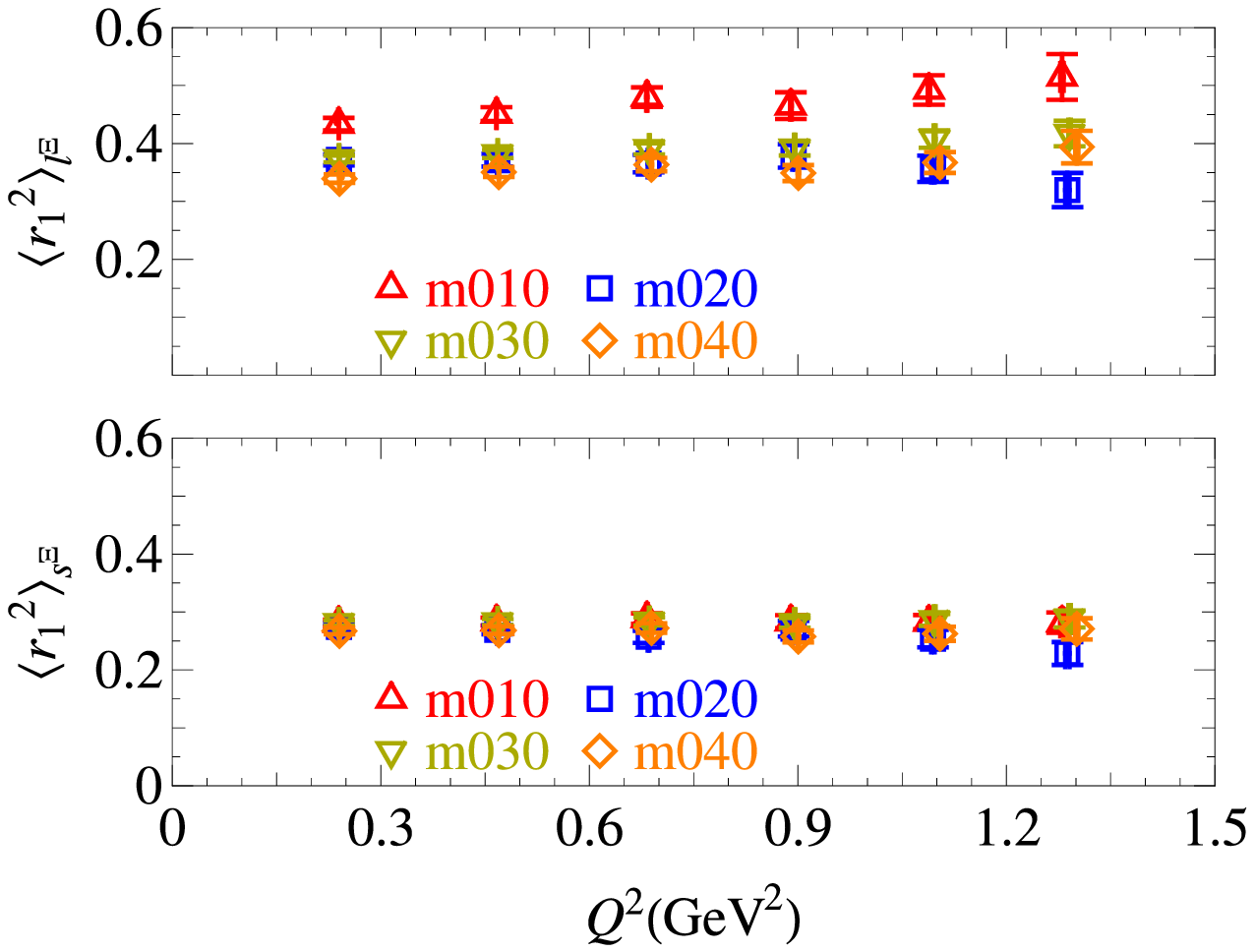}
\caption{Dirac mean-squared radii (in units of $\mbox{fm}^2$) as defined in Eq.~\ref{eq:DipoleConstantForm} as a functions of $Q^2$ (in units of $\mbox{GeV}^2$) for $N$ (left), $\Sigma$ (middle) and $\Xi$ (right) for each $V^\phi$
}\label{fig:allB-r1_Q2}
\end{figure}

\begin{figure}
\includegraphics[width=0.32\textwidth]{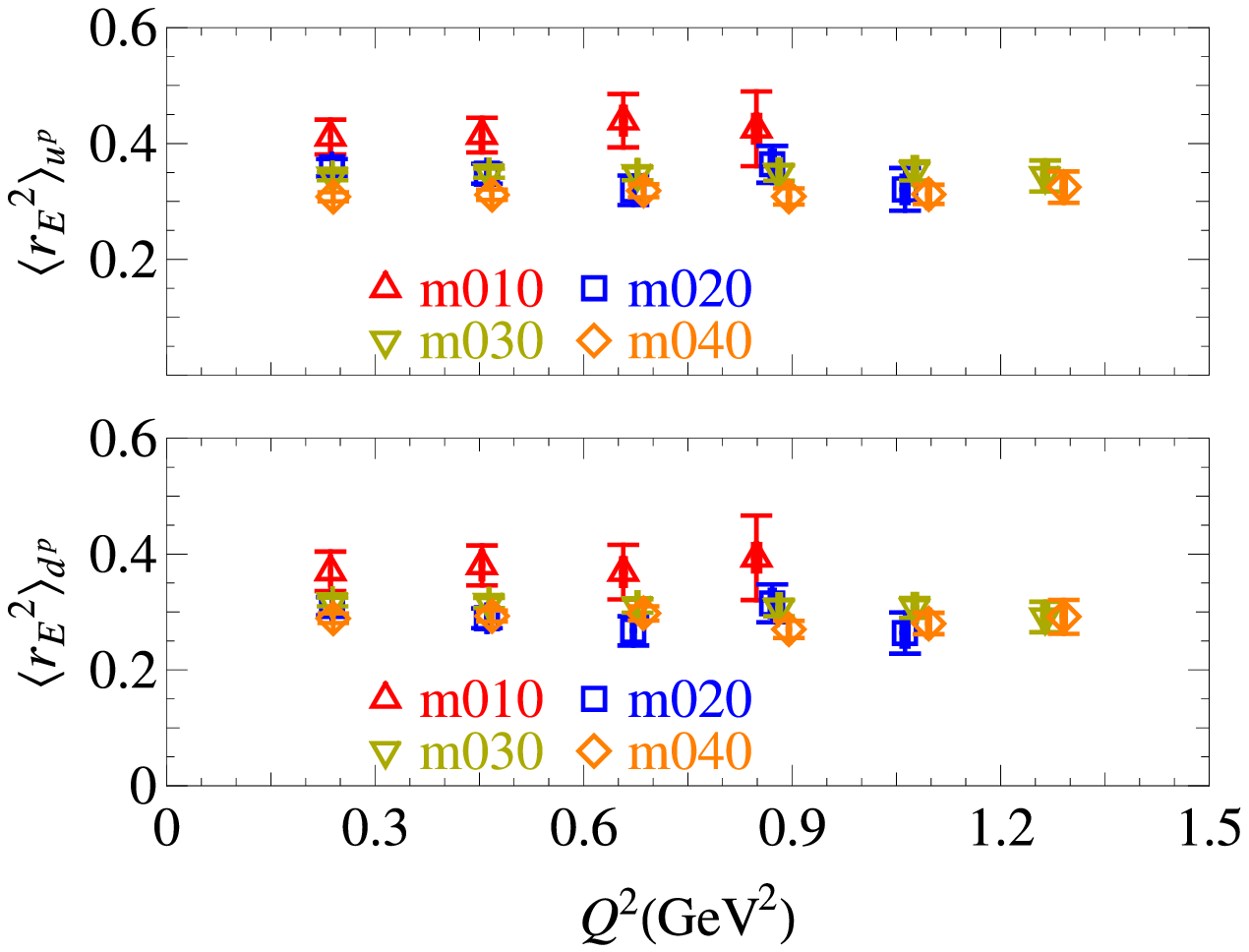}
\includegraphics[width=0.32\textwidth]{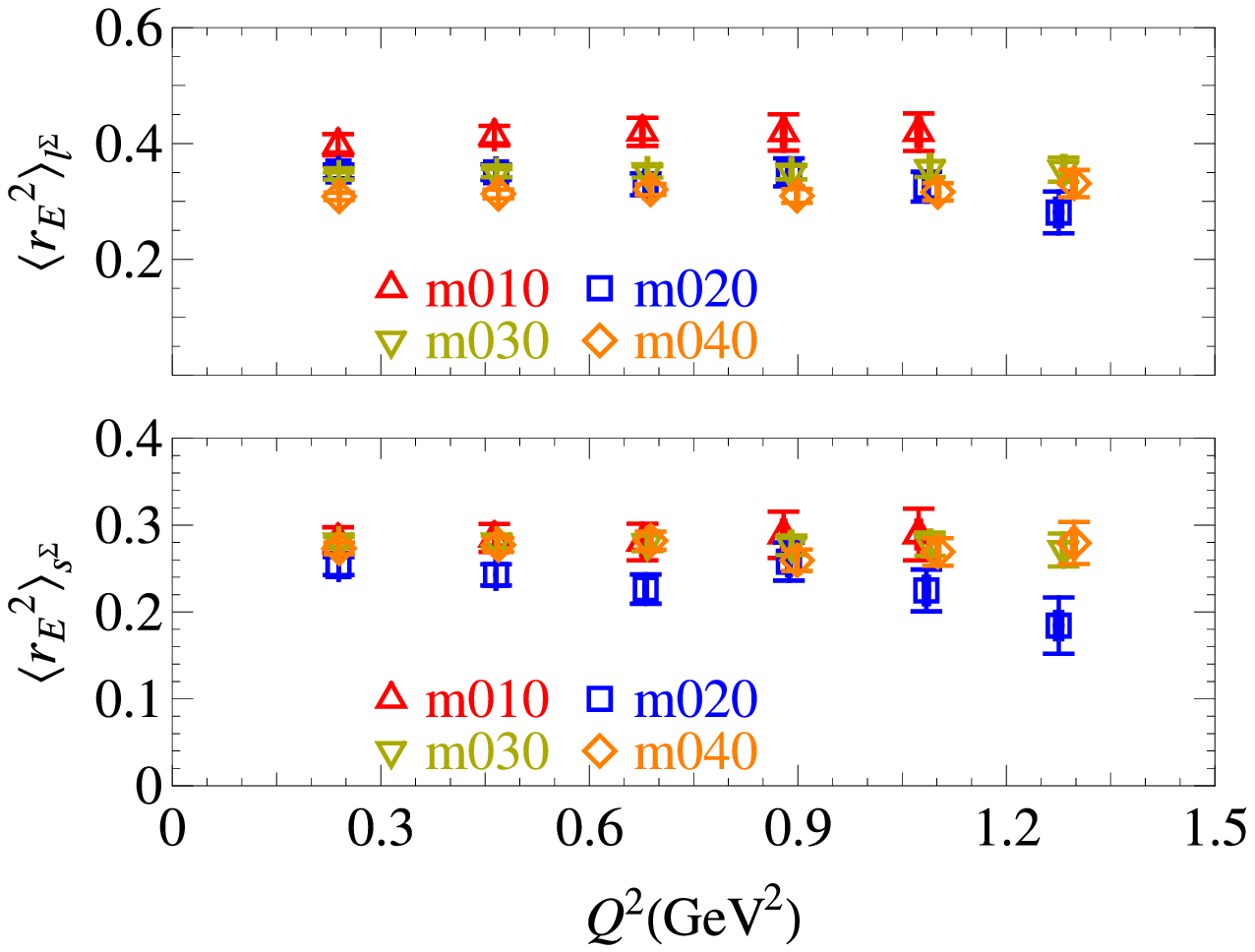}
\includegraphics[width=0.32\textwidth]{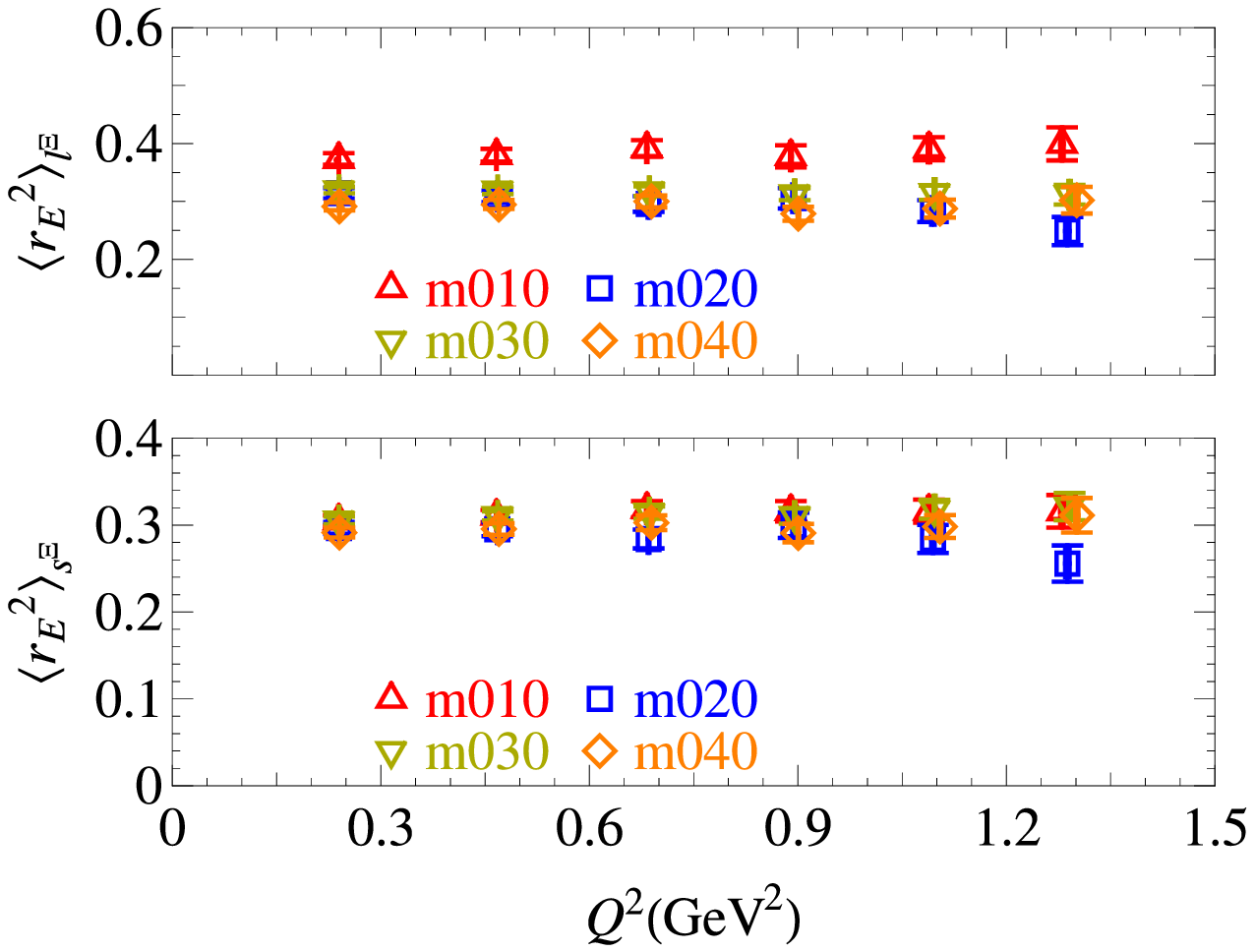}
\caption{Electric mean-squared radii (in units of $\mbox{fm}^2$) as a functions of $Q^2$ (in units of $\mbox{GeV}^2$) for $N$ (left), $\Sigma$ (middle) and $\Xi$ (right) for each $V^\phi$
}\label{fig:allB-rE_Q2}
\end{figure}

We first concentrate on the case of Dirac form factors, where ${\cal
  F}(0)=F_{1} (0)$ is calculated directly on the lattice (unlike
$F_2(0)$, which would require
extrapolation). Figure~\ref{fig:allB-r1_Q2} shows the results from our
$N$, $\Sigma$ and $\Xi$ baryons for each inserted quark current. We
see almost no momentum dependence of  $r^\prime$ for all of the six matrix elements in
Figure~\ref{fig:allB-r1_Q2}. This is an indication that the   dipole-form
is a good description of the data. There are a few cases, such as the ``m020'' set that seem
to deviate from the dipole form at large $Q^2$, but a dipole
fit still goes through most of the points. One exception is
 the light-quark current for the $\Xi$ baryon matrix element in the ``m010'' set. In this case the central value of $r^\prime$  changes about 10\% as one goes to large momentum. When quark components are combined to form the full baryon form factors, this discrepancy does not occur, possibly due to
cancellation between different quark components. Overall, we observe
that $r^\prime$ for all baryons and most of the quark contributions
within the baryons is in good agreement with a constant with respect
to $Q^2$. In the $F_2$ case, where $F_2(0)$ is an unknown constant depending on the matrix element, we find that the dipole description is also reasonable for
both quark components and baryons. 
It is easy to extend Eq.~\ref{eq:DipoleConstantForm} to the electric
form factor $G_E$, and the results are shown in
Figure~\ref{fig:allB-rE_Q2}. As before, the largest discrepancy from the
dipole description occurs for most of the ``m020'' set's quark matrix elements.
However, in this case the dipole form seems that it does not describe the data
well. Although for $G_M$ no such discrepancy is observed, we need to look
for alternative forms to fit the data.

One simple extension is
\begin{eqnarray}
G_E = \frac{1}{(1+Q^2/M_e^2)^p}
\end{eqnarray}
with $p=1,2,3,4$ for monopole, dipole, tripole and quadrupole. We are
also inspired by fit forms used in
Refs.~\cite{Kelly:2004hm,Arrington:2007ux} to describe the experimental
data:
\begin{eqnarray}\label{eq:KelleyFitForm}
G(Q^2) &=& \frac{\sum_{k=0}^{n}a_k \tau^k}{1+\sum_{k=1}^{n+2}b_k \tau^k}
\end{eqnarray}
with $\tau=\frac{Q^2}{4M^2}$. Unfortunately, we do not have comparable amounts of data (and as wide range of $Q^2$) as experiment to adopt the same number of parameters for the fit. Therefore, we constrain the fit form to go asymptotically to $1/Q^4$ at large $Q^2$ and to have $G_E(Q^2=0)=1$:
\begin{eqnarray}
G_E = \frac{A Q^2+1}{C Q^2+1}\frac{1}{(1+Q^2/M_d^2)^2},
\end{eqnarray}
where $C$ may be constrained to be positive to avoid putting unphysical poles within our $Q^2$ range. The choice of $C=0$ and $A$ free we call ``dipoleV0''; $C>0$ and $A$ free we call ``dipoleV1''; and $C>0$ and $A>0$, ``dipoleV2''.  Since the $\Xi$ matrix elements deviate from the dipole form the most, we chose them to
test for these new fit forms.  Table~\ref{tab:Xi-Radii-fitformTest} summarizes the fitted $\chi^2/\mbox{dof}$ using these different fit forms and the electric mean-squared charge radii obtained from the fits to the light-quark component of the cascade. We find that using ``dipoleV1'', we get a coefficient $C$ that is consistent with zero; the fitted results are very close to the result from ``dipoleV0''. The extracted electric mean-squared charge radii obtained from dipole fit forms are as much as 8\% larger than those from the fit form that describes the lattice data well. Therefore, for the rest of this paper, we will use ``dipoleV0'' to fit quark-component $G_E$ and use the standard dipole for all the other form factors.

\begin{table}
\begin{center}
\begin{tabular}{c|ccccccc}
\hline\hline
$m_\pi^2(\mbox{GeV}^2)$ & monopole & dipole & tripole & quadrupole & dipoleV0 & dipoleV1 & dipoleV2 \\
\hline
  0.1256(15) &  0.462(16) &  0.381(12) &  0.357(11) &  0.345(10) &  0.370(10) &  0.368(11) &  0.381(12) \\
  & [7.64] & [0.31] & [0.28] & [0.9] & [0.12] & [0.15] & [0.47] \\
  0.246(2) &  0.353(14) &  0.305(11) &  0.289(10) &  0.282(10) &  0.328(10) &  0.328(8) &  0.305(11) \\
  & [0.87] & [1.17] & [2.63] & [3.7] & [0.43] & [0.54] & [1.75] \\
  0.3493(17) &  0.370(9) &  0.318(7) &  0.301(6) &  0.293(6) &  0.322(5) &  0.314(16) &  0.318(7) \\
  & [6.74] & [0.09] & [1.89] & [3.82] & [0.02] & [0.02] & [0.14] \\
  0.463(3) &  0.338(11) &  0.292(9) &  0.278(8) &  0.271(8) &  0.294(7) &  0.27(2) &  0.292(9) \\
  & [4.15] & [0.36] & [1.03] & [1.85] & [0.43] & [0.48] & [0.54] \\
\hline\hline
\end{tabular}
\end{center}
\caption{\label{tab:Xi-Radii-fitformTest}Cascade electric mean-squared charge radii, $\langle r_E^2 \rangle_{l^\Xi}$, in units of $\mbox{fm}^2$ from different fit forms.  The square brackets indicate the $\chi^2/{\rm dof}$ for each fit.}
\end{table}

\subsection{Charge Radii}\label{subsec:ChargeRadii}

\begin{figure}
\includegraphics[width=0.45\textwidth]{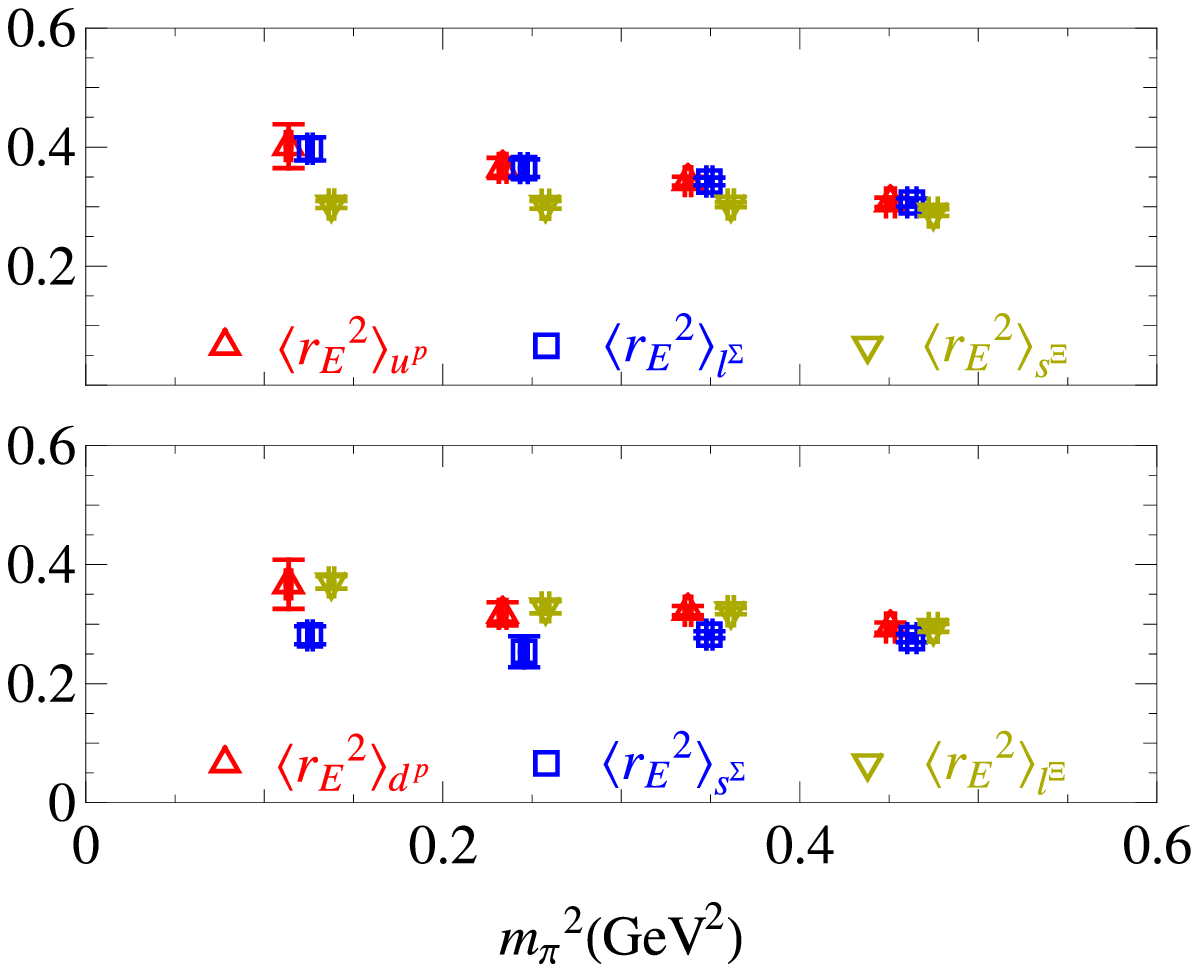}
\includegraphics[width=0.45\textwidth]{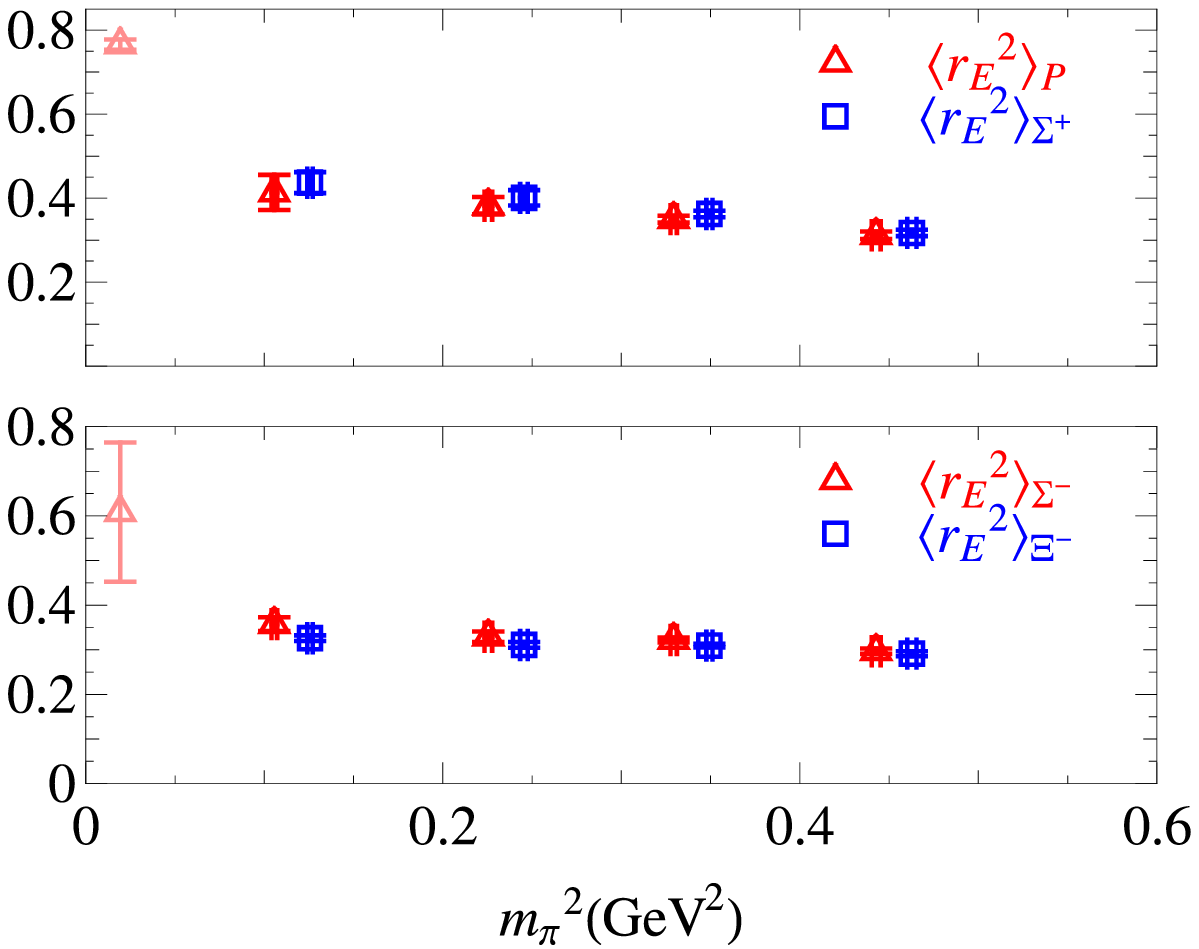}
\caption{The electric mean-squared radii in units of $\mbox{fm}^2$ as functions of $m_\pi^2$ (in GeV$^2$) from each quark contribution (left) and baryon (right).  The leftmost triangles in both figures are the extrapolated values at the physical pion mass.
}\label{fig:allB-r2E_mpi2}
\end{figure}

The mean-squared electric charge radii can be extracted from electric form factor $G_E$ via
\begin{eqnarray}\label{eq:GEradii}
\langle r_{E}^2\rangle &=& (-6)\frac{d}{dQ^2}\left(\frac{G_{E}(Q^2)}{G_{E}(0)}\right)\Big|_{Q^2=0}.
\end{eqnarray}
Similar definitions can be used to find the Dirac and Pauli radii, $r_{1,B}$ and $r_{2,B}$ respectively, where the relations are $r_{E,B}^2=r_{1,B}^2+\frac{3}{2}\frac{\kappa_B}{2m_B^2}$ with $\kappa_B = F_{2,B}(Q^2=0)$. In Subsec.~\ref{subsec:Dipole}, we discuss an alternative fit form, ``dipoleV0'', that works better in our kinematic region. Therefore, we will use this form to extract mean-squared electric radii; the numbers are summarized in Table~\ref{tab:Baryon-r2E} and Figure~\ref{fig:allB-r2E_mpi2}.

In the left-hand panel of Figure~\ref{fig:allB-r2E_mpi2}, we plot the quark contributions to the nucleon, Sigma and cascade baryons. The upper plot displays the dominant (two-valence) quark contributions in the $p$, $\Sigma$ and $\Xi$, while the lower one shows the single-quark contributions. These charge radii tend to increase in the light sector, while the strange sector is relatively flat as one changes the pion mass. The rightmost points are closest to the SU(3) point, where the quark-contribution differences are the smallest. As one increases the difference in light and strange-quark masses, only the strange-quark contribution in the baryon starts to show differences depending on the baryon species, while the light-quark contribution seems to be independent of its surrounding environment for pion masses as light as 350~MeV. It is expected that the strange contribution displays relatively smaller charge radius, since the strange quark is heavier and thus has shorter Compton wavelength.

\begin{figure}
\includegraphics[width=0.5\textwidth]{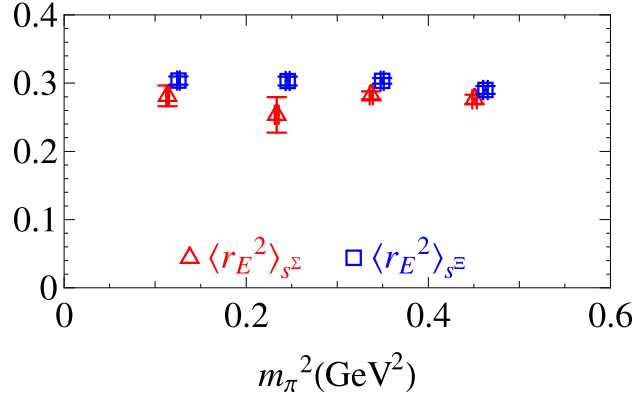}
\caption{The strange-quark contribution to the electric mean-squared radii in units of $\mbox{fm}^2$ as functions of $m_\pi^2$ (in GeV$^2$)}\label{fig:S-r2E_mpi2}
\end{figure}

We further compare the strange-quark contributions in the $\Sigma$ and $\Xi$, as shown in Figure~\ref{fig:S-r2E_mpi2}. The strange quark in the cascade has slightly larger contribution to the charge radii than in the Sigma, but overall they agree within $1.5\sigma$. This also shows that the quark contribution is not much affected by the environmental baryon, at least down to 350-MeV pion mass.

In the right-hand panel of Figure~\ref{fig:allB-r2E_mpi2} we plot the electric charge radii with the neutron and $\Xi^0$ omitted, since they are neutral particles and $G_{E,\{n,\Xi^0\}}(Q^2) \approx 0$ within our statistical errors. Firstly, we see that there is small SU(3) symmetry breaking between the SU(3) partners, $p$ and $\Sigma^+$ (or $\Sigma^-$ and $\Xi^-$); their charge radii are consistent  within statistical errors. A similar observation can also be made for the previous lattice quenched study, Ref.~\cite{Boinepalli:2006xd}. We observe roughly the same slope for the charge radii of $p$ and $\Sigma^{+/-}$ baryons; however, our $\Xi^-$ has about half the increase with decreasing pion mass; this difference could be caused by the quenched approximation. Overall, the SU(3) symmetry breaking in the charge radii is much smaller than what we observed in our study of the axial coupling constants\cite{Lin:2007ap}; this suggests that different physical observables can have substantially different responses to the replacement of one quark by another in a baryon. For charge radii, the effect is negligible.

%%%%%%%%%%%%%%%%%%%%%%%%%%%%%%%%%%%%%%%%%%%%%%%%%%% extrapolation %%%%%%%%%%%%%5
For the extrapolation of charge radii\footnote{In this work, we use HBXPT to perform the mass extrapolation for charge radii. Other extrapolations, such as Finite-Range Regulation (FRR), are used for octet charge radii on quenched lattice data in Ref.~\cite{Wang:2008vb}, applying corrections to the effective field theory to account for quenching.} to physical masses, we adopt continuum three-flavor heavy-baryon chiral perturbation theory (HBXPT)\cite{Jenkins:1990jv,Jenkins:1991ts,Kubis:1999xb,Kubis:2000aa,Arndt:2003ww}:
\begin{eqnarray}\label{eq:ChPT-radii}
\langle r_E^2 \rangle &=& -\frac{6}{\left(f_\pi^{\rm lat}\right)^2}c^\prime + \frac{3}{2m_B^2}b^\prime
-\frac{1}{16\pi^2 \left(f_\pi^{\rm lat}\right)^2}\sum_X [{\cal G}(m_X,\delta,f_\pi^{\rm lat})+{\cal H}(m_X,\delta,f_\pi^{\rm lat})]\\
{\cal G}(m,\delta,\mu) &=& (\gamma_X -5\beta_X)\ln \left(\frac {m^2}{\mu^2} \right)\\
{\cal H}(m,\delta,\mu) &=& 10 \beta_X^\prime {\cal F}(m,\delta,\mu)\\
{\cal F}(m,\delta,\mu) &=& \ln \left(\frac {m^2}{\mu^2} \right) -\frac{\delta}{\sqrt{\delta^2-m^2}} \ln \frac{\delta -\sqrt{\delta^2-m^2+i\epsilon}}{\delta +\sqrt{\delta^2-m^2+i\epsilon}}.
\end{eqnarray}
In the above, $c^\prime=(Qc_-+\alpha_Dc_+)$ and $b^\prime=(Q\mu_F+\alpha_D\mu_D)$ as in Ref.~\cite{Arndt:2003ww} with coefficients $Q$ and $\alpha_D$ listed in Table~1, and $\gamma_X$ and $\beta^{(\prime)}_X$ are given in Tables~II--VIII for various baryon flavors.
The sum over $X$ includes all possible meson-loop contributions to the electric charge radii, $\delta$ is the mass difference between an octet baryon and its decuplet partner, and $m_B$ is the octet baryon mass; these numbers are taken from LHPC\cite{WalkerLoud:2008bp}.
The octet axial couplings $D$ and $F$ are set to $0.715(50)$ and $0.453(50)$ respectively from a numerical determination in the same mixed-action calculation\cite{Lin:2007ap}; the coupling $C$ which is related to $g_{\pi N \Delta}$ is set to be 0 (when ignoring decuplet contributions) or $1.2(2)$ from an axial $\Delta-N$ transition form factor calculation\cite{Alexandrou:2007zz}.
Note that we replace the scale
$\mu$ in the original formulation with $f_\pi^{\rm lat}$\cite{Beane:2006kx,Edwards:2006qx,Lin:2008uz}.
To next-to-leading order in our calculation, such a replacement is consistent with the original formulation.

\begin{figure}
\includegraphics[width=0.45\textwidth]{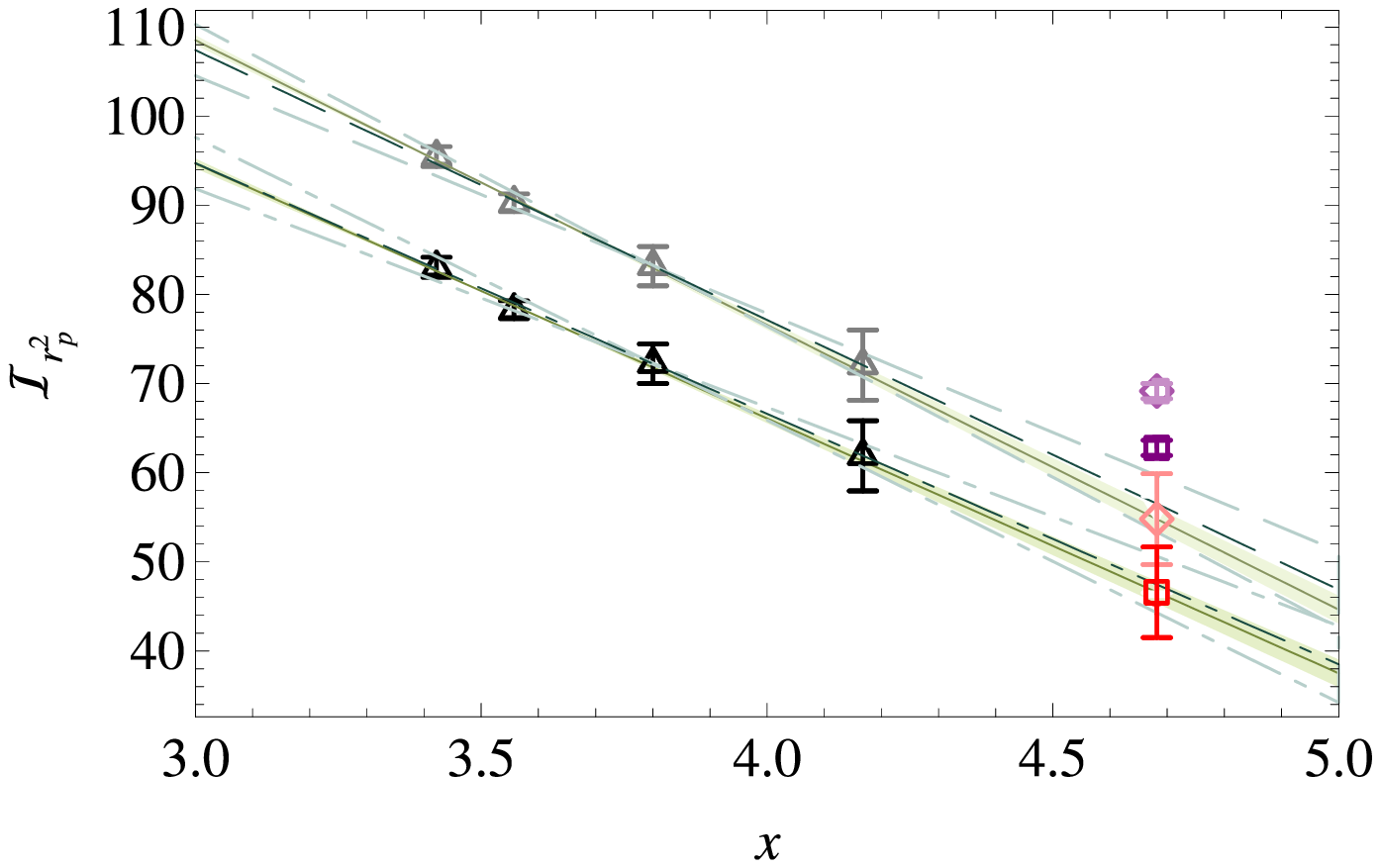}
\includegraphics[width=0.45\textwidth]{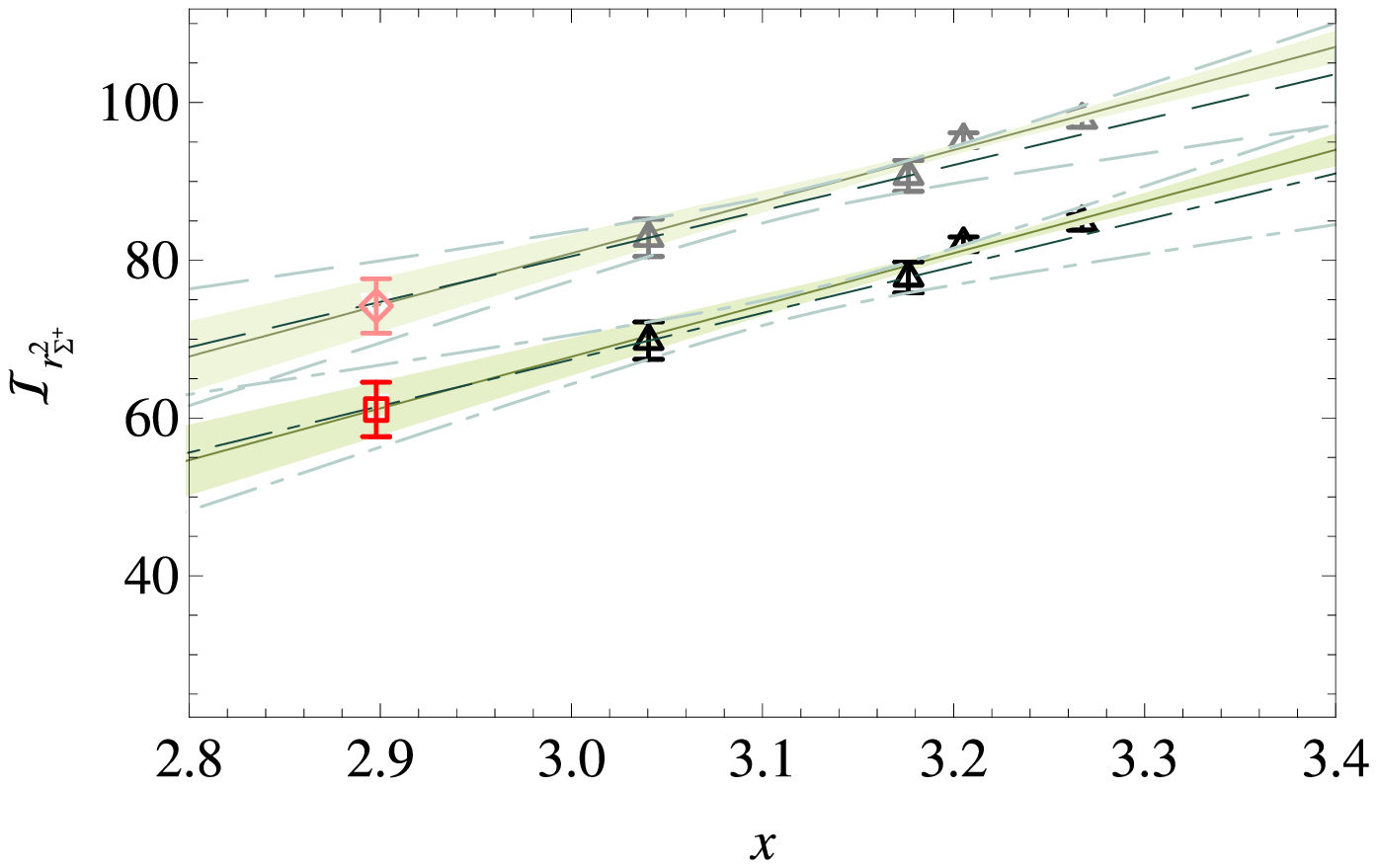}
\includegraphics[width=0.45\textwidth]{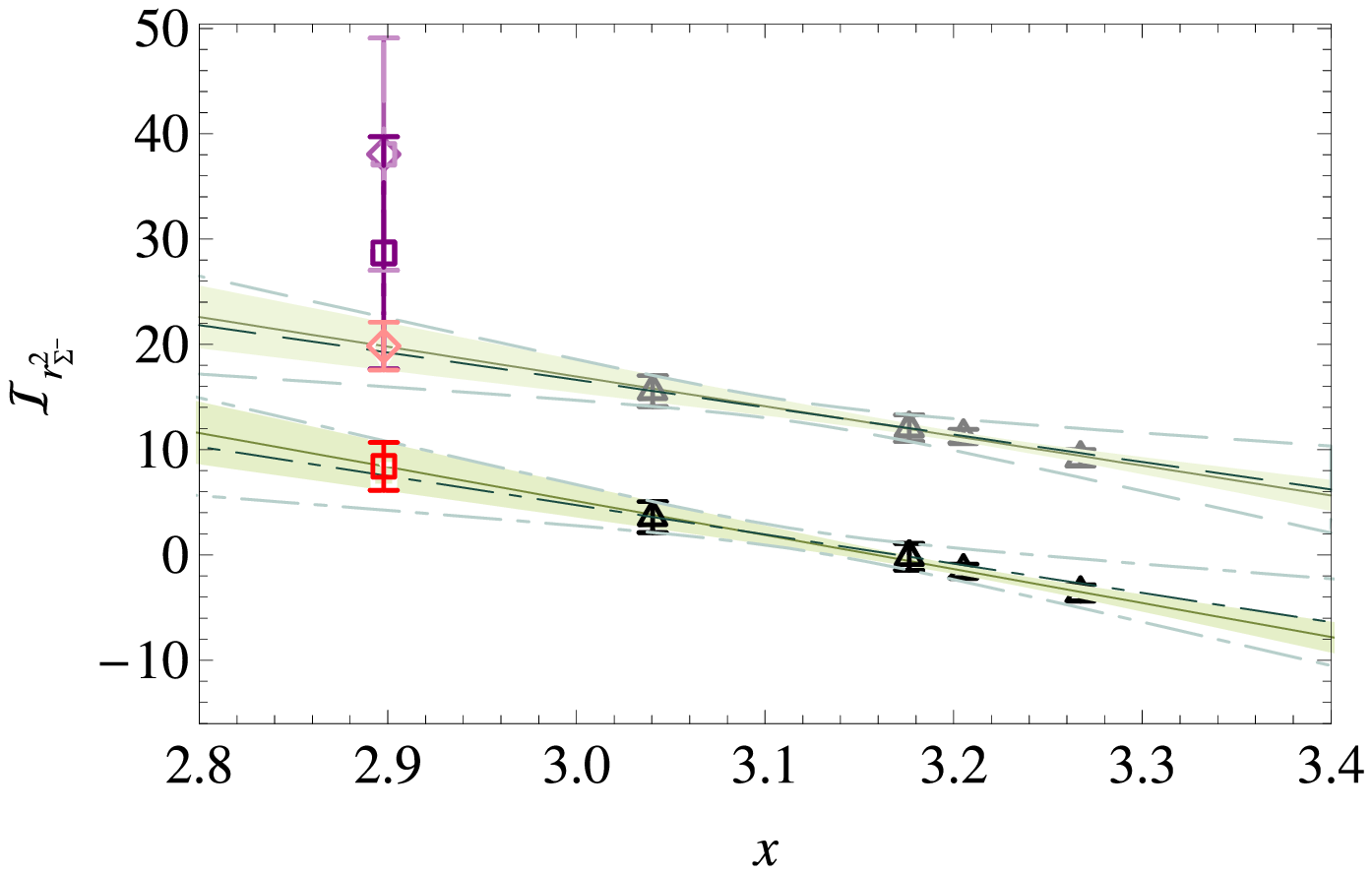}
\includegraphics[width=0.45\textwidth]{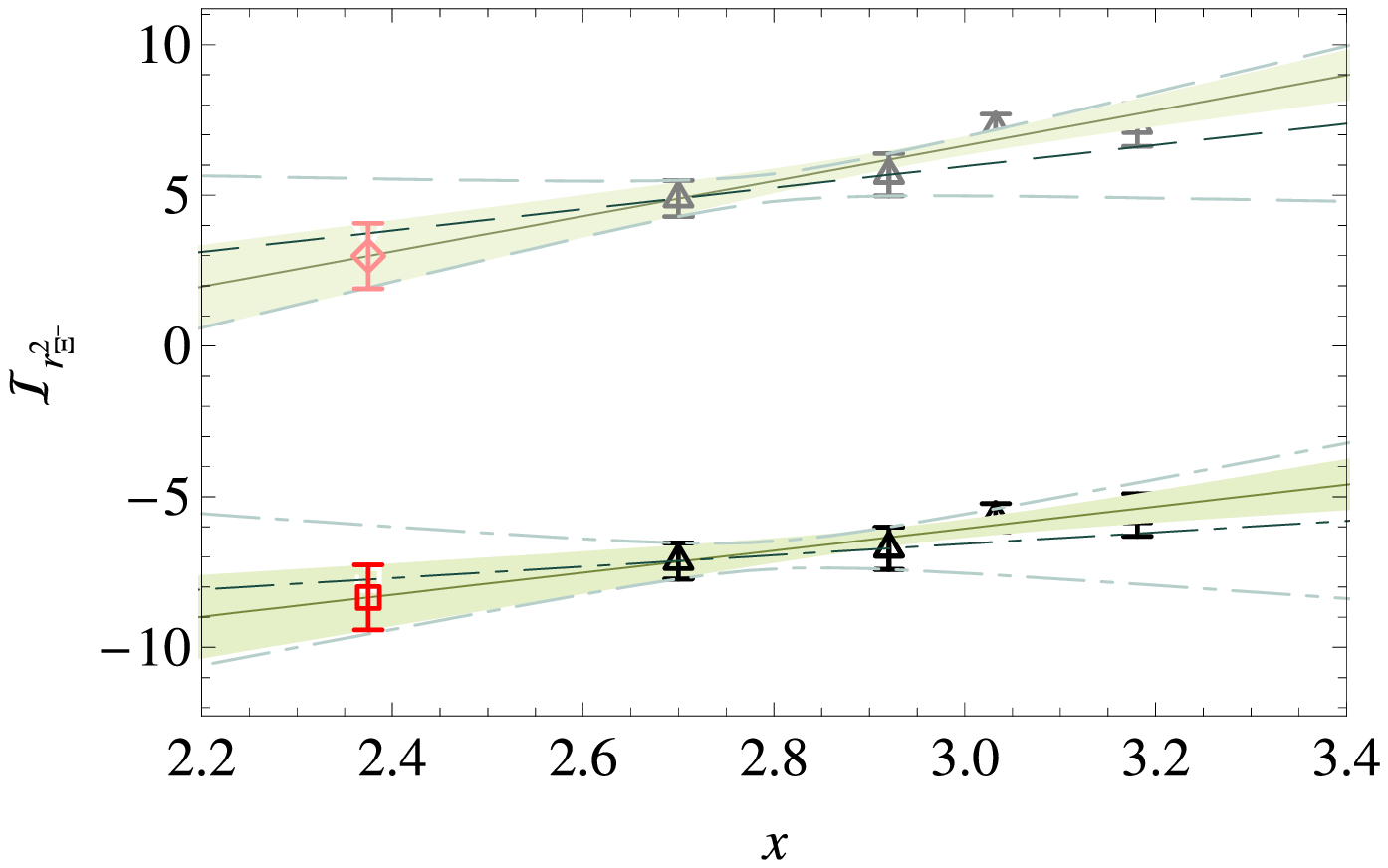}
%\vspace{-0.3in}
\caption{Chiral extrapolations using $C=1.2(2)$ and $C=0$ in Eq.~\ref{eq:ChPT-radii-linear}. The triangles are the lattice data points ${\cal I}_{r^2}$; black points indicate $C=1.2$ values and the gray are $C=0$. The extrapolation using all ensembles is shown as a band and the red (and pink) symbols indicate the
the extrapolated value at physical limit. The (dot-)dashed lines show the extrapolations using the lightest two ensembles only. The purple squares and diamonds are the ${\cal I}_{r^2}$ with experimental values of the charge radii and masses for $C=0$ and $C=1.2(2)$ respectively.
}\label{fig:allB-r2-c1p2-x}
\end{figure}

We reorganize Eq.~\ref{eq:ChPT-radii} as follows:
\begin{eqnarray}\label{eq:ChPT-radii-linear}
{\cal I}_{r^2} &=&16\pi^2 \left(f_\pi^{\rm lat}\right)^2 \langle r_E^2 \rangle - \sum_X [{\cal G}(m_X,\delta,f_\pi^{\rm lat})+{\cal H}(m_X,\delta,f_\pi^{\rm lat})]\\
&=& 96 \pi^2 c^\prime + \left[24 \pi^2\left(\frac{f_\pi^{\rm lat}}{m_B}\right)^2\right] b^\prime. \label{eq:ChPT-radii-linear-fit}
\end{eqnarray}
Such an arrangement converts potential chiral-log terms and a multidimensional parametrization into a simple linear extrapolation. Similar procedures have been adopted by NPLQCD collaboration for $f_K/f_\pi$ extrapolation\cite{Beane:2006kx,Orginos:2006zz}. We study the effects of adding decuplet baryons as dynamical degrees of freedom by setting $C=0$ and see how they affect our final extrapolation results.
Figure~\ref{fig:allB-r2-c1p2-x} shows ${\cal I}_{r^2}$ as functions of $x=\left[24 \pi^2\left(\frac{f_\pi^{\rm lat}}{m_B}\right)^2\right]$ for the $p$, $\Sigma^+$, $\Sigma^-$ and $\Xi^-$ baryons; $C=1.2(2)$ and $C=0$ are shown as black and gray points respectively.
We find that our lattice data fall onto a straight line for both $C$, as Eq.~\ref{eq:ChPT-radii-linear-fit} predicts; thus, the extrapolation is straightforward.
We summarize the extrapolations in Table~\ref{tab:Baryon-r2E} and display fits in Figure~\ref{fig:allB-r2-c1p2-x}.

Using the data from all ensembles we find consistent (within one standard deviation) results for both $C=0$ and $C=1.2(2)$. This indicates that for our data, the effects of the decuplet intermediate states are relatively mild.
The contributions from the ${\cal H}$ function are larger for all channels than those from ${\cal G}$. We observe that the values of ${\cal I}_{r^2}$ with $C$ zero and non-zero are very different; however, it turns out that the difference is absorbed into the $c^\prime$ parameter, while $b^\prime$ remains consistent in both cases. As shown in Fig.~\ref{fig:allB-r2-c1p2-x}, there is only an overall constant shift between the two choices of $C$. Thus, we find that there is not much impact on the final extrapolated charge radii from including decuplet-baryon effects.

To better understand the systematic error associated with our extrapolation, we try restricting our fit to the lightest two ensembles, since the convergence of NLO HBXPT formulations could be poor at higher masses. The results are summarized in Table~\ref{tab:Baryon-r2E}; they are consistent with the fits using all ensembles. The errorbar of the extrapolated values are larger, which is expected since we have fewer data points with larger statistical errors  to constrain the fit. The addition of decuplet degrees of freedom is generally negligible, and we find that HBXPT at NLO has been working well in our extrapolation of charge radii.
Comparing to experiment, we get electric mean-squared charge radii for the proton and $\Sigma^-$ $0.54(7)$ and $0.32(2)$~fm$^2$ which are 3.4 and 4.3$\sigma$ away from the PDG values\cite{PDBook} (0.766(12) and 0.61(16)~fm$^2$ respectively), if all the data are included in the fit. If we concentrate on the results obtained from fits to the lightest two ensembles, the deviation is between 1 and 2$\sigma$. These deviations may be caused by lattice artifacts, such as the omission of lattice spacing and volume extrapolation and the
higher-order effects from HBXPT. In the case of the proton, it has been observed in previous lattice calculations (for example, Refs.~\cite{Gockeler:2007hj,Yamazaki:2007mk,Lin:2008uz}) that the lattice numbers even at 350-MeV pion mass are smaller than experiment. It seems likely that pion-loop contributions are quite large, boosting these values for pion masses smaller than 300~MeV.  Resolving this will be a challenge for future lattice calculations.

In addition to the known radii, we can make predictions for electric charge radii of the $\Sigma^+$ and $\Xi^-$: 0.67(5) and 0.306(15)~fm$^2$ respectively. However, judging from our comparison with the known result for $p$ and $\Sigma^-$.  Since these quantities should have similar systematics as the ones known experimentally  we should add  a systematic error that is 2 to 3 times the statistical error.

%%%%%%%%%%%%%%%%%%%%%%%%%%%%%%%%%%%%%%%%%%%%%%%%%%%%%%%%%%%%%%%%%%%%%%%%%%%

\begin{table}
\begin{center}
\begin{tabular}{c|cccccccc}
\hline\hline
$m_\pi^2(\mbox{GeV}^2)$ & $p$ & $\Sigma^+$ & $\Sigma^-$ & $\Xi^-$ \\
\hline
  0.1256(15) &  0.41(4) &  0.44(3) &  0.357(16) &  0.326(6) \\
  0.246(2) &  0.382(21) &  0.401(18) &  0.329(12) &  0.311(7) \\
  0.3493(17) &  0.350(8) &  0.362(8) &  0.322(5) &  0.309(4) \\
  0.463(3) &  0.312(9) &  0.317(8) &  0.297(6) &  0.291(5) \\
    \hline
$C=0$ & 0.56(7)[0.19]&  0.69(5)[1.09]&  0.35(3)[0.06]&  0.329(15)[0.72]\\
$C=1.2(2)$ & 0.54(7)[0.21]&  0.67(5)[0.86]&  0.32(3)[0.07]&  0.306(15)[0.31]\\
\hline
$C=0$ (2pts) &0.59(14)[n/a]&  0.70(7)[n/a]&  0.34(5)[n/a]&  0.34(3)[n/a]\\
$C=1.2(2)$ (2pts) &0.55(14)[n/a]&  0.67(7)[n/a]&  0.31(5)[n/a]&  0.31(3)[n/a]\\
\hline
 Exp't &  0.766(12) & n/a &  0.61(16) & n/a \\
\hline\hline
\end{tabular}
\end{center}
\caption{\label{tab:Baryon-r2E}Mean-squared charge radii for octet baryons. The numbers in square brackets indicate the $\chi^2/\mbox{dof}$ of the fits.}
\end{table}

%%%%%%%%%%%%%%%%%%%%%%%%%%%%%%%%%%%%%%%%%%%%%%%%%%%%%%%%%%%%%%%%%%%%%%%%%%%%%%%%%%%%%%%

\subsection{Magnetic Moments and Magnetic Radii}\label{subsec:MagneticMoments}

Studying the momentum-transfer dependence of magnetic form factors gives us the magnetic moment via
\begin{eqnarray}\label{eq:magMoment}
\mu_B &=&  G_{M}^B(Q^2=0)
\end{eqnarray}
with natural units $\frac{e}{2M_B}$, where $M_B$ are the baryon ($B\in\{N,\Sigma,\Xi\}$) masses. To compare among different baryons, we convert these natural units into nuclear magneton units $\mu_N=\frac{e}{2M_N}$; therefore, we convert the magnetic moments with factors of $\frac{M_N}{M_B}$.
The magnetic radii are also obtained through
\begin{eqnarray}\label{eq:GM-radii}
\langle r_{M}^2\rangle &=& (-6)\frac{d}{dQ^2}\left(\frac{G_{M}(Q^2)}{G_{M}(0)}\right)\Big|_{Q^2=0}.
\end{eqnarray}
The magnetic moment is linked to the Pauli magnetic moment (or anomalous magnetic moment, $\kappa_B=F_{2,B}(Q^2=0)$) by $\mu_B=\kappa_B+e_B$, where $e_B$ is the charge of baryon. The magnetic radii are related to the Pauli radii, $r_{2,B}$ through  $r_{M,B}^2=r_{2,B}^2+\frac{3}{2}\frac{\kappa_B}{2M_B^2}$.

\begin{figure}
\includegraphics[width=0.45\textwidth]{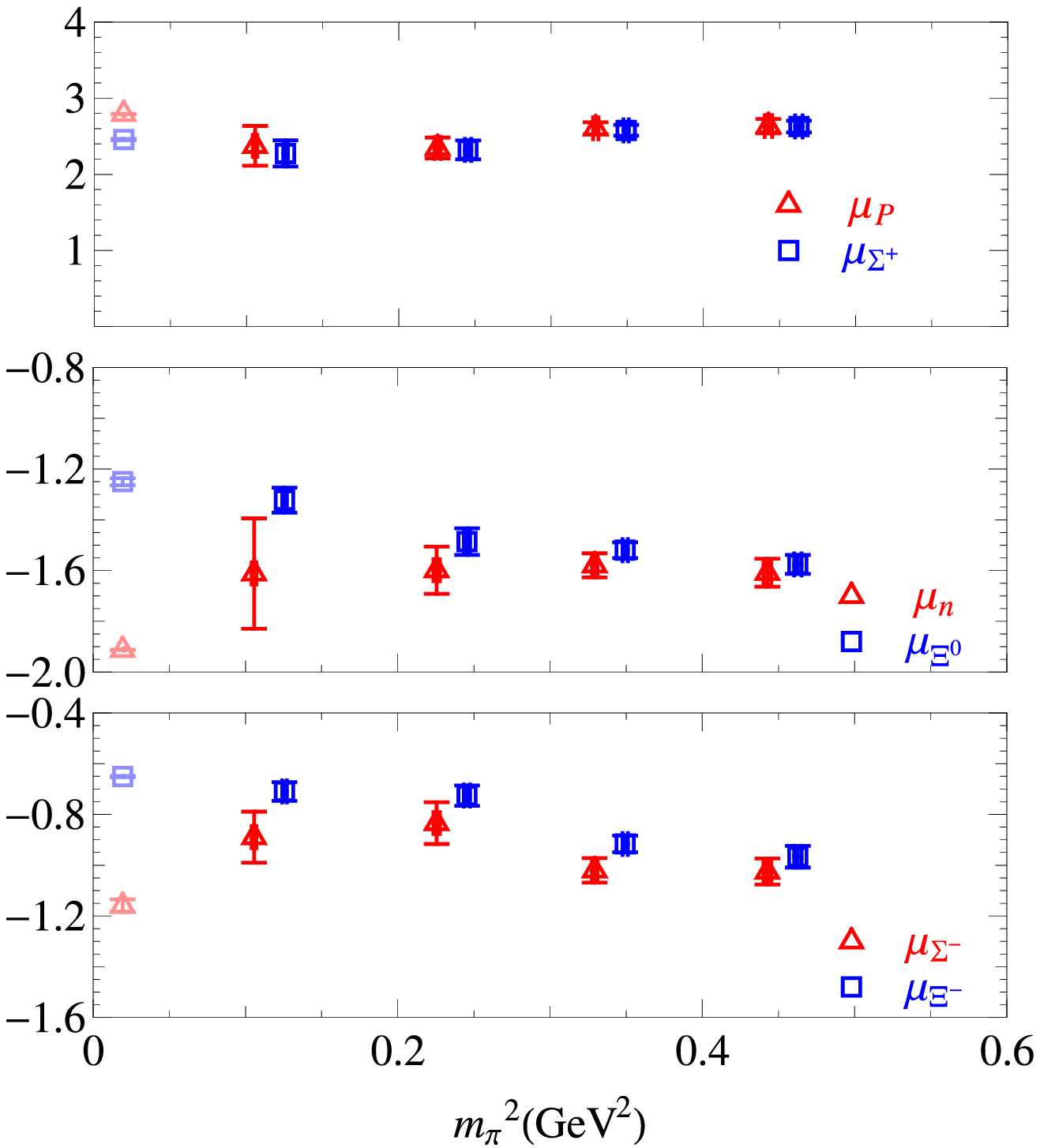}
\includegraphics[width=0.45\textwidth]{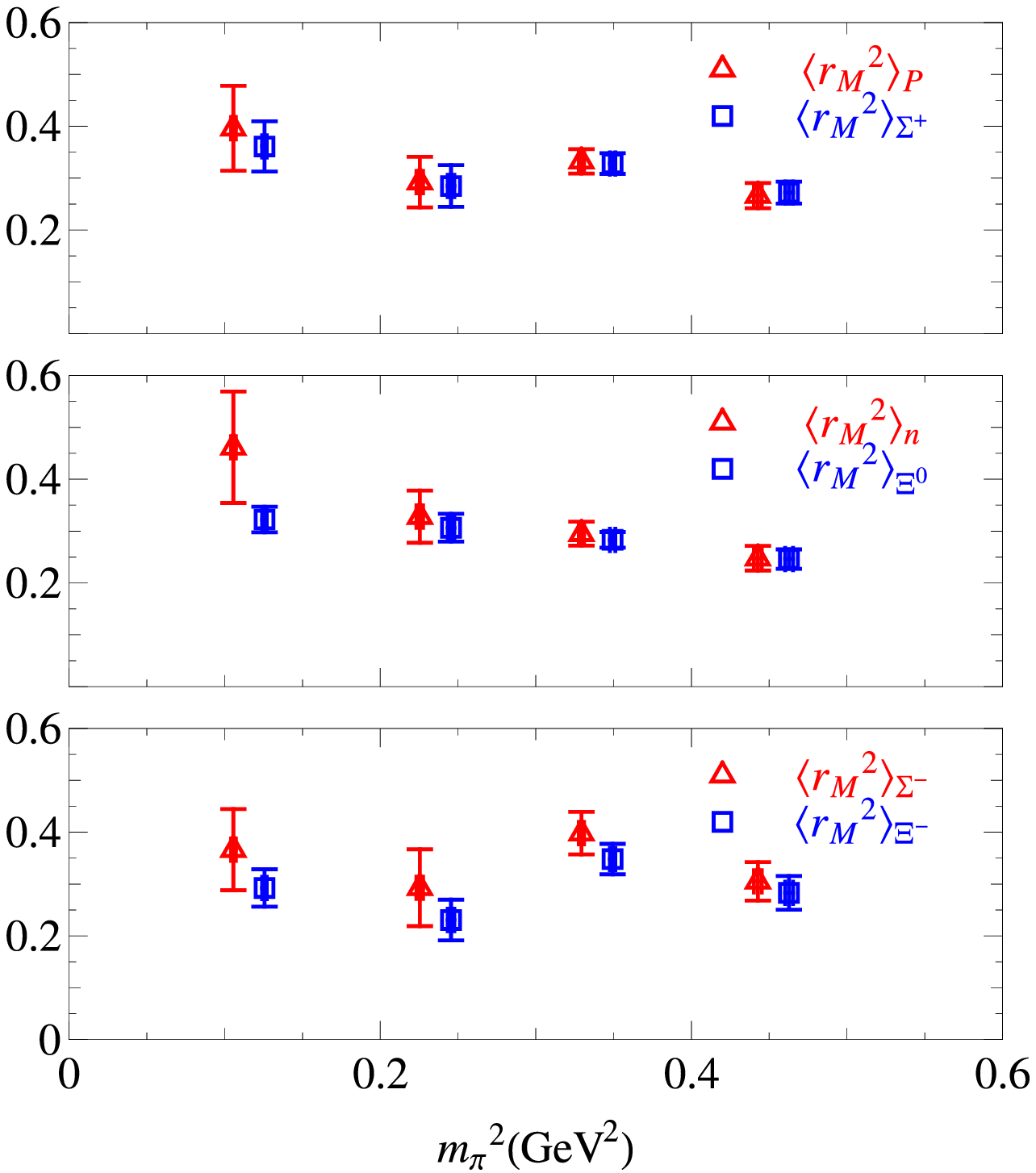}
\caption{Left: Baryon magnetic moments in units of $\mu_N$ as functions of $m_\pi^2$ (in GeV$^2$).  The leftmost points are the experimental numbers.\\
Right: The magnetic mean-squared radii, in units of $\mbox{fm}^2$, as functions of $m_\pi^2$ (in GeV$^2$)
}\label{fig:allB-muMrErM_mpi2}
\end{figure}

It is straightforward to extrapolate the magnetic form factor using the dipole form, and the results are summarized in Tables~\ref{tab:Baryon-mus}
and \ref{tab:Baryon-r2m}).
If SU(3) symmetry is exact, we expect the octet magnetic moments to obey the Coleman-Glashow relations\cite{Coleman:1961jn}:
\begin{eqnarray}\label{eq:magMomentSU3}
\mu_{\Sigma^+} = \mu_p;\,\,
\mu_{\Xi^0} = \mu_n;\,\,
\mu_{\Xi^-} = \mu_{\Sigma^-}.
\end{eqnarray}
In nature, such an SU(3) symmetry approximately holds between the proton and $\Sigma^+$, but is broken by more than 50\% in the case of $\{n,\Xi^0\}$ and $\{\Sigma^-,\Xi^-\}$. The left panel of Figure~\ref{fig:allB-muMrErM_mpi2} shows the magnetic moments of each baryon compared with its SU(3) partner, as paired up in Eq.~\ref{eq:magMomentSU3}.
We find that as seen in experiment, the SU(3) breaking on the magnetic moments are rather small. As we go to larger pion masses (that is, as the light mass goes to the strange mass), the discrepancy gradually goes to zero as SU(3) is restored. But even at our lightest pion mass, around~350 MeV, the effects of SU(3) symmetry breaking effect can be ignored.
In all the baryon magnetic moments calculated in this work, we find only small changes as we decrease the pion mass to around 350~MeV.

The right panel of Figure~\ref{fig:allB-muMrErM_mpi2} shows magnetic radii from a dipole fit to the magnetic form factors. Once again, we see small SU(3) flavor breaking even on our lightest pion-mass ensembles. Comparing with Ref.~\cite{Boinepalli:2006xd}, where the quenched approximation is used, we observe that $\langle r_M^2 \rangle_p \approx \langle r_M^2 \rangle_{\Sigma^+}$; however, a larger rate of increase with decreasing pion mass is shown in their data. $\langle r_M^2 \rangle_n$ becomes larger than $\langle r_M^2 \rangle_{\Xi^0}$ only at the lightest pion mass. Overall, our data suggests that the magnetic radius's dependence on the quark content is mild and may only start to dominate at very light pion mass, a result which is quite different from quenched calculations.

\begin{figure}
\includegraphics[width=0.45\textwidth]{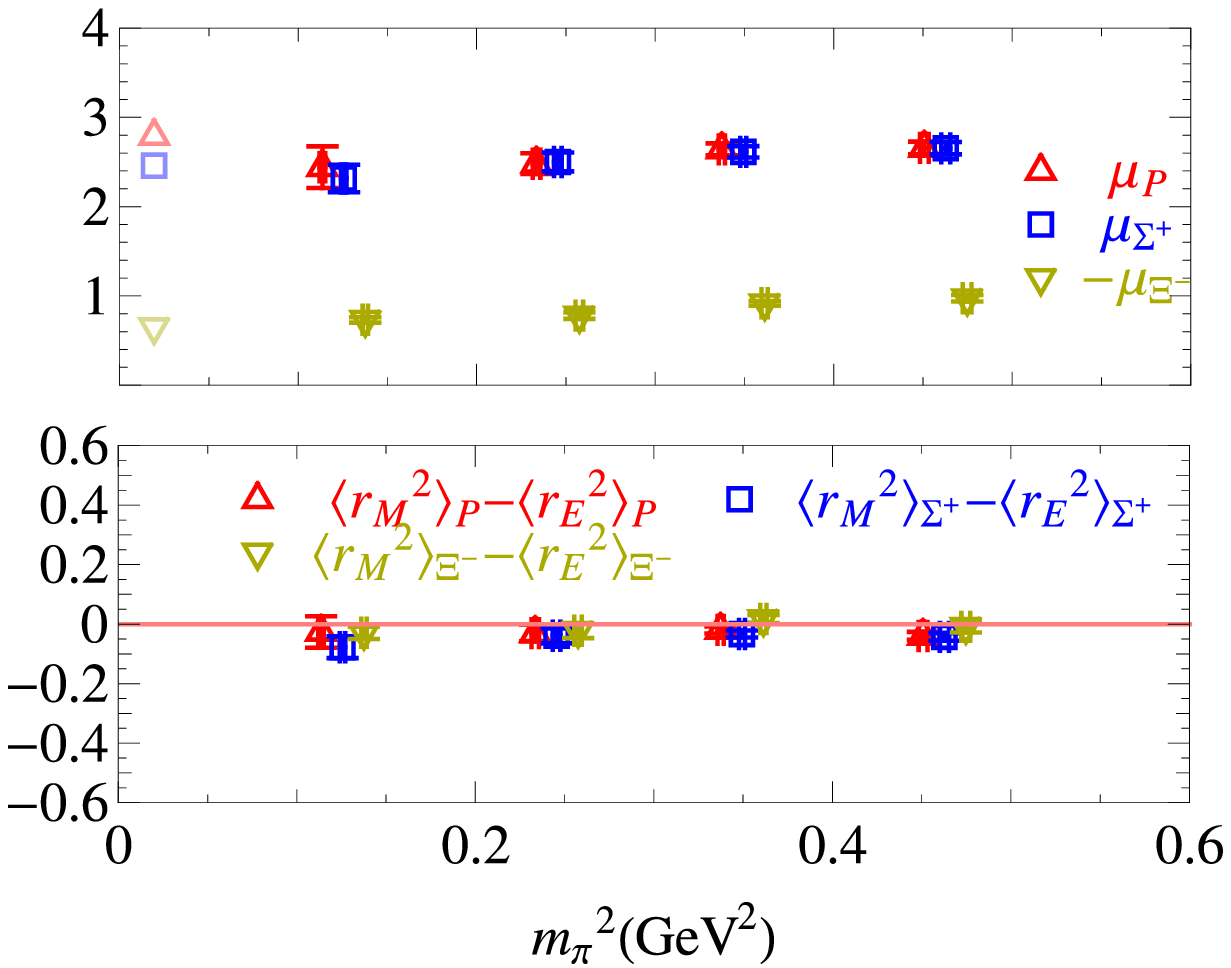}
\includegraphics[width=0.45\textwidth]{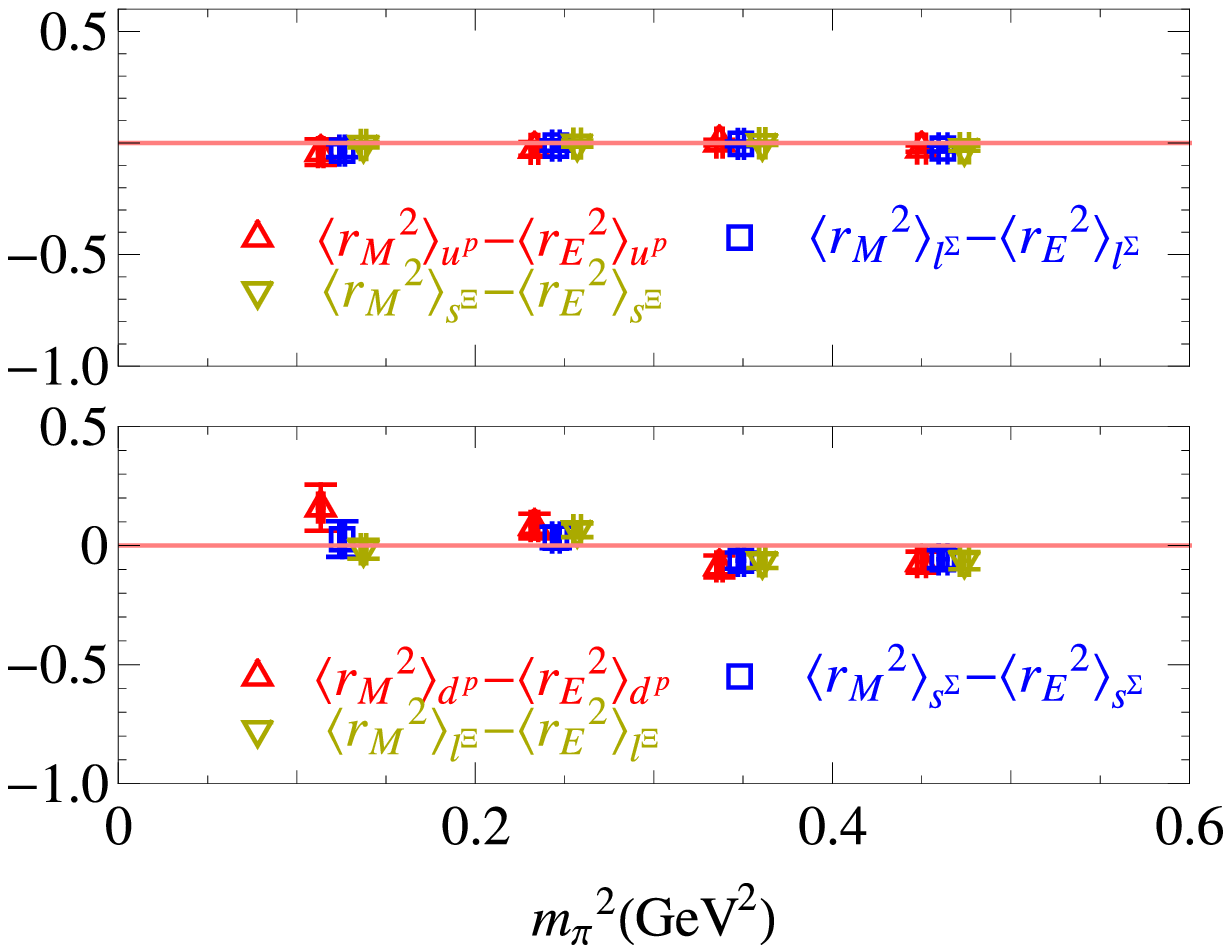}
\caption{Left: Magnetic moments (top, in units of $\mu_N$) and the differences between magnetic and electric mean-squared radii (bottom, in units of $\mbox{fm}^2$) as functions of $m_\pi^2$ (in GeV$^2$) for the proton, $\Sigma^+$ and $\Xi^-$ from fitting over $G_M/G_E$\\
Right: The differences between magnetic and electric mean-squared radii quark contributions from fitting over $G_M/G_E$
}\label{fig:allB-muMrErM_mpi2-GEM}
\end{figure}

Alternatively, we can obtain the magnetic moments and radii from polynomial fitting to the ratio of magnetic and electric form factors, $G_M/G_E$. From the definition of the electric and magnetic radii, Eqs.~\ref{eq:GEradii} and \ref{eq:GM-radii}, we expect that $G_E^B/G_M^B \approx \frac{1}{\mu_p} \frac{Q^2}{6}(\langle r_E^B \rangle_p - \langle r_M^2 \rangle_B)$.
From Figures~\ref{fig:N_Gs}, \ref{fig:Sig_Gs} and \ref{fig:Xi_Gs}, we expect that ratio to be around 1 with small deviations; thus, we fit $G_M/G_E \approx A (1+B Q^2)$ where magnetic moment is $\mu=A G_E(0)$, and $B$ is proportional to $\langle r_M^2 \rangle-\langle r_E^2 \rangle$. In the case of $n$ and $\Xi^0$, we use $G_{E}^p$ and $G_E^{\Xi^-}$ in the ratio instead of $G_E^{n}$ and $G_E^{\Xi^0}$.
Table~\ref{tab:Baryon-mus-GEM} summarizes the magnetic moments, which appear consistent with the dipole extrapolation approach. The left panel of Figure~\ref{fig:allB-muMrErM_mpi2-GEM} shows the magnetic moment (top) and the difference between the electric and magnetic radii for the proton, $\Sigma^+$ and $\Xi^-$. Both results are consistent with what we obtained from the dipole extrapolations. We examine the radii differences from the quark contributions, as shown in the right panel of Figure~\ref{fig:allB-muMrErM_mpi2-GEM} and observe less than 10\% discrepancy.
The ratio approach also benefits from cancellation of noise due to the gauge fields, and thus it has smaller statistical error. Therefore, we will concentrate on the results from this approach for the rest of this work.

%%%%%%%%%%%%%%%%%%%%%%%%%% Extrapolation %%%%%%%%%%%%%%%%%%%%%%%%%%%%%%%%%%%5

\begin{figure}
\includegraphics[width=0.45\textwidth]{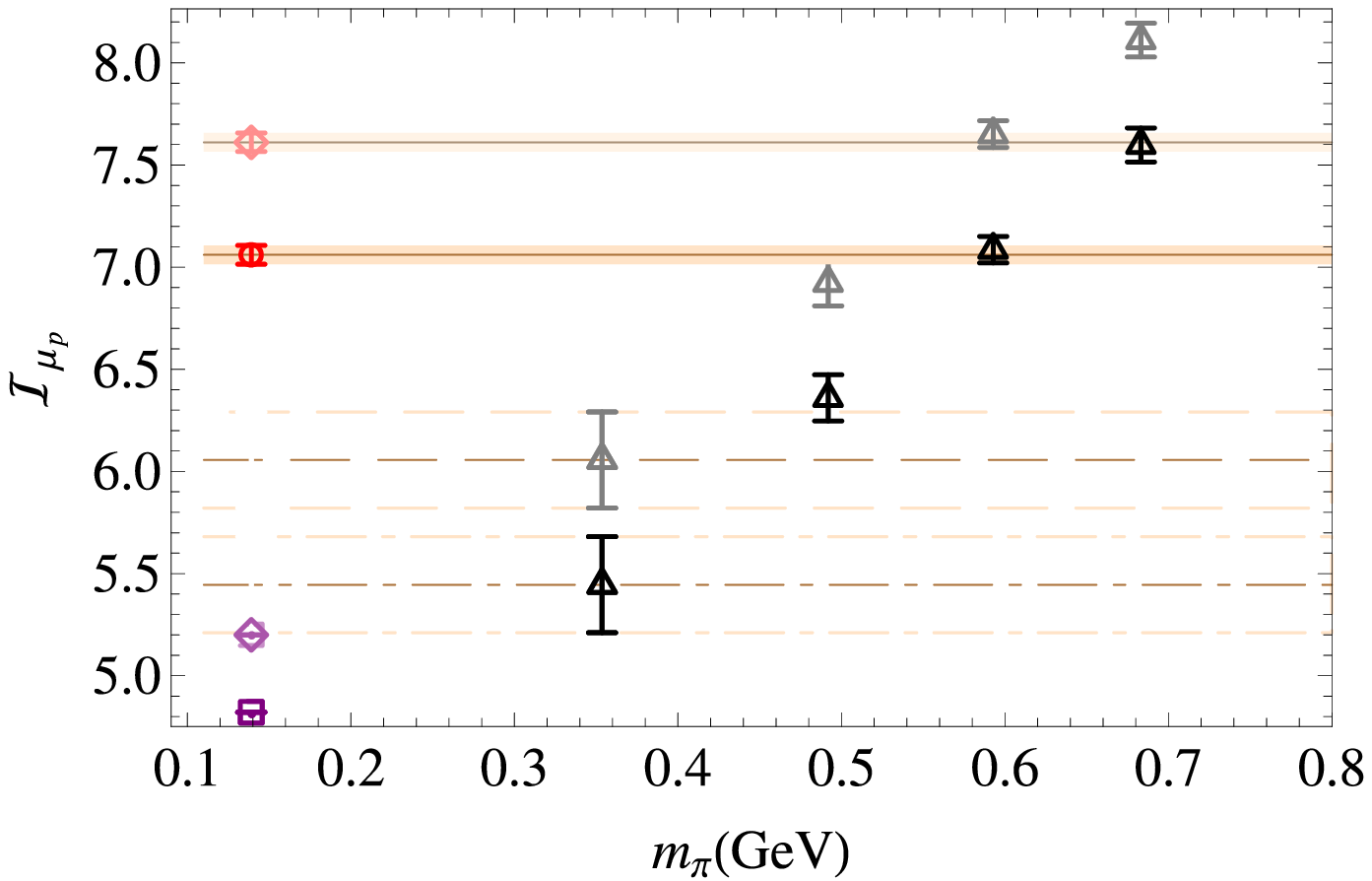}
\includegraphics[width=0.45\textwidth]{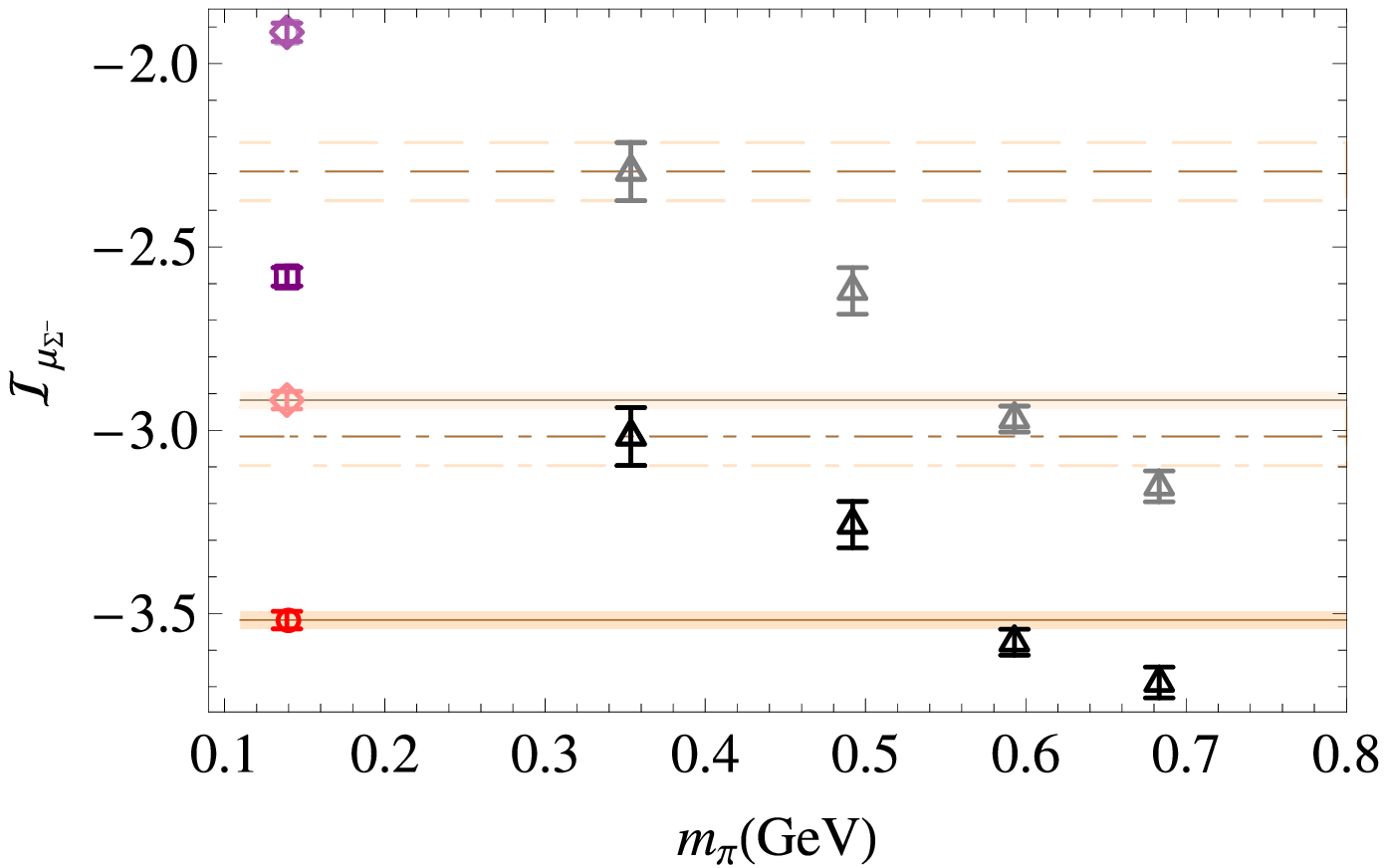}
\caption{Examples of ${\cal I}_{\mu_B}$ as functions of pion mass. The lines and bands indicate the chiral extrapolations using Eq.~\ref{eq:ChPT-mu-linear}. The symbols are as in Fig.~\ref{fig:allB-r2-c1p2-x}.
}\label{fig:allB-mu-c1p2-mpi}
\end{figure}

The HBXPT for octet magnetic moments has been derived in Refs.~\cite{Jenkins:1992pi,Kubis:1999xb,Chen:2001yi,Tiburzi:2005is}. The most general form including the dynamical decuplet degrees of freedom is
\begin{eqnarray}\label{eq:ChPT-mu}
\mu_B &=& b^\prime + \frac{M_B}{4\pi\left(f_\pi^{\rm lat}\right)^2}\sum_X \left\{\zeta_X M_X + \zeta^{\prime}_X \frac{C^2}{\pi} \left[-{\cal F}(M_X,\delta,f_\pi^{\rm lat}) +\frac{5}{3}\right]\right\},
\end{eqnarray}
where $b^\prime$, $\delta$, $m_B$, $X$, $C$ and ${\cal F}$ are defined in Sec.~\ref{subsec:ChargeRadii}, the scale $\mu$ is replaced by $f_\pi^{\rm lat}$, and $\zeta^{(\prime)}_X$ can be found in Tables~II--IX of Ref.~\cite{Tiburzi:2005is}. %$d^\prime=(Q\mu_F+\alpha_D\mu_D)$
As before, we can rewrite Eq.~\ref{eq:ChPT-mu} as
\begin{eqnarray}\label{eq:ChPT-mu-linear}
{\cal I}_{\mu_B} &=& \mu_B - \frac{M_B}{4\pi\left(f_\pi^{\rm lat}\right)^2}\sum_X \left\{\zeta_X M_X + \zeta^{\prime}_X \frac{C^2}{\pi} \left[-{\cal F}(M_X,\delta,f_\pi^{\rm lat}) +\frac{5}{3}\right]\right\}\\
&=& b^\prime;
\end{eqnarray}
therefore, NLO SU(3) HBXPT predicts the lattice data to be flat over the various pion masses. We plot ${\cal I}_{\mu_B}$ from our calculations in Fig.~\ref{fig:allB-mu-c1p2-mpi} (with decuplet dynamical freedom (black points) and without (gray)) as triangles for the proton and $\Sigma^-$.  Our data do not show a flat plateau, but instead are linearly dependent on the pion mass.

There are a couple of possibilities which might explain this behavior:
\begin{itemize}
\item
There might be systematic errors entering our magnetic moments during the extrapolation to zero momentum. However, this can be excluded since we compare both dipole extrapolation of magnetic moments and linear extrapolation to $Q^2=0$ of the ratios $G_M/G_E$, finding consistent magnetic moments.
\item
Or our pion masses may still be too heavy for the NLO formulation to work. We might simply need to go to NNLO to find a better description.
\end{itemize}

To better extrapolate our data, we examine for the possibility of adding a selection of terms from NNLO HBXPT. Following suggestions in Ref.~\cite{Tiburzi:2008}, we expect corrections from terms proportional to $m_\pi^2$ and $m_\pi^2 \ln \frac{m_\pi^2}{\mu^2}$ and that there is a parameter which depends on $\mu$ to absorb the divergence of the second term. %The first term should have a fixed coefficient.
Therefore, we propose a modified ${\cal I}_{\mu_B}$:
\begin{eqnarray}\label{eq:ChPT-mu-linear2}
{\cal I}^\prime_{\mu_B}={\cal I}_{\mu_B} - \omega m_\pi^2 \ln \frac{m_\pi^2}{\mu^2} = b^\prime + e^\prime m_\pi^2.
\end{eqnarray}
However, the coefficient, $\omega$, has not yet been calculated.
Note that in NLO HBXPT for both charge radii and magnetic moments, there are also log terms when one includes the decuplet degree of freedom. In both cases, we only see an overall constant shift throughout our pion-mass regions which does not affect the
charge radii or magnetic moments at the physical point. Therefore, we will neglect such contribution and extrapolate ${\cal I}_{\mu_B}$ in terms of $m_\pi^2$.

From now on, we concentrate on the $C=1.2$ case only. Tables~\ref{tab:Baryon-mus-GEM} and Figure~\ref{fig:allB-mu-c1p2-mpi} summarize the fitted results when including the NNLO $m_\pi^2$ terms. We first note that the fit is dramatically improved due to the introduction of the additional free parameter. When we include the pion masses all the way up to 700 MeV, we find that the magnetic moments for the 6 octet baryons are less than $3\sigma$ away from experiment. However, all of the fitted $\chi^2/{\rm dof}$ are larger than one, which is still pretty poor compared with the charge-radii case. The pion masses in our calculations may still be too large to apply HBXPT. We try extrapolating only the lowest two pion masses to check systematic error due to heavy pions. We find that the magnetic moments are consistent within errorbars regardless of the number of points included. It is possibly significant that as higher masses are excluded, the extrapolated central values move toward the experimental values, except for the cascades. In the cases of $p$, $n$ and $\Sigma^+$, the magnetic moments are consistent with experimental ones.

%%%%%%%%%%%%%%%%%%%%%%%%%%%%%%%%%%%%%%%%%%%%%%%%%%
SU(6) symmetry predicts the ratio $\mu_{d^p}/\mu_{u^p}$ should be around $-1/2$. Compared with what we obtain in this work, as shown in the left panel of Figure~\ref{fig:mu-special-case}, the ratio agrees within $2\sigma$ for all the pion-mass points. The heaviest two pion points have roughly the same magnitude as in the quenched calculation\cite{Boinepalli:2006xd}. However, at the lightest two pion masses, they are consistent with the $-1/2$ value. The difference could be due to sea-quark effects, which become larger as the pion mass becomes smaller. A naive linear extrapolation through all the points gives $-0.50(10)$, consistent with the SU(6) symmetry expectations.
%, since it is dominated by the heaviest two pion-mass points.
% is preserved in the lattice calculations.

We also check the sum of the magnetic moments of the proton and neutron, $\mu_{p}+\mu_{n}$, which should be about 1 from isospin symmetry. The right panel of Figure~\ref{fig:mu-special-case} shows our lattice calculation as a function of squared pion mass. Again, the values from different pion masses are consistent with each other within 2 standard deviations and differ from 1 by about the same amount. A naive linear extrapolation suggests the sum is 0.78(13), which is consistent with experiment but about $2\sigma$ away from 1. This symmetry is softly broken, possibly due to finite lattice-spacing effects. Finer lattice-spacing calculations would be needed to confirm this.

\begin{figure}
\includegraphics[width=0.45\textwidth]{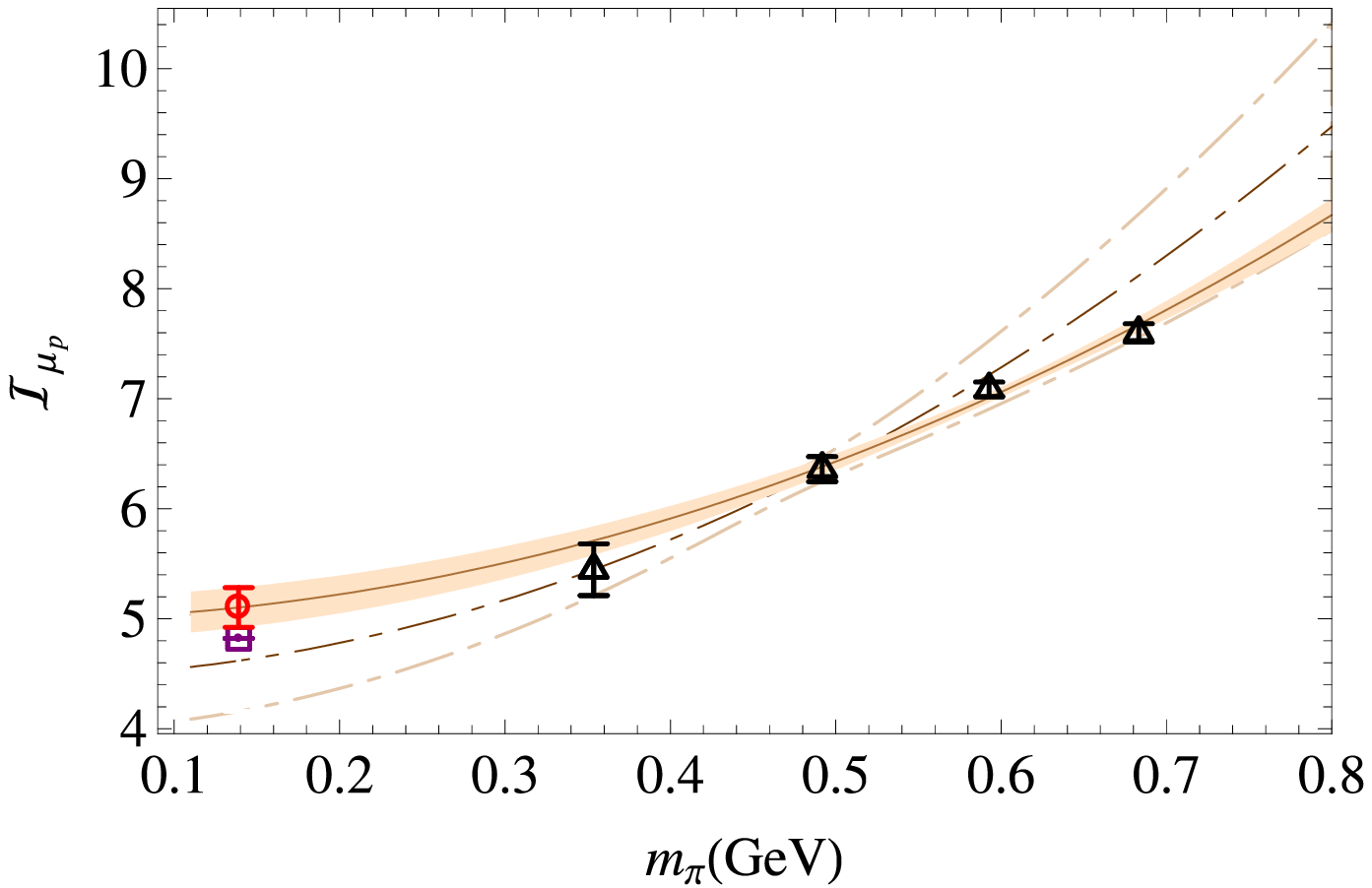}
\includegraphics[width=0.45\textwidth]{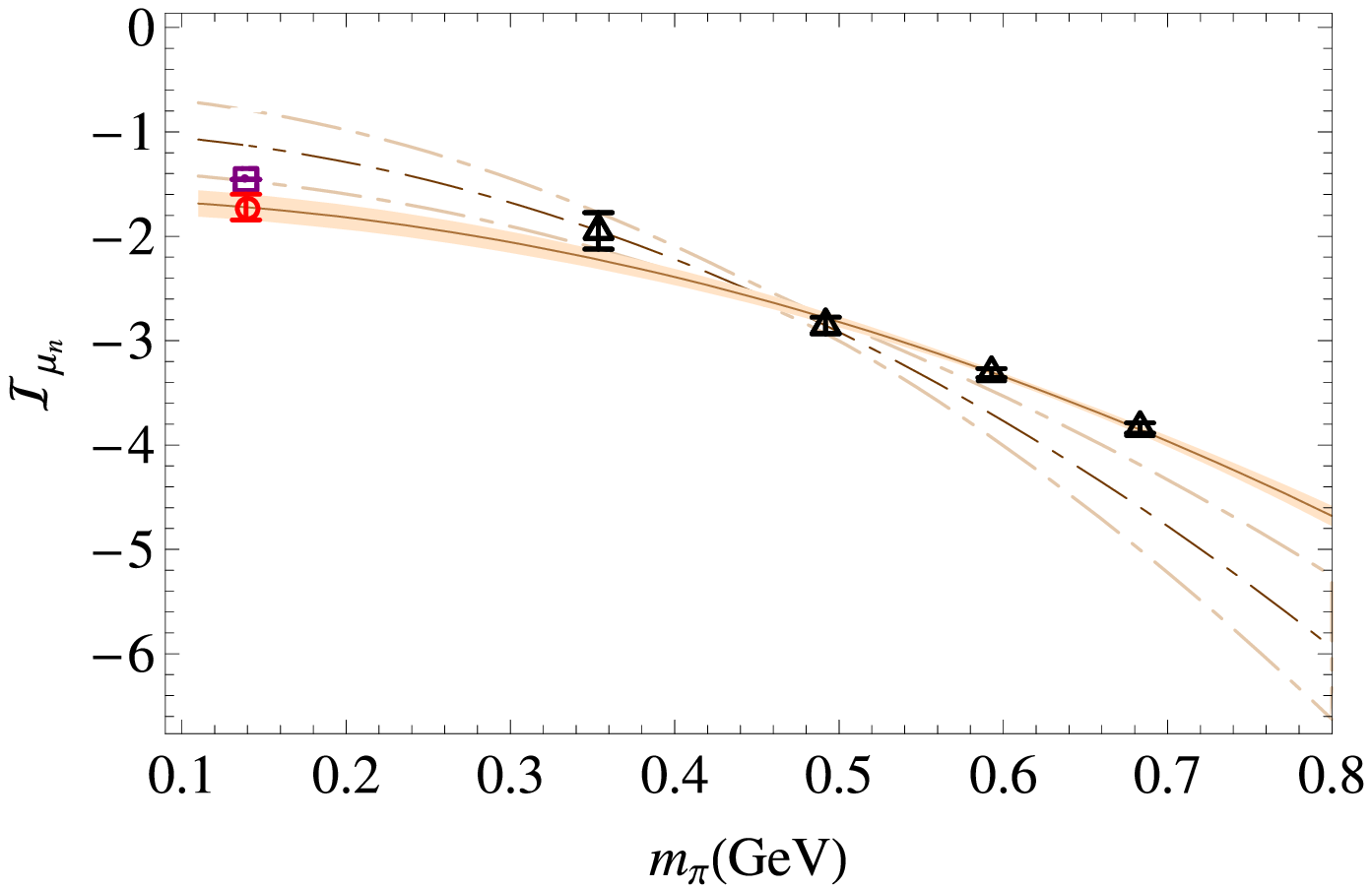}
\includegraphics[width=0.45\textwidth]{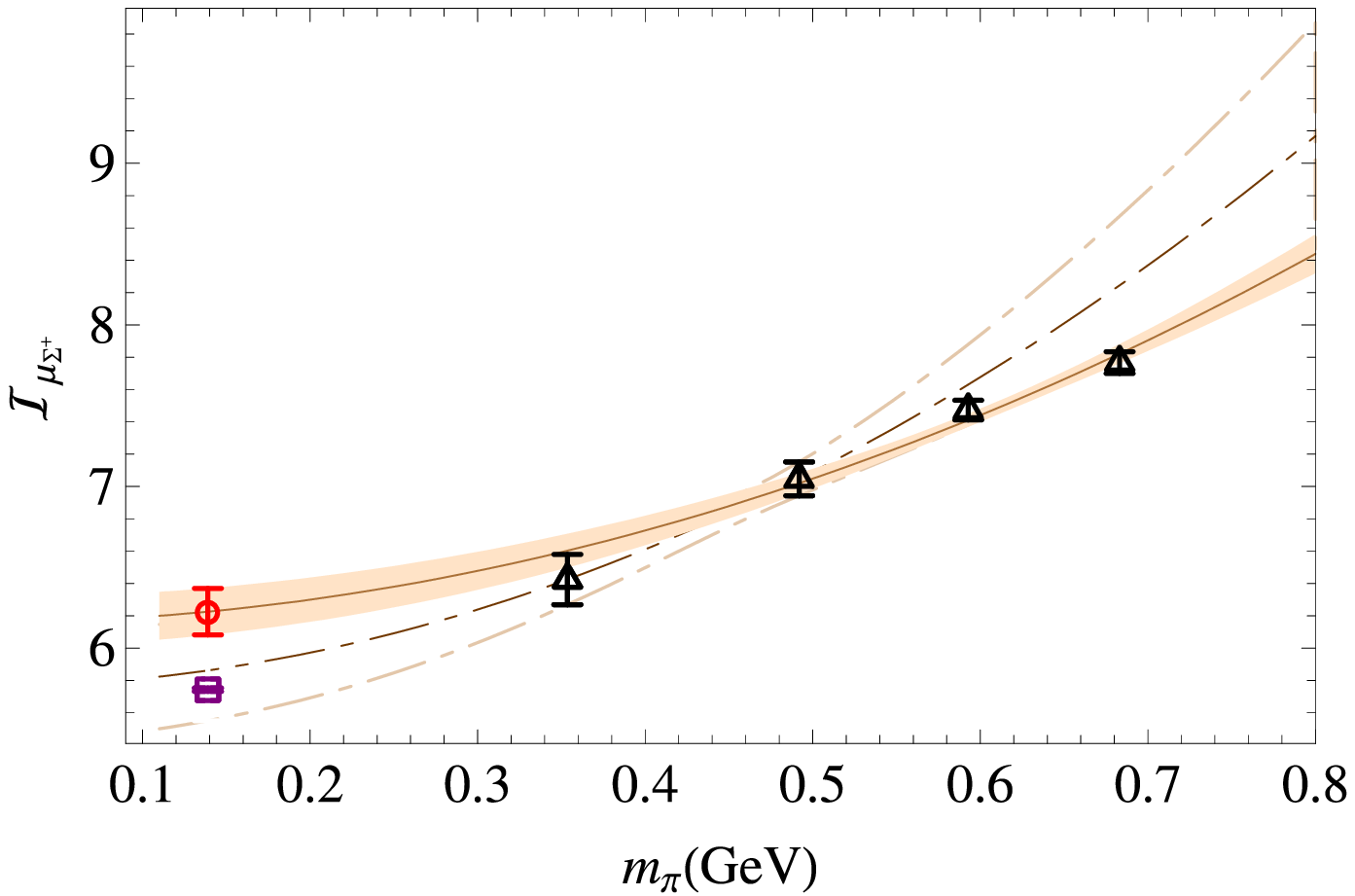}
\includegraphics[width=0.45\textwidth]{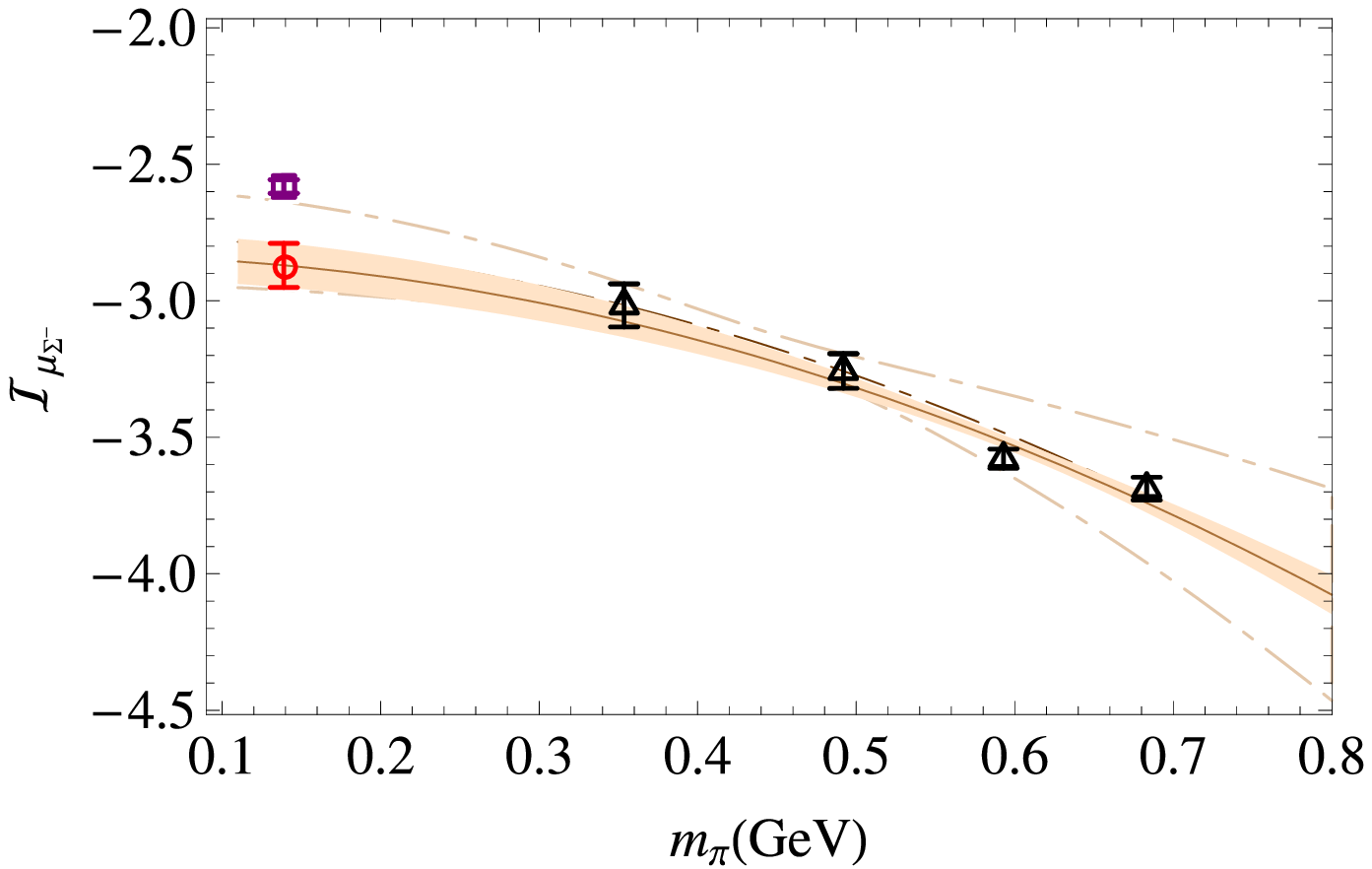}
\includegraphics[width=0.45\textwidth]{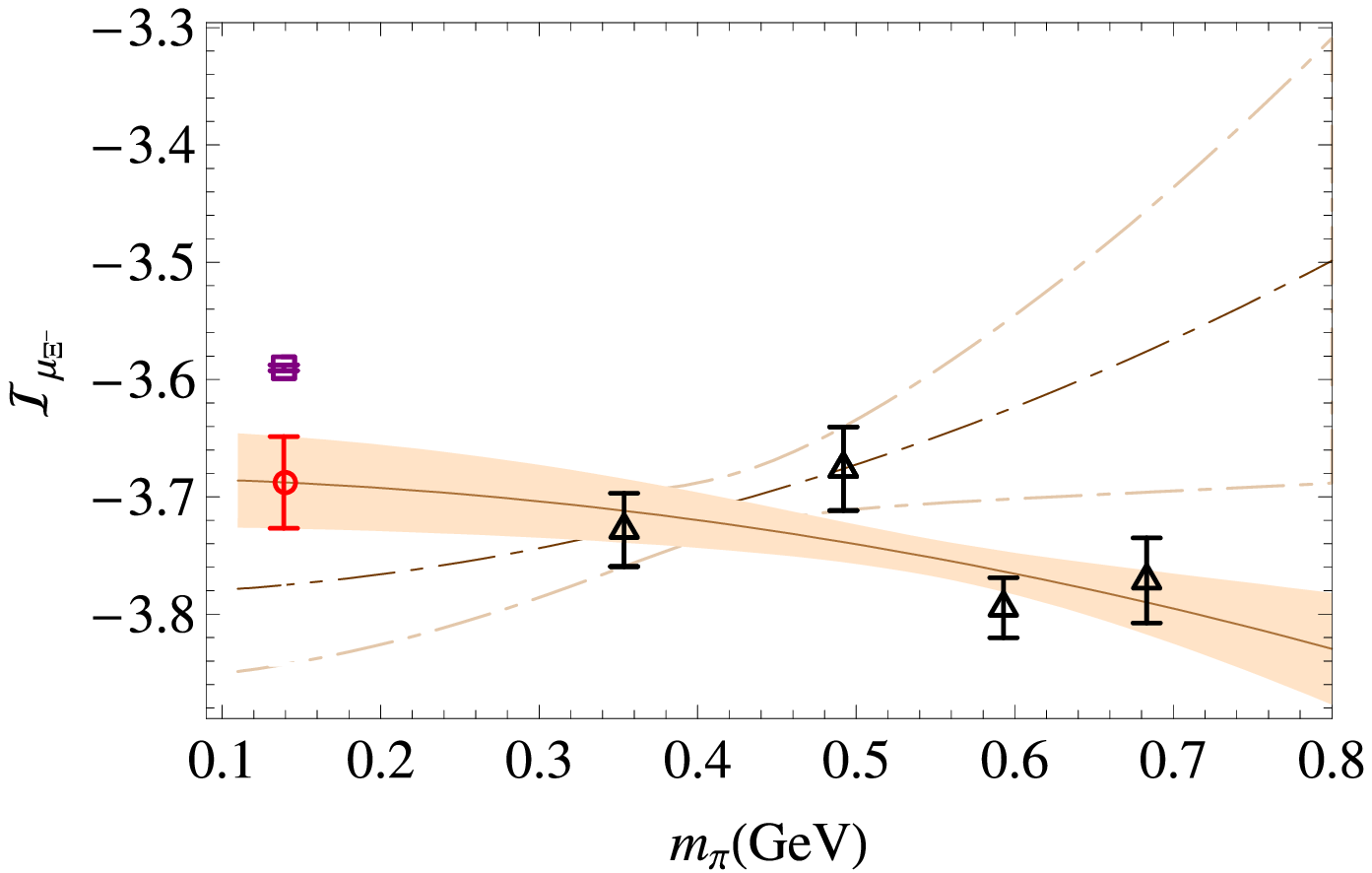}
\includegraphics[width=0.45\textwidth]{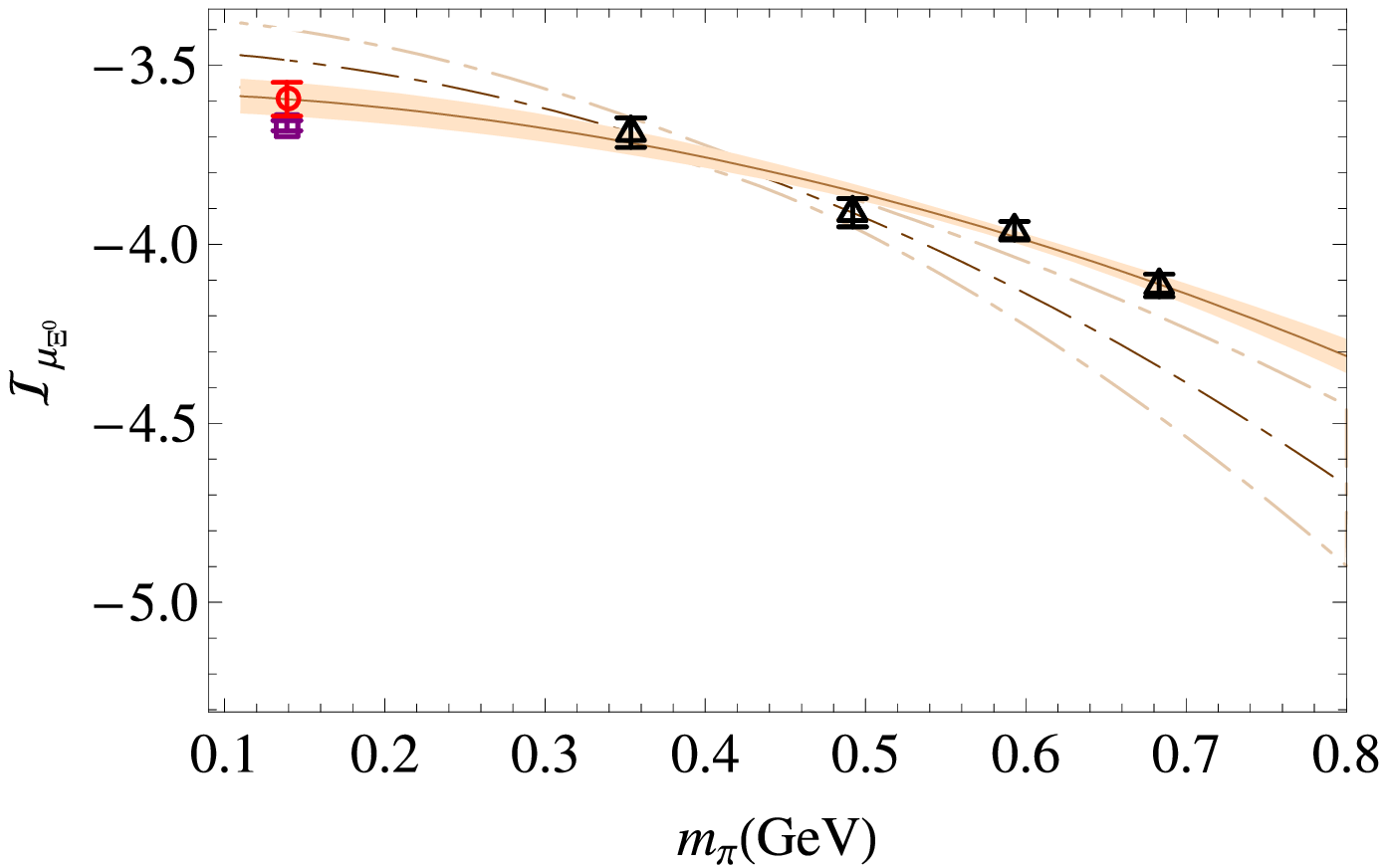}
\caption{Chiral extrapolations of ${\cal I}_{\mu_B}$ with $C=1.2(2)$ according to Eq.~\ref{eq:ChPT-mu-linear2}. The triangles are the lattice data for ${\cal I}_{\mu_B}$, the circles are the extrapolations to the physical point, and the squares are the corresponding experimental values in terms of ${\cal I}_{\mu_B}$.
The solid band is the extrapolation using all ensembles while the dot-dashed lines use the lightest two pion masses only.
}\label{fig:allB-mu-c1p2-mpi2}
\end{figure}

\begin{figure}
\includegraphics[width=0.45\textwidth]{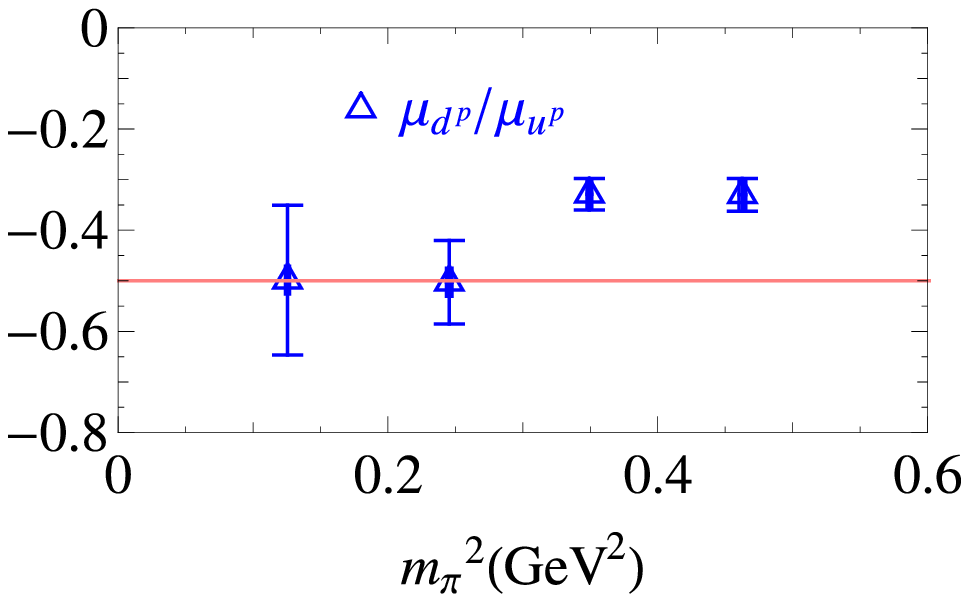}
\includegraphics[width=0.45\textwidth]{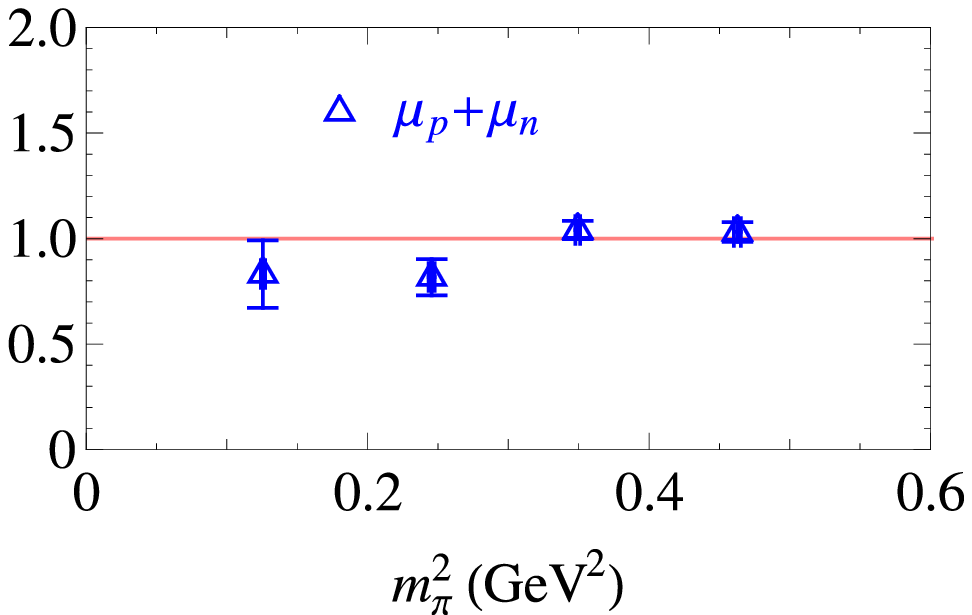}
\caption{Left: Magnetic moment ratios of up and down-quark contributions inside the proton. The straight line indicates the SU(6) prediction of $-1/2$.\\
Right: The sum of the magnetic moments of the proton and neutron. The line indicates 1, as predicted by isospin symmetry
}\label{fig:mu-special-case}
\end{figure}

%%%%%%%%%%%%%%%% from dipole extraplations %%%%%%%%
\begin{table}
\begin{center}
\begin{tabular}{c|ccccccccc}
\hline\hline
$m_\pi^2({\rm GeV}^2)$ & $p$ & $n$ & $\Sigma^+$ & $\Sigma^-$ & $\Xi^-$ & $\Xi^0$ \\
\hline
  0.1256(15) &  2.4(3) &  $-$1.6(2) &  2.27(17) & $-$0.89(10) & $-$0.71(4) &  $-$1.32(5) \\
  0.246(2) &  2.35(14) &  $-$1.60(9) &  2.32(12) & $-$0.83(8) & $-$0.73(4) &  $-$1.49(5) \\
  0.3493(17) &  2.60(8) &  $-$1.58(5) &  2.58(7) &  $-$1.02(5) & $-$0.92(3) &  $-$1.52(3) \\
  0.463(3) &  2.63(10) &  $-$1.61(6) &  2.63(8) &  $-$1.03(5) & $-$0.97(4) &  $-$1.58(4) \\
\hline\hline
\end{tabular}
\end{center}
\caption{\label{tab:Baryon-mus}$\mu_B$ (in units of $\mu_N$) for octet baryons from dipole-fitted magnetic form factors}
\end{table}

\begin{table}
\begin{center}
\begin{tabular}{c|cccccc}
\hline\hline
$m_\pi^2({\rm GeV}^2)$ & $p$ & $\Sigma^+$ & $n$ & $\Xi^0$ & $\Sigma^-$ & $\Xi^-$ \\
\hline
  0.1254(15) &  0.40(8) &  0.36(5) &  0.46(11) &  0.32(2) &  0.37(8) &  0.29(4) \\
  0.245(2) &  0.29(5) &  0.28(4) &  0.33(5) &  0.31(3) &  0.29(7) &  0.23(4) \\
  0.3487(17) &  0.33(2) &  0.328(20) &  0.29(2) &  0.283(15) &  0.40(4) &  0.35(3) \\
  0.462(2) &  0.27(2) &  0.27(2) &  0.25(2) &  0.246(19) &  0.31(4) &  0.28(3) \\
\hline\hline
\end{tabular}
\end{center}
\caption{\label{tab:Baryon-r2m}Mean-squared magnetic radii for octet baryons}
\end{table}

%%%%%%%%%%%%%%%%%%%%%%%
%%%%%%%%%% G_E/G_M %%%%%%%%%%%%%

\begin{table}[h]
\begin{center}
\begin{tabular}{c|ccccccccc}
\hline\hline
$m_\pi^2({\rm GeV}^2)$ & $p$ & $n$ & $\Sigma^+$ & $\Sigma^-$ & $\Xi^-$ & $\Xi^0$ \\
\hline
  0.1256(15) &  2.4(2) &  $-$1.59(17) &  2.27(16) & $-$0.88(8) & $-$0.71(3) &  $-$1.32(4) \\
 0.246(2) &  2.46(11) &  $-$1.66(8) &  2.47(10) & $-$0.87(6) & $-$0.77(4) &  $-$1.52(4) \\
  0.3493(17) &  2.61(7) &  $-$1.59(4) &  2.58(6) & $-$0.99(4) & $-$0.90(3) &    $-$1.53(3) \\
  0.463(3) &  2.64(8) &  $-$1.62(5) &  2.63(7) &  $-$1.02(4) & $-$0.97(4) &  $-$1.59(3) \\
  \hline
\hline & &&NNLO\\  \hline
$C=1.2$& 3.07(18)[1.72]&  $-$2.18(12)[1.75]&  2.94(14)[1.54]&  $-$1.45(8)[2.85]& $-$0.75(4)[2.54]&  $-$1.18(5)[1.59]\\
$C=1.2$ 2pt& 2.6(5)[n/a]&  $-$1.6(3)[n/a]&  2.6(3)[n/a]&  $-$1.38(16)[n/a]& $-$0.84(7)[n/a]&  $-$1.07(9)[n/a]\\
\hline
Exp't &  2.7928474(3) &  $-$1.9130427(5) &  2.458(10) &  $-$1.16(3) & $-$0.651(3) &  $-$1.250(14) \\
\hline\hline
\end{tabular}
\end{center}
\caption{\label{tab:Baryon-mus-GEM}$\mu_B$ (in units of $\mu_N$) for octet baryons from fitting the ratio $\frac{G_M}{G_E}$. The extrapolations are done with HBXPT according to Eq.~\ref{eq:ChPT-mu-linear2} with $C=1.2(2)$ and $C=0$ (if specified).}
\end{table}

\section{Conclusions}\label{sec:conclusion}

In this work, we study  of the electromagnetic form factors of the nucleon, Sigma and cascade baryons and discuss their momentum dependence and the effects of SU(3) flavor symmetry.

We re-examine the dipole fitting form as candidate to describe the momentum dependence of these form factors. In most cases this is adequate to fit the data,
however in the case of of the electric form factor $G_E$  a new fit form motivated by the phenomenological fit forms used by Refs.~\cite{Kelly:2004hm,Arrington:2007ux} is introduced. This form has only one new parameter relative to the standard dipole fit.

We study the $Q^2$ dependence of the form-factor ratios $\mu_B G_{E,B}/G_{M,B}$ and the individual form factors $G_{E,B}$ and $G_{M,B}/\mu_B$. In most cases, the pion-mass dependence is small throughout the kinematic region of this calculation. Most of the ratios are below 1, except for the $\Xi^{-}$ case. The values for $\Sigma$ and $\Xi$ hyperons are higher than for the nucleon, and also higher than the phenomenological fits to the experimental nucleon form factor data.

The charge radii are obtained from the modified dipole fit form in Subsec.~\ref{subsec:ChargeRadii}. SU(3) symmetry breaking in the quark sector is relatively small, and there is only mild dependence on the baryon species.
Similar relations can be seen between the SU(3) partners $p$ and $\Sigma^+$ (or $\Sigma^-$ and $\Xi^-$). We use NLO HBXPT to extrapolate the charge radii for the $p$, $\Sigma^+$, $\Sigma^-$ and $\Xi^-$ baryons to the physical limit. The fits work very well for all baryon flavors and are consistent regardless of whether the largest-pion mass ensembles are included. We find that including the decuplet degrees of freedom has no significant effect on the final extrapolated charge radii. The extrapolated electric mean-squared charge radii for the proton and $\Sigma^-$ ($0.54(7)$ and $0.32(2)$~fm$^2$) are 3.4 and 4.3$\sigma$ away from the experimental ones. The electric charge radii of the $\Sigma^+$ and $\Xi^-$ are predicted to be 0.67(5) and 0.306(15)~fm$^2$ respectively.

The magnetic moments are obtained by two different approaches: traditional dipole fits to the magnetic form factors to get magnetic moments, and a linear-ansatz fit to the form factor ratio $G_{M,B}/G_{E,B}$. We find the extracted magnetic moments are consistent between these two methods indicating that systematic errors due to $Q^2$ extrapolation are under control. The magnetic radii are approximately the same as the electric radii; the maximum deviation is about 10\%. For the chiral extrapolation of the magnetic moments we use NLO HBXPT and  find that such fits work only if we add terms with the functional form dictated by NNLO
but with free coefficients. Using all ensembles, we find that the magnetic moments for the 6 octet baryons we studied are within a few standard deviations of the experimental values.

In all cases an unknown systematic error needs to be assigned due
to volume, lattices spacing, and chiral extrapolations\footnote{Mixed-action HBXPT formulas can be obtained using PQCD results~\cite{Arndt:2003ww,Tiburzi:2004mv}, following the suggestion in Ref.~\cite{Chen:2007ug}. However, given the computed value of the mixed pion mass~\cite{Orginos:2007tw}, which is found to be smaller than our pion masses, such effects are expected to be sub-leading. Such behavior was also observed in the LHPC mixed-action spectroscopy calculation on these lattices~\cite{WalkerLoud:2008bp}.}.
In addition, in all cases
we have ignored the coupling of the electromagnetic current to vacuum polarization loops. This last omission may be justified on the basis of recent
calculations of such disconnected diagrams for the proton form factors\cite{Babich:2007jg,Lewis:2002ix}. Addressing the above systematics is the focus of future work
using high statistics improved Wilson fermion calculations on anisotropic lattices\cite{Edwards:2008ja,Lin:2008pr}.

\section*{Acknowledgements}
We thank the LHPC and NPLQCD collaborations for their light- and strange-quark forward and (some of the) backward propagators. We would also like to thank Brian C. Tiburzi for detailed discussions on SU(3) heavy-baryon chiral perturbation theory for charge radii and magnetic moments. These calculations were performed using the Chroma software suite\cite{Edwards:2004sx} on clusters at Jefferson Laboratory using time awarded under the SciDAC Initiative. This work is supported by Jefferson Science Associates, LLC under U.S. DOE Contract No. DE-AC05-06OR23177. The U.S. Government retains a non-exclusive, paid-up, irrevocable, world-wide license to publish or reproduce this manuscript for U.S. Government purposes. KO is supported in part by the Jeffress Memorial Trust grant J-813, DOE OJI grant DE-FG02-07ER41527 and DOE grant DE-FG02-04ER41302.
%%%%%%%%%%%%%%%%%%%%%%%%%%%%%%%%%%%%%%%%%%%%%%%%%%%%%%%%%%%%%%%%%%%%%%%%%%%%%%%%%%%%%%%%%%%%%%%%%%%%%%%%%%%%%%%%%%%%%%%%%

\section*{Appendix}
In this section, we collect the details of the bare form factors for each pion-mass ensemble and momentum transfer in the Tables. Note that the magnetic form factors are naturally converted into units of $\frac{e}{2m_B}$, where the $m_B$ is the baryon mass calculated in its corresponding pion sea.
\begin{table}[h]
\begin{center}
\begin{tabular}{c|ccccccccc}
\hline\hline
$Q^2(\mbox{GeV}^2)$ & $F_1^u$ & $F_1^d$ & $F_1^p$ & $F_1^n$ & $F_2^u$ & $F_2^d$ & $F_2^p$ & $F_2^n$ \\
\hline
 0 &  1.80(6) &  0.91(3) &  0.90(3) &  0.0074(11) & n/a & n/a & n/a & n/a \\
  0.2358(5) &  1.27(4) &  0.60(2) &  0.65(2) & $-$0.022(10) &  0.73(15) &  $-$1.04(10) &  0.83(10) & $-$0.94(7) \\
  0.4537(16) &  0.98(5) &  0.43(2) &  0.51(3) & $-$0.040(13) &  0.68(11) & $-$0.68(7) &  0.68(7) & $-$0.68(5) \\
  0.657(3) &  0.77(6) &  0.33(3) &  0.40(4) & $-$0.036(18) &  0.62(14) & $-$0.53(9) &  0.59(9) & $-$0.56(7) \\
  0.849(5) &  0.60(8) &  0.26(4) &  0.32(5) & $-$0.03(2) &  0.18(16) & $-$0.33(13) &  0.23(11) & $-$0.28(9) \\
%  1.030(7) &    1(7) &    0.2(17) &    0(4) &    $-$0.1(15) &    0.3(19) &      0(110) &    0(3) &    0(7) \\
%  1.203(9) &    1(5) &    0(2) &    0(2) &    $-$0.1(10) &    0(5) &    0(3) &    0(4) &    0(3) \\
\hline\hline
\end{tabular}
\end{center}
\caption{\label{tab:NuclFsM1}Bare nucleon form factors as a function of $Q^2$ at $m_\pi=$ 0.354(2) GeV}
\end{table}

\begin{table}[h]
\begin{center}
\begin{tabular}{c|ccccccccc}
\hline\hline
$Q^2(\mbox{GeV}^2)$ & $F_1^u$ & $F_1^d$ & $F_1^p$ & $F_1^n$ & $F_2^u$ & $F_2^d$ & $F_2^p$ & $F_2^n$ \\
\hline
 0 &  1.807(2) &  0.9159(14) &  0.8997(13) &  0.0081(8) & n/a & n/a & n/a & n/a \\
  0.2379(3) &  1.323(17) &  0.642(9) &  0.668(10) & $-$0.013(6) &  0.84(10) &  $-$1.16(7) &  0.95(6) &  $-$1.05(4) \\
  0.4609(11) &  1.05(3) &  0.495(14) &  0.535(15) & $-$0.020(8) &  0.70(9) & $-$0.86(6) &  0.75(5) & $-$0.80(4) \\
  0.671(2) &  0.90(5) &  0.40(2) &  0.47(3) & $-$0.032(12) &  0.55(10) & $-$0.73(7) &  0.61(7) & $-$0.67(5) \\
  0.872(3) &  0.70(5) &  0.30(3) &  0.37(3) & $-$0.035(15) &  0.45(13) & $-$0.51(7) &  0.47(8) & $-$0.49(5) \\
  1.062(5) &  0.68(7) &  0.29(3) &  0.36(4) & $-$0.035(15) &  0.45(12) & $-$0.46(7) &  0.45(8) & $-$0.45(6) \\
%  1.245(6) &    1(3) &    0.3(11) &    0.4(18) &   0.0(4) &    1(7) &    0(3) &    1(4) &    0(2) \\
\hline\hline
\end{tabular}
\end{center}
\caption{\label{tab:NuclFsM2}Bare nucleon form factors as a function of $Q^2$ at $m_\pi=$ 0.495(2) GeV}
\end{table}

\begin{table}[h]
\begin{center}
\begin{tabular}{c|ccccccccc}
\hline\hline
$Q^2(\mbox{GeV}^2)$ & $F_1^u$ & $F_1^d$ & $F_1^p$ & $F_1^n$ & $F_2^u$ & $F_2^d$ & $F_2^p$ & $F_2^n$ \\
\hline
 0 &  1.7884(14) &  0.9057(8) &  0.8903(8) &  0.0077(4) & n/a & n/a & n/a & n/a \\
  0.23874(19) &  1.327(8) &  0.636(4) &  0.672(5) & $-$0.018(3) &  1.00(5) &  $-$1.07(3) &  1.03(3) &  $-$1.05(2) \\
  0.4639(7) &  1.037(12) &  0.472(7) &  0.534(7) & $-$0.031(4) &  0.80(4) & $-$0.85(3) &  0.82(3) & $-$0.832(19) \\
  0.6776(14) &  0.848(18) &  0.367(9) &  0.443(10) & $-$0.038(6) &  0.64(5) & $-$0.70(3) &  0.66(3) & $-$0.68(2) \\
  0.881(2) &  0.70(2) &  0.298(11) &  0.368(12) & $-$0.035(6) &  0.43(5) & $-$0.55(4) &  0.47(3) & $-$0.51(3) \\
  1.077(3) &  0.60(2) &  0.239(12) &  0.319(14) & $-$0.040(6) &  0.37(4) & $-$0.50(4) &  0.41(3) & $-$0.45(2) \\
  1.264(4) &  0.54(4) &  0.202(18) &  0.29(2) & $-$0.044(9) &  0.33(6) & $-$0.48(5) &  0.38(4) & $-$0.43(4) \\
\hline\hline
\end{tabular}
\end{center}
\caption{\label{tab:NuclFsM3}Bare nucleon form factors as a function of $Q^2$ at $m_\pi=$ 0.5911(15) GeV}
\end{table}

\begin{table}[h]
\begin{center}
\begin{tabular}{c|ccccccccc}
\hline\hline
$Q^2(\mbox{GeV}^2)$ & $F_1^u$ & $F_1^d$ & $F_1^p$ & $F_1^n$ & $F_2^u$ & $F_2^d$ & $F_2^p$ & $F_2^n$ \\
\hline
 0 &  1.79(3) &  0.904(16) &  0.889(16) &  0.0072(5) & n/a & n/a & n/a & n/a \\
  0.23991(12) &  1.36(3) &  0.656(13) &  0.690(13) & $-$0.017(3) &  1.16(6) &  $-$1.10(4) &  1.14(4) &  $-$1.12(3) \\
  0.4681(4) &  1.09(2) &  0.494(12) &  0.559(12) & $-$0.032(4) &  0.91(5) & $-$0.87(4) &  0.90(3) & $-$0.89(3) \\
  0.6862(9) &  0.89(2) &  0.384(12) &  0.463(13) & $-$0.040(5) &  0.75(5) & $-$0.70(4) &  0.73(3) & $-$0.71(3) \\
  0.8953(15) &  0.77(3) &  0.329(14) &  0.406(16) & $-$0.038(6) &  0.67(7) & $-$0.63(4) &  0.66(5) & $-$0.64(4) \\
  1.097(2) &  0.66(3) &  0.264(15) &  0.354(17) & $-$0.045(7) &  0.56(6) & $-$0.53(4) &  0.55(4) & $-$0.54(3) \\
  1.291(3) &  0.57(5) &  0.211(21) &  0.31(2) & $-$0.048(8) &  0.48(7) & $-$0.45(5) &  0.47(5) & $-$0.46(4) \\
\hline\hline
\end{tabular}
\end{center}
\caption{\label{tab:NuclFsM4}Bare nucleon form factors as a function of $Q^2$ at $m_\pi=$ 0.6803(18) GeV}
\end{table}

\begin{table}[h]
\begin{center}
\begin{tabular}{c|ccccccccc}
\hline\hline
$Q^2(\mbox{GeV}^2)$ & $F_1^l$ & $F_1^s$ & $F_1^{\Sigma^+}$ & $F_1^{\Sigma^-}$ & $F_2^l$ & $F_2^s$ & $F_2^{\Sigma^+}$ & $F_2^{\Sigma^-}$ \\
\hline
 0 &  1.80(4) &  0.895(20) &  0.902(21) & $-$0.898(21) & n/a & n/a & n/a & n/a \\
  0.23857(20) &  1.28(3) &  0.650(17) &  0.637(18) & $-$0.644(16) &  1.14(13) &  $-$1.01(6) &  1.10(8) & $-$0.04(5) \\
  0.4633(7) &  0.96(3) &  0.492(17) &  0.478(18) & $-$0.485(15) &  0.91(9) & $-$0.81(6) &  0.88(6) & $-$0.03(4) \\
  0.6764(15) &  0.76(4) &  0.389(20) &  0.38(2) & $-$0.384(16) &  0.74(11) & $-$0.69(6) &  0.72(7) & $-$0.02(4) \\
  0.879(2) &  0.63(4) &  0.32(2) &  0.31(3) & $-$0.316(19) &  0.54(12) & $-$0.46(7) &  0.51(8) & $-$0.02(5) \\
  1.074(3) &  0.54(4) &  0.26(2) &  0.27(3) & $-$0.267(19) &  0.48(9) & $-$0.40(7) &  0.45(6) & $-$0.02(4) \\
%  1.260(4) &      1(100) &    0(4) &    0(6) &    0(5) &      1(110) &    0(7) &    1(6) &    0(5) \\
\hline\hline
\end{tabular}
\end{center}
\caption{\label{tab:SigFsM1}Bare Sigma form factors as a function of $Q^2$ at $m_\pi=$ 0.354(2) GeV}
\end{table}

\begin{table}[h]
\begin{center}
\begin{tabular}{c|ccccccccc}
\hline\hline
$Q^2(\mbox{GeV}^2)$ & $F_1^l$ & $F_1^s$ & $F_1^{\Sigma^+}$ & $F_1^{\Sigma^-}$ & $F_2^l$ & $F_2^s$ & $F_2^{\Sigma^+}$ & $F_2^{\Sigma^-}$ \\
\hline
 0 &  1.801(2) &  0.9021(14) &  0.8999(13) & $-$0.9010(11) & n/a & n/a & n/a & n/a \\
  0.2392(2) &  1.326(14) &  0.672(7) &  0.660(8) & $-$0.666(6) &  1.09(9) &  $-$1.17(5) &  1.11(6) &  0.03(4) \\
  0.4657(8) &  1.039(21) &  0.530(11) &  0.516(12) & $-$0.523(9) &  0.85(8) & $-$0.94(5) &  0.88(5) &  0.03(3) \\
  0.6812(16) &  0.88(3) &  0.441(17) &  0.437(18) & $-$0.439(15) &  0.65(9) & $-$0.80(5) &  0.70(6) &  0.05(4) \\
  0.887(3) &  0.71(4) &  0.340(21) &  0.36(2) & $-$0.351(17) &  0.57(11) & $-$0.60(6) &  0.58(7) &  0.01(5) \\
  1.085(4) &  0.66(4) &  0.31(2) &  0.33(2) & $-$0.32(2) &  0.53(10) & $-$0.57(7) &  0.54(6) &  0.01(4) \\
  1.275(5) &  0.64(7) &  0.32(4) &  0.32(4) & $-$0.32(4) &  0.42(13) & $-$0.52(9) &  0.45(9) &  0.03(6) \\
\hline\hline
\end{tabular}
\end{center}
\caption{\label{tab:SigFsM2}Bare Sigma form factors as a function of $Q^2$ at $m_\pi=$ 0.495(2) GeV}
\end{table}

\begin{table}[h]
\begin{center}
\begin{tabular}{c|ccccccccc}
\hline\hline
$Q^2(\mbox{GeV}^2)$ & $F_1^l$ & $F_1^s$ & $F_1^{\Sigma^+}$ & $F_1^{\Sigma^-}$ & $F_2^l$ & $F_2^s$ & $F_2^{\Sigma^+}$ & $F_2^{\Sigma^-}$ \\
\hline
 0 &  1.7853(14) &  0.8953(8) &  0.8917(8) & $-$0.8935(6) & n/a & n/a & n/a & n/a \\
  0.23954(11) &  1.322(7) &  0.653(4) &  0.664(4) & $-$0.658(3) &  1.13(5) &  $-$1.12(3) &  1.12(3) &  0.00(2) \\
  0.4668(4) &  1.032(11) &  0.497(6) &  0.522(6) & $-$0.510(5) &  0.90(4) & $-$0.89(2) &  0.90(3) & $-$0.002(18) \\
  0.6834(8) &  0.837(15) &  0.391(8) &  0.428(9) & $-$0.409(7) &  0.71(4) & $-$0.75(3) &  0.72(3) &  0.011(19) \\
  0.8909(13) &  0.697(18) &  0.321(10) &  0.358(11) & $-$0.339(8) &  0.51(5) & $-$0.59(3) &  0.54(3) &  0.03(2) \\
  1.0902(19) &  0.590(21) &  0.261(11) &  0.306(12) & $-$0.284(10) &  0.43(4) & $-$0.53(3) &  0.47(3) &  0.030(19) \\
  1.282(3) &  0.52(3) &  0.218(15) &  0.272(16) & $-$0.245(14) &  0.38(5) & $-$0.49(4) &  0.42(3) &  0.04(2) \\
\hline\hline
\end{tabular}
\end{center}
\caption{\label{tab:SigFsM3}Bare Sigma form factors as a function of $Q^2$ at $m_\pi=$ 0.5911(15) GeV}
\end{table}

\begin{table}[h]
\begin{center}
\begin{tabular}{c|ccccccccc}
\hline\hline
$Q^2(\mbox{GeV}^2)$ & $F_1^l$ & $F_1^s$ & $F_1^{\Sigma^+}$ & $F_1^{\Sigma^-}$ & $F_2^l$ & $F_2^s$ & $F_2^{\Sigma^+}$ & $F_2^{\Sigma^-}$ \\
\hline
 0 &  1.7744(16) &  0.8930(9) &  0.8853(9) & $-$0.8891(8) & n/a & n/a & n/a & n/a \\
  0.24018(11) &  1.353(8) &  0.658(4) &  0.683(5) & $-$0.670(4) &  1.20(6) &  $-$1.11(3) &  1.17(4) & $-$0.03(2) \\
  0.4691(4) &  1.075(12) &  0.502(7) &  0.550(7) & $-$0.526(6) &  0.95(5) & $-$0.89(3) &  0.93(3) & $-$0.021(20) \\
  0.6882(8) &  0.876(17) &  0.393(9) &  0.453(9) & $-$0.423(8) &  0.78(5) & $-$0.71(3) &  0.76(3) & $-$0.02(2) \\
  0.8987(13) &  0.76(2) &  0.336(11) &  0.397(13) & $-$0.366(10) &  0.68(6) & $-$0.64(4) &  0.67(4) & $-$0.01(3) \\
  1.1014(19) &  0.65(3) &  0.271(13) &  0.343(14) & $-$0.307(12) &  0.57(5) & $-$0.53(4) &  0.56(4) & $-$0.01(2) \\
  1.297(3) &  0.55(4) &  0.219(18) &  0.294(20) & $-$0.257(17) &  0.49(7) & $-$0.46(5) &  0.48(5) & $-$0.01(3) \\
\hline\hline
\end{tabular}
\end{center}
\caption{\label{tab:SigFsM4}Bare Sigma form factors as a function of $Q^2$ at $m_\pi=$ 0.6803(18) GeV}
\end{table}

\begin{table}[h]
\begin{center}
\begin{tabular}{c|ccccccccc}
\hline\hline
$Q^2(\mbox{GeV}^2)$ & $F_1^l$ & $F_1^s$ & $F_1^{\Xi^-}$ & $F_1^{\Xi^0}$ & $F_2^l$ & $F_2^s$ & $F_2^{\Xi^-}$ & $F_2^{\Xi^0}$ \\
\hline
 0 &  0.914(14) &  1.78(3) & $-$0.897(14) &  0.0169(5) & n/a & n/a & n/a & n/a \\
  0.23945(10) &  0.612(11) &  1.36(2) & $-$0.657(11) & $-$0.045(3) &  $-$1.11(4) &  1.01(5) &  0.04(2) &  $-$1.08(3) \\
  0.4665(4) &  0.435(10) &  1.080(20) & $-$0.505(9) & $-$0.070(5) & $-$0.81(3) &  0.78(4) &  0.012(19) & $-$0.80(2) \\
  0.6828(7) &  0.316(10) &  0.88(2) & $-$0.398(10) & $-$0.083(6) & $-$0.66(3) &  0.62(4) &  0.013(20) & $-$0.64(2) \\
  0.8898(12) &  0.257(13) &  0.75(2) & $-$0.336(11) & $-$0.079(8) & $-$0.48(4) &  0.53(5) & $-$0.02(2) & $-$0.50(3) \\
  1.0886(17) &  0.198(11) &  0.65(2) & $-$0.281(11) & $-$0.083(7) & $-$0.40(3) &  0.43(5) & $-$0.013(20) & $-$0.41(2) \\
  1.280(2) &  0.157(14) &  0.57(3) & $-$0.241(14) & $-$0.084(9) & $-$0.33(3) &  0.36(5) & $-$0.01(2) & $-$0.34(3) \\
\hline\hline
\end{tabular}
\end{center}
\caption{\label{tab:XiFsM1}Bare cascade form factors as a function of $Q^2$ at $m_\pi=$ 0.354(2) GeV}
\end{table}

\begin{table}[h]
\begin{center}
\begin{tabular}{c|ccccccccc}
\hline\hline
$Q^2(\mbox{GeV}^2)$ & $F_1^l$ & $F_1^s$ & $F_1^{\Xi^-}$ & $F_1^{\Xi^0}$ & $F_2^l$ & $F_2^s$ & $F_2^{\Xi^-}$ & $F_2^{\Xi^0}$ \\
\hline
 0 &  0.9110(9) &  1.7743(16) & $-$0.8951(8) &  0.0159(5) & n/a & n/a & n/a & n/a \\
  0.23978(13) &  0.643(5) &  1.364(8) & $-$0.669(4) & $-$0.026(3) &  $-$1.19(4) &  0.99(6) &  0.07(3) &  $-$1.13(3) \\
  0.4676(5) &  0.484(8) &  1.101(14) & $-$0.528(7) & $-$0.044(4) & $-$0.88(3) &  0.81(5) &  0.02(2) & $-$0.86(2) \\
  0.6852(10) &  0.386(11) &  0.936(21) & $-$0.441(10) & $-$0.054(6) & $-$0.72(4) &  0.65(6) &  0.02(3) & $-$0.70(2) \\
  0.8937(16) &  0.306(14) &  0.77(3) & $-$0.358(12) & $-$0.052(8) & $-$0.55(5) &  0.51(8) &  0.01(3) & $-$0.54(3) \\
  1.094(2) &  0.271(16) &  0.70(3) & $-$0.323(15) & $-$0.052(8) & $-$0.46(4) &  0.46(7) &   0.00(3) & $-$0.46(3) \\
  1.288(3) &  0.26(2) &  0.67(4) & $-$0.31(2) & $-$0.051(10) & $-$0.43(5) &  0.39(8) &  0.01(3) & $-$0.41(4) \\
\hline\hline
\end{tabular}
\end{center}
\caption{\label{tab:XiFsM2}Bare cascade form factors as a function of $Q^2$ at $m_\pi=$ 0.495(2) GeV}
\end{table}

\begin{table}[h]
\begin{center}
\begin{tabular}{c|ccccccccc}
\hline\hline
$Q^2(\mbox{GeV}^2)$ & $F_1^l$ & $F_1^s$ & $F_1^{\Xi^-}$ & $F_1^{\Xi^0}$ & $F_2^l$ & $F_2^s$ & $F_2^{\Xi^-}$ & $F_2^{\Xi^0}$ \\
\hline
 0 &  0.9034(6) &  1.7660(11) & $-$0.8898(5) &  0.0136(3) & n/a & n/a & n/a & n/a \\
  0.23991(8) &  0.636(3) &  1.348(5) & $-$0.662(3) & $-$0.0253(17) &  $-$1.13(2) &  1.06(4) &  0.023(16) &  $-$1.110(17) \\
  0.4681(3) &  0.473(5) &  1.072(9) & $-$0.515(4) & $-$0.042(3) & $-$0.879(21) &  0.86(3) &  0.007(14) & $-$0.872(14) \\
  0.6862(6) &  0.365(7) &  0.880(13) & $-$0.415(6) & $-$0.050(4) & $-$0.72(2) &  0.69(3) &  0.008(15) & $-$0.708(16) \\
  0.8954(10) &  0.295(8) &  0.746(16) & $-$0.347(7) & $-$0.052(4) & $-$0.58(3) &  0.51(4) &  0.023(18) & $-$0.557(19) \\
  1.0966(14) &  0.236(9) &  0.632(18) & $-$0.289(8) & $-$0.053(4) & $-$0.51(3) &  0.43(3) &  0.026(16) & $-$0.482(18) \\
  1.2909(19) &  0.195(11) &  0.55(2) & $-$0.248(11) & $-$0.053(6) & $-$0.46(3) &  0.36(4) &  0.033(18) & $-$0.42(2) \\
\hline\hline
\end{tabular}
\end{center}
\caption{\label{tab:XiFsM3}Bare cascade form factors as a function of $Q^2$ at $m_\pi=$ 0.5911(15) GeV}
\end{table}

\begin{table}[h]
\begin{center}
\begin{tabular}{c|ccccccccc}
\hline\hline
$Q^2(\mbox{GeV}^2)$ & $F_1^l$ & $F_1^s$ & $F_1^{\Xi^-}$ & $F_1^{\Xi^0}$ & $F_2^l$ & $F_2^s$ & $F_2^{\Xi^-}$ & $F_2^{\Xi^0}$ \\
\hline
 0 &  0.8977(9) &  1.7636(15) & $-$0.8871(7) &  0.0106(4) & n/a & n/a & n/a & n/a \\
  0.24035(9) &  0.651(4) &  1.363(7) & $-$0.671(3) & $-$0.021(2) &  $-$1.12(3) &  1.16(5) & $-$0.01(2) &  $-$1.14(2) \\
  0.4697(3) &  0.491(6) &  1.094(11) & $-$0.528(5) & $-$0.037(3) & $-$0.89(3) &  0.93(4) & $-$0.013(19) & $-$0.899(20) \\
  0.6895(7) &  0.380(9) &  0.898(15) & $-$0.426(7) & $-$0.046(4) & $-$0.71(3) &  0.76(4) & $-$0.019(20) & $-$0.73(2) \\
  0.9007(11) &  0.321(10) &  0.787(21) & $-$0.369(10) & $-$0.048(6) & $-$0.63(4) &  0.67(6) & $-$0.01(2) & $-$0.64(3) \\
  1.1044(16) &  0.257(12) &  0.67(2) & $-$0.310(11) & $-$0.052(6) & $-$0.53(4) &  0.56(5) & $-$0.01(2) & $-$0.54(3) \\
  1.301(2) &  0.204(15) &  0.57(3) & $-$0.259(16) & $-$0.055(7) & $-$0.45(4) &  0.49(6) & $-$0.02(3) & $-$0.46(4) \\
\hline\hline
\end{tabular}
\end{center}
\caption{\label{tab:XiFsM4}Bare cascade form factors as a function of $Q^2$ at $m_\pi=$ 0.6803(18) GeV}
\end{table}

%%%%%%%%%%%%%%%%%%%%%%%%%%%%%%%%%%%%%%%%%%%%%%%%%%%%%%%%%%%%%%%%%%%%%%%%%%%%%%%%%%%%
\begin{table}[h]
\begin{center}
\begin{tabular}{c|ccccccccc}
\hline\hline
$Q^2(\mbox{GeV}^2)$ & $G_E^u$ & $G_E^d$ & $G_E^p$ & $G_M^u$ & $G_M^d$ & $G_M^p$ & $G_M^n$ \\
\hline
 0 &  1.80(6) &  0.91(3) &  0.90(3) & n/a & n/a & n/a & n/a \\
  0.2358(5) &  1.30(5) &  0.555(21) &  0.61(2) &  2.00(16) & $-$0.44(10) &  1.48(11) & $-$0.96(8) \\
  0.4537(16) &  1.03(5) &  0.37(2) &  0.45(2) &  1.66(14) & $-$0.26(7) &  1.19(9) & $-$0.72(6) \\
  0.657(3) &  0.84(7) &  0.26(3) &  0.33(3) &  1.39(18) & $-$0.20(9) &  0.99(11) & $-$0.60(7) \\
  0.849(5) &  0.63(9) &  0.20(4) &  0.28(4) &  0.78(20) & n/a &  0.55(13) & $-$0.31(10) \\
\hline\hline
\end{tabular}
\end{center}
\caption{\label{tab:NuclGsM1}Bare nucleon form factors as a function of $Q^2$ at $m_\pi=$ 0.354(2) GeV}
\end{table}

\begin{table}[h]
\begin{center}
\begin{tabular}{c|ccccccccc}
\hline\hline
$Q^2(\mbox{GeV}^2)$ & $G_E^u$ & $G_E^d$ & $G_E^p$ & $G_M^u$ & $G_M^d$ & $G_M^p$ & $G_M^n$ \\
\hline
 0 &  1.807(2) &  0.9159(14) &  0.8997(13) & n/a & n/a & n/a & n/a \\
  0.2379(3) &  1.353(18) &  0.600(10) &  0.634(10) &  2.16(10) & $-$0.52(7) &  1.62(6) &  $-$1.07(4) \\
  0.4609(11) &  1.10(3) &  0.436(14) &  0.483(14) &  1.75(10) & $-$0.36(6) &  1.29(6) & $-$0.82(4) \\
  0.671(2) &  0.96(5) &  0.33(2) &  0.41(2) &  1.46(13) & $-$0.33(7) &  1.08(8) & $-$0.70(5) \\
  0.872(3) &  0.76(6) &  0.23(3) &  0.31(3) &  1.15(15) & $-$0.21(7) &  0.83(10) & $-$0.52(6) \\
  1.062(5) &  0.75(8) &  0.21(3) &  0.28(3) &  1.13(16) & $-$0.17(7) &  0.81(11) & $-$0.49(7) \\
%  1.245(6) & n/a & n/a & n/a & n/a & n/a & n/a & n/a \\
\hline\hline
\end{tabular}
\end{center}
\caption{\label{tab:NuclGsM2}Bare nucleon form factors as a function of $Q^2$ at $m_\pi=$ 0.495(2) GeV}
\end{table}

\begin{table}[h]
\begin{center}
\begin{tabular}{c|ccccccccc}
\hline\hline
$Q^2(\mbox{GeV}^2)$ & $G_E^u$ & $G_E^d$ & $G_E^p$ & $G_M^u$ & $G_M^d$ & $G_M^p$ & $G_M^n$ \\
\hline
 0 &  1.7884(14) &  0.9057(8) &  0.8903(8) & n/a & n/a & n/a & n/a \\
  0.23874(19) &  1.359(9) &  0.602(4) &  0.639(4) &  2.33(5) & $-$0.44(3) &  1.70(3) &  $-$1.07(2) \\
  0.4639(7) &  1.087(13) &  0.420(7) &  0.483(6) &  1.84(5) & $-$0.37(3) &  1.35(3) & $-$0.863(21) \\
  0.6776(14) &  0.906(19) &  0.303(9) &  0.383(9) &  1.48(5) & $-$0.34(3) &  1.10(4) & $-$0.72(2) \\
  0.881(2) &  0.75(2) &  0.233(11) &  0.312(11) &  1.14(6) & $-$0.25(4) &  0.84(4) & $-$0.55(3) \\
  1.077(3) &  0.65(3) &  0.167(12) &  0.259(12) &  0.97(6) & $-$0.26(3) &  0.73(4) & $-$0.49(3) \\
  1.264(4) &  0.59(5) &  0.122(17) &  0.225(18) &  0.86(8) & $-$0.27(5) &  0.67(6) & $-$0.47(4) \\
\hline\hline
\end{tabular}
\end{center}
\caption{\label{tab:NuclGsM3}Bare nucleon form factors as a function of $Q^2$ at $m_\pi=$ 0.5911(15) GeV}
\end{table}

\begin{table}[h]
\begin{center}
\begin{tabular}{c|ccccccccc}
\hline\hline
$Q^2(\mbox{GeV}^2)$ & $G_E^u$ & $G_E^d$ & $G_E^p$ & $G_M^u$ & $G_M^d$ & $G_M^p$ & $G_M^n$ \\
\hline
 0 &  1.79(3) &  0.904(16) &  0.889(16) & n/a & n/a & n/a & n/a \\
  0.23991(12) &  1.39(3) &  0.626(12) &  0.659(13) &  2.52(7) & $-$0.45(3) &  1.83(5) &  $-$1.14(3) \\
  0.4681(4) &  1.13(2) &  0.448(11) &  0.512(12) &  2.00(6) & $-$0.38(3) &  1.46(4) & $-$0.92(3) \\
  0.6862(9) &  0.94(3) &  0.330(12) &  0.407(12) &  1.63(6) & $-$0.31(3) &  1.19(4) & $-$0.75(3) \\
  0.8953(15) &  0.84(3) &  0.266(13) &  0.339(13) &  1.45(9) & $-$0.30(4) &  1.06(6) & $-$0.68(4) \\
  1.097(2) &  0.73(4) &  0.199(13) &  0.287(14) &  1.22(8) & $-$0.26(4) &  0.90(5) & $-$0.58(4) \\
  1.291(3) &  0.63(5) &  0.145(18) &  0.239(20) &  1.04(11) & $-$0.24(5) &  0.78(7) & $-$0.51(5) \\
\hline\hline
\end{tabular}
\end{center}
\caption{\label{tab:NuclGsM4}Bare nucleon form factors as a function of $Q^2$ at $m_\pi=$ 0.6803(18) GeV}
\end{table}

\begin{table}[h]
\begin{center}
\begin{tabular}{c|ccccccccc}
\hline\hline
$Q^2(\mbox{GeV}^2)$ & $G_E^l$ & $G_E^s$ & $G_E^{\Sigma^+}$ & $G_E^{\Sigma^-}$ & $G_M^l$ & $G_M^s$ & $G_M^{\Sigma^+}$ & $G_M^{\Sigma^-}$ \\
\hline
 0 &  1.80(4) &  0.895(20) &  0.902(21) & $-$0.898(21) & n/a & n/a & n/a & n/a \\
  0.23857(20) &  1.24(3) &  0.683(18) &  0.601(17) & $-$0.642(16) &  2.42(14) & $-$0.36(6) &  1.73(9) & $-$0.69(5) \\
  0.4633(7) &  0.91(3) &  0.544(18) &  0.422(16) & $-$0.483(15) &  1.87(11) & $-$0.31(5) &  1.35(7) & $-$0.52(4) \\
  0.6764(15) &  0.70(3) &  0.45(2) &  0.313(20) & $-$0.383(16) &  1.50(12) & $-$0.30(6) &  1.10(8) & $-$0.40(5) \\
  0.879(2) &  0.56(4) &  0.38(3) &  0.25(2) & $-$0.313(19) &  1.16(14) & $-$0.14(7) &  0.82(10) & $-$0.34(6) \\
  1.074(3) &  0.47(4) &  0.32(3) &  0.20(2) & $-$0.263(19) &  1.01(12) & $-$0.14(7) &  0.72(8) & $-$0.29(5) \\
\hline\hline
\end{tabular}
\end{center}
\caption{\label{tab:SigGsM1}Bare Sigma form factors as a function of $Q^2$ at $m_\pi=$ 0.354(2) GeV}
\end{table}

\begin{table}[h]
\begin{center}
\begin{tabular}{c|ccccccccc}
\hline\hline
$Q^2(\mbox{GeV}^2)$ & $G_E^l$ & $G_E^s$ & $G_E^{\Sigma^+}$ & $G_E^{\Sigma^-}$ & $G_M^l$ & $G_M^s$ & $G_M^{\Sigma^+}$ & $G_M^{\Sigma^-}$ \\
\hline
 0 &  1.801(2) &  0.9021(14) &  0.8999(13) & $-$0.9010(11) & n/a & n/a & n/a & n/a \\
  0.2392(2) &  1.293(14) &  0.707(7) &  0.626(8) & $-$0.667(6) &  2.41(10) & $-$0.50(5) &  1.77(6) & $-$0.64(4) \\
  0.4657(8) &  0.989(20) &  0.585(12) &  0.465(11) & $-$0.525(9) &  1.89(9) & $-$0.41(5) &  1.39(5) & $-$0.49(4) \\
  0.6812(16) &  0.82(3) &  0.510(19) &  0.378(17) & $-$0.444(15) &  1.53(10) & $-$0.36(5) &  1.14(7) & $-$0.39(4) \\
  0.887(3) &  0.65(4) &  0.41(2) &  0.298(21) & $-$0.352(18) &  1.29(13) & $-$0.26(6) &  0.94(8) & $-$0.34(5) \\
  1.085(4) &  0.58(4) &  0.39(3) &  0.26(2) & $-$0.32(2) &  1.18(12) & $-$0.25(6) &  0.87(8) & $-$0.31(5) \\
  1.275(5) &  0.58(6) &  0.40(5) &  0.25(3) & $-$0.33(4) &  1.06(19) & $-$0.21(8) &  0.78(12) & $-$0.29(7) \\
\hline\hline
\end{tabular}
\end{center}
\caption{\label{tab:SigGsM2}Bare Sigma form factors as a function of $Q^2$ at $m_\pi=$ 0.495(2) GeV}
\end{table}

\begin{table}[h]
\begin{center}
\begin{tabular}{c|ccccccccc}
\hline\hline
$Q^2(\mbox{GeV}^2)$ & $G_E^l$ & $G_E^s$ & $G_E^{\Sigma^+}$ & $G_E^{\Sigma^-}$ & $G_M^l$ & $G_M^s$ & $G_M^{\Sigma^+}$ & $G_M^{\Sigma^-}$ \\
\hline
 0 &  1.7853(14) &  0.8953(8) &  0.8917(8) & $-$0.8935(6) & n/a & n/a & n/a & n/a \\
  0.23954(11) &  1.290(7) &  0.684(4) &  0.632(4) & $-$0.658(3) &  2.45(5) & $-$0.46(3) &  1.79(3) & $-$0.66(2) \\
  0.4668(4) &  0.982(10) &  0.546(6) &  0.472(6) & $-$0.509(5) &  1.93(5) & $-$0.40(2) &  1.42(3) & $-$0.512(19) \\
  0.6834(8) &  0.779(15) &  0.451(9) &  0.369(8) & $-$0.410(7) &  1.55(5) & $-$0.36(3) &  1.15(3) & $-$0.399(21) \\
  0.8909(13) &  0.643(17) &  0.384(11) &  0.300(10) & $-$0.342(8) &  1.21(6) & $-$0.27(3) &  0.90(4) & $-$0.31(2) \\
  1.0902(19) &  0.533(19) &  0.329(13) &  0.246(10) & $-$0.287(10) &  1.02(5) & $-$0.27(3) &  0.77(3) & $-$0.25(2) \\
  1.282(3) &  0.46(3) &  0.294(18) &  0.208(14) & $-$0.251(14) &  0.90(7) & $-$0.28(4) &  0.69(4) & $-$0.21(3) \\
\hline\hline
\end{tabular}
\end{center}
\caption{\label{tab:SigGsM3}Bare Sigma form factors as a function of $Q^2$ at $m_\pi=$ 0.5911(15) GeV}
\end{table}

\begin{table}[h]
\begin{center}
\begin{tabular}{c|ccccccccc}
\hline\hline
$Q^2(\mbox{GeV}^2)$ & $G_E^l$ & $G_E^s$ & $G_E^{\Sigma^+}$ & $G_E^{\Sigma^-}$ & $G_M^l$ & $G_M^s$ & $G_M^{\Sigma^+}$ & $G_M^{\Sigma^-}$ \\
\hline
 0 &  1.7744(16) &  0.8930(9) &  0.8853(9) & $-$0.8891(8) & n/a & n/a & n/a & n/a \\
  0.24018(11) &  1.322(8) &  0.687(4) &  0.652(4) & $-$0.669(4) &  2.56(6) & $-$0.45(3) &  1.86(4) & $-$0.70(2) \\
  0.4691(4) &  1.027(12) &  0.546(7) &  0.503(7) & $-$0.525(6) &  2.02(5) & $-$0.38(3) &  1.48(3) & $-$0.55(2) \\
  0.6882(8) &  0.818(16) &  0.446(10) &  0.397(8) & $-$0.421(8) &  1.66(6) & $-$0.32(3) &  1.21(4) & $-$0.45(2) \\
  0.8987(13) &  0.697(20) &  0.397(13) &  0.332(11) & $-$0.365(10) &  1.44(8) & $-$0.30(4) &  1.06(5) & $-$0.38(3) \\
  1.1014(19) &  0.58(2) &  0.334(15) &  0.277(12) & $-$0.305(12) &  1.22(7) & $-$0.26(4) &  0.90(5) & $-$0.32(3) \\
  1.297(3) &  0.48(3) &  0.28(2) &  0.227(16) & $-$0.255(17) &  1.04(9) & $-$0.24(4) &  0.77(6) & $-$0.27(3) \\
\hline\hline
\end{tabular}
\end{center}
\caption{\label{tab:SigGsM4}Bare Sigma form factors as a function of $Q^2$ at $m_\pi=$ 0.6803(18) GeV}
\end{table}

\begin{table}[h]
\begin{center}
\begin{tabular}{c|ccccccccc}
\hline\hline
$Q^2(\mbox{GeV}^2)$ & $G_E^l$ & $G_E^s$ & $G_E^{\Xi^-}$ & $G_E^{\Xi^0}$ & $G_M^l$ & $G_M^s$ & $G_M^{\Xi^-}$ & $G_M^{\Xi^0}$ \\
\hline
 0 &  0.914(14) &  1.78(3) & $-$0.897(14) &  0.0169(5) & n/a & n/a & n/a & n/a \\
  0.23945(10) &  0.644(11) &  1.33(2) & $-$0.658(11) & $-$0.014(3) & $-$0.50(4) &  2.36(6) & $-$0.62(2) &  $-$1.12(3) \\
  0.4665(4) &  0.481(10) &  1.036(19) & $-$0.505(9) & $-$0.025(4) & $-$0.38(3) &  1.86(5) & $-$0.49(2) & $-$0.87(2) \\
  0.6828(7) &  0.370(11) &  0.828(20) & $-$0.399(10) & $-$0.029(6) & $-$0.34(3) &  1.50(5) & $-$0.39(2) & $-$0.72(3) \\
  0.8898(12) &  0.309(14) &  0.69(2) & $-$0.334(11) & $-$0.025(7) & $-$0.23(4) &  1.28(7) & $-$0.35(3) & $-$0.58(3) \\
  1.0886(17) &  0.250(13) &  0.59(2) & $-$0.280(11) & $-$0.030(7) & $-$0.20(3) &  1.08(6) & $-$0.29(2) & $-$0.49(3) \\
  1.280(2) &  0.208(16) &  0.51(3) & $-$0.240(14) & $-$0.031(8) & $-$0.17(4) &  0.93(7) & $-$0.25(3) & $-$0.42(3) \\
\hline\hline
\end{tabular}
\end{center}
\caption{\label{tab:XiGsM1}Bare cascade form factors as a function of $Q^2$ at $m_\pi=$ 0.354(2) GeV}
\end{table}

\begin{table}[h]
\begin{center}
\begin{tabular}{c|ccccccccc}
\hline\hline
$Q^2(\mbox{GeV}^2)$ & $G_E^l$ & $G_E^s$ & $G_E^{\Xi^-}$ & $G_E^{\Xi^0}$ & $G_M^l$ & $G_M^s$ & $G_M^{\Xi^-}$ & $G_M^{\Xi^0}$ \\
\hline
 0 &  0.9110(9) &  1.7743(16) & $-$0.8951(8) &  0.0159(5) & n/a & n/a & n/a & n/a \\
  0.23978(13) &  0.675(5) &  1.336(8) & $-$0.671(4) &  0.005(3) & $-$0.55(4) &  2.36(6) & $-$0.60(3) &  $-$1.15(3) \\
  0.4676(5) &  0.532(8) &  1.057(14) & $-$0.530(7) &  0.002(4) & $-$0.40(3) &  1.91(6) & $-$0.50(2) & $-$0.90(2) \\
  0.6852(10) &  0.443(12) &  0.884(20) & $-$0.443(10) &  0.001(6) & $-$0.34(4) &  1.59(6) & $-$0.42(3) & $-$0.75(3) \\
  0.8937(16) &  0.363(15) &  0.72(3) & $-$0.360(13) &  0.003(8) & $-$0.24(5) &  1.28(9) & $-$0.34(4) & $-$0.59(3) \\
  1.094(2) &  0.329(18) &  0.64(3) & $-$0.323(15) &  0.006(8) & $-$0.19(4) &  1.16(8) & $-$0.32(3) & $-$0.52(3) \\
  1.288(3) &  0.32(3) &  0.61(4) & $-$0.31(2) &  0.010(10) & $-$0.17(5) &  1.06(10) & $-$0.30(4) & $-$0.47(4) \\
\hline\hline
\end{tabular}
\end{center}
\caption{\label{tab:XiGsM2}Bare cascade form factors as a function of $Q^2$ at $m_\pi=$ 0.495(2) GeV}
\end{table}

\begin{table}[h]
\begin{center}
\begin{tabular}{c|ccccccccc}
\hline\hline
$Q^2(\mbox{GeV}^2)$ & $G_E^l$ & $G_E^s$ & $G_E^{\Xi^-}$ & $G_E^{\Xi^0}$ & $G_M^l$ & $G_M^s$ & $G_M^{\Xi^-}$ & $G_M^{\Xi^0}$ \\
\hline
 0 &  0.9034(6) &  1.7660(11) & $-$0.8898(5) &  0.0136(3) & n/a & n/a & n/a & n/a \\
  0.23991(8) &  0.667(3) &  1.320(5) & $-$0.662(3) &  0.0047(17) & $-$0.50(2) &  2.41(4) & $-$0.639(17) &  $-$1.135(17) \\
  0.4681(3) &  0.519(5) &  1.027(8) & $-$0.515(4) &  0.003(3) & $-$0.406(21) &  1.93(3) & $-$0.508(15) & $-$0.914(15) \\
  0.6862(6) &  0.420(7) &  0.826(12) & $-$0.415(6) &  0.004(4) & $-$0.35(2) &  1.57(4) & $-$0.407(16) & $-$0.758(17) \\
  0.8954(10) &  0.353(9) &  0.694(15) & $-$0.349(7) &  0.004(4) & $-$0.29(3) &  1.26(5) & $-$0.324(20) & $-$0.609(21) \\
  1.0966(14) &  0.299(10) &  0.579(17) & $-$0.292(8) &  0.006(4) & $-$0.27(2) &  1.06(4) & $-$0.263(18) & $-$0.535(20) \\
  1.2909(19) &  0.261(13) &  0.50(2) & $-$0.252(11) &  0.009(5) & $-$0.26(3) &  0.91(5) & $-$0.21(2) & $-$0.48(2) \\
\hline\hline
\end{tabular}
\end{center}
\caption{\label{tab:XiGsM3}Bare cascade form factors as a function of $Q^2$ at $m_\pi=$ 0.5911(15) GeV}
\end{table}

\begin{table}[h]
\begin{center}
\begin{tabular}{c|ccccccccc}
\hline\hline
$Q^2(\mbox{GeV}^2)$ & $G_E^l$ & $G_E^s$ & $G_E^{\Xi^-}$ & $G_E^{\Xi^0}$ & $G_M^l$ & $G_M^s$ & $G_M^{\Xi^-}$ & $G_M^{\Xi^0}$ \\
\hline
 0 &  0.8977(9) &  1.7636(15) & $-$0.8871(7) &  0.0106(4) & n/a & n/a & n/a & n/a \\
  0.24035(9) &  0.679(4) &  1.334(7) & $-$0.671(3) &  0.0080(20) & $-$0.47(3) &  2.52(5) & $-$0.68(2) &  $-$1.16(2) \\
  0.4697(3) &  0.534(7) &  1.048(11) & $-$0.528(6) &  0.007(3) & $-$0.40(3) &  2.02(5) & $-$0.541(20) & $-$0.937(21) \\
  0.6895(7) &  0.431(9) &  0.843(15) & $-$0.425(7) &  0.007(4) & $-$0.33(3) &  1.66(5) & $-$0.45(2) & $-$0.77(2) \\
  0.9007(11) &  0.380(12) &  0.724(19) & $-$0.368(10) &  0.012(5) & $-$0.31(4) &  1.46(7) & $-$0.38(3) & $-$0.69(3) \\
  1.1044(16) &  0.318(14) &  0.61(2) & $-$0.308(11) &  0.010(6) & $-$0.27(3) &  1.24(6) & $-$0.32(2) & $-$0.59(3) \\
  1.301(2) &  0.265(19) &  0.51(3) & $-$0.257(16) &  0.008(7) & $-$0.24(4) &  1.07(8) & $-$0.27(3) & $-$0.52(4) \\
\hline\hline
\end{tabular}
\end{center}
\caption{\label{tab:XiGsM4}Bare cascade form factors as a function of $Q^2$ at $m_\pi=$ 0.6803(18) GeV}
\end{table}

\clearpage
%\bibliography{nuc_ref}

\end{document}